\RequirePackage{fix-cm}
\documentclass[twocolumn,epjc3]{svjour3}  
\smartqed  
\RequirePackage{graphicx}
 \RequirePackage{mathptmx}      
\usepackage[utf8x]{inputenc}

\usepackage{subfigure,color,dsfont,slashed,multirow}
\usepackage{fontenc}
\usepackage{pbox}
\usepackage{feynmp}
\usepackage{mathtools}
\usepackage{nccmath}%
\usepackage{ etoolbox, xparse} 
\usepackage{widetext}

\usepackage{array}
\newcolumntype{P}[1]{>{\centering\arraybackslash}p{#1}}

\def\EmissT{\not \! \!  E_{T}}

\def\twoleps{2\ell+\EmissT}
\def\fourleps{4\ell+\EmissT}
\def\sixleps{6\ell+\EmissT}

\newcommand{\newc}{\newcommand}

\newc{\lcal}{\int {\cal L}dt}

\newc{\lsp}{{\widetilde{\chi}^0_1}}
\newc{\niki}{{\widetilde{\chi}^0_2}}
\newc{\nuc}{{\widetilde{\chi}^0_3}}
\newc{\ndort}{{\widetilde{\chi}^0_4}}
\newc{\nbes}{{\widetilde{\chi}^0_5}}
\newc{\nalti}{{\widetilde{\chi}^0_6}}
\newc{\mnbir}{{M_{\widetilde{\chi}^0_1}}}
\newc{\mniki}{{M_{\widetilde{\chi}^0_2}}}
\newc{\mnuc}{{M_{\widetilde{\chi}^0_3}}}
\newc{\mndort}{{M_{\widetilde{\chi}^0_4}}}
\newc{\mnbes}{{M_{\widetilde{\chi}^0_5}}}
\newc{\mnalti}{{M_{\widetilde{\chi}^0_6}}}
\newc{\stauR}{{\widetilde{\tau}_R}}
\newc{\stau}{{\widetilde{\tau}_1}}
\newc{\staut}{{\widetilde{\tau}_2}}
\newc{\mstop}{m_{\widetilde{t}}}
\newc{\mHpm}{m_{H^\pm}}
\newc{\simgt}{\lower.7ex\hbox{$\;\stackrel{\textstyle>}{\sim}\;$}}
\newc{\simlt}{\lower.7ex\hbox{$\;\stackrel{\textstyle<}{\sim}\;$}}
\newc{\ie}{{\it i.e.}}
\newc{\etal}{{\it et al.}}
\newc{\eg}{{\it e.g.}}
\newc{\kev}{\hbox{\rm\,keV}}
\newc{\mev}{\hbox{\rm\,MeV}}
\newc{\gev}{\hbox{\rm\,GeV}}
\newc{\tev}{\hbox{\rm\,TeV}}
\newc{\xpb}{\hbox{\rm\, pb}}
\newc{\xfb}{\hbox{\rm\, fb}}

\newcommand{\blap}[1]{\smash[t]{\begin{tabular}[c]{@{}c@{}}#1\end{tabular}}}

\def\lsim{\mathrel{\raise.3ex\hbox{$<$\kern-.75em\lower1ex\hbox{$\sim$}}}}
\def\gsim{\mathrel{\raise.3ex\hbox{$>$\kern-.75em\lower1ex\hbox{$\sim$}}}}

\def\slashchar#1{\setbox0=\hbox{$#1$}           
	\dimen0=\wd0                                 
	\setbox1=\hbox{/} \dimen1=\wd1               
	\ifdim\dimen0>\dimen1                        
	\rlap{\hbox to \dimen0{\hfil/\hfil}}      
	#1                                        
	\else                                        
	\rlap{\hbox to \dimen1{\hfil$#1$\hfil}}   
	/                                         
	\fi}                                         %
\catcode`@=11
\long\def\@caption#1[#2]#3{\par\addcontentsline{\csname
		ext@#1\endcsname}{#1}{\protect\numberline{\csname
			the#1\endcsname}{\ignorespaces #2}}\begingroup
	\small
	\@parboxrestore
	\@makecaption{\csname fnum@#1\endcsname}{\ignorespaces #3}\par
	\endgroup}
\catcode`@=12

\journalname{Eur. Phys. J. C}
\begin{document}

\title{Heavy $Z^\prime$ Bosons in the Secluded $U(1)^\prime$ Model at Hadron Colliders 
}


\author{Mariana Frank\thanksref{em1,am1}
		\and	
        Levent Selbuz\thanksref{el1,al1}
        \and
        Ismail Turan\thanksref{ei1,ai1}
}

\thankstext{em1}{e-mail: mariana.frank@concordia.ca}
\thankstext{el1}{e-mail: Levent.Selbuz@eng.ankara.edu.tr}
\thankstext{ei1}{e-mail: ituran@metu.edu.tr}

\institute{Department of Physics,  Concordia University, 7141 Sherbrooke St. West,
	Montreal, Quebec, Canada H4B 1R6 \label{am1}
\and
Department of Engineering Physics, Ankara	University, TR06100 Ankara, Turkey \label{al1}
\and
Department of Physics, Middle East Technical University, TR06800 Ankara, Turkey \label{ai1}
}




\date{Received:  / Accepted: }

\maketitle

\begin{abstract}
We study $Z^{\prime}$ phenomenology at hadron
colliders in an $U(1)^{\prime}$
extended MSSM. We choose a $U(1)^{\prime}$ model with a secluded sector, where the tension
between the electroweak scale and developing a large enough mass for
$Z^{\prime}$ is resolved by incorporating three additional 
singlet superfields into the model.
We perform a detailed analysis of the production, followed by decays, including  into
supersymmetric particles, of a $Z^{\prime}$
boson  with mass between 4 and 5.2 TeV, with particular emphasis on its possible discovery.  
We select three different scenarios
consistent with the latest available experimental data and relic
density constraints, and  concentrate on final signals with $\twoleps$,
$\fourleps$ and $\sixleps$. Including  the SM background  from processes with two, three or four vector bosons, we show the likelihood of observing a $Z^\prime$ boson is not promising for the HL-LHC at 14 TeV.  While at 27 and 100 TeV,    the situation is more optimistic, and we devise specific benchmark scenarios which could be observed.

\keywords{Supersymmetry \and Heavy Z$^\prime$ boson \and LHC \and FCC-hh}
\PACS{12.60.Cn,12.60.Jv,14.80.Ly.}
\end{abstract}

\section{Introduction}
\label{sec:intro}

After the discovery of the Higgs boson, the missing piece of the Standard Model (SM), the quest for physics beyond the SM (BSM) has intensified, both from theorists and experimentalists. The searches and analyses are motivated by the fact that while    precise
theoretical calculations within the SM have been confirmed by a wide range of experimental data establishing SM  as a well-tested physics theory, it still lacks explanation for some of the fundamental phenomena, such as matter-antimatter asymmetry  of the universe, dark matter, or neutrino oscillations. It is also plagued by theoretical inconsistencies, so it is at best incomplete (for instance, it does not include gravity). In fact, the discovery of the Higgs boson, with mass of the order of the electroweak scale, as expected,  points towards a higher structure, because in the SM, it is unclear why the Higgs boson is so much lighter than the Planck mass, as one expects that the large quantum contributions to the square of the Higgs boson mass would inevitably make the mass huge, and comparable to the scale at which new physics appears, unless there is an incredible amount of fine-tuning cancellation between the quadratic radiative corrections and the bare mass.
One can explore BSM physics taking a model independent approach, by assuming an effective field theoretic approach \cite{Buchmuller:1985jz},  which provides a general framework where higher order interactions of independent operators are built and one would be able to match them to explicit ultraviolet complete  models in a systematic way.  Or, one can enlarge the particle and/or gauge symmetry of the model. 

Of the latter,  the addition of supersymmetry (SUSY) to the SM is the most popular BSM scenario. It resolves the Higgs mass/gauge hierarchy problem, and provides, in its simplest scenario, the Minimal Supersymmetric 
Standard  Model (MSSM), a natural dark matter (DM) candidate. It does not, however, explain neutrino masses and it provides no resolution for the $\mu$ problem \cite{Cvetic:1996mf,Suematsu:1994qm,Lee:2007fw,Demir:1998dm}. The $\mu$ parameter, so-called higgsino mass term,  enters the supersymmetric Lagrangian as $\mu \hat{H}_u \hat{H}_d$, to give masses to the fermionic components of the Higgs bosons, and thus it is expected to be of the order of the SUSY-breaking scale.   But the $\mu$ term also enters in the scalar potential, so,  for successful electroweak symmetry breaking, its value should be at electroweak scale.  Adding an  $U(1)^\prime$ gauge group to the SM/MSSM symmetry group resolves  this inconsistency. An additional singlet Higgs field $S$ develops a vacuum expectation value (VEV) which breaks the $U(1)^\prime$ symmetry and generates the $\mu$ term dynamically, with $\mu \sim {\cal O}(\langle S \rangle)$. In addition, the model contains three singlet right-handed neutrinos that yield masses for the left-handed neutrinos (Dirac or Majorana). 

Models with additional $U(1)^\prime$ groups extend the spectrum of MSSM minimally: in addition to the right-handed neutrino and Higgs field $S$, they include another neutral gauge boson $Z^\prime$ (as well as  theirs supersymmetric partners).
This gauge field, a consequence of the additional $U(1)^\prime$ group,  is predicted by many
extensions of the SM. String-inspired models \cite{Cvetic:1996mf,Barr:1985qs}  and
grand-unification (GUT) models usually contain a number of extra
$U(1)$ symmetries.  The GUT group $SO(10)$ \cite{Rizzo:1998ut} and
exceptional group $E_6$ \cite{Hewett:1988xc,Athron:2016gor,Langacker:1998tc} are some examples. Here the 
$U(1)^\prime$ symmetries are broken at some intermediate energy scales \cite{Cleaver:1997nj}, 
between the GUT and electroweak scales. Phenomenologically, the
most interesting option is the breaking around  TeV scales, giving rise to extra neutral $Z^\prime$ gauge bosons observable at colliders.

The physics of $Z^{\prime}$ bosons has been extensively studied in
the literature, in models without supersymmetry \cite{DelAguila:1993px,Leike:1998wr,Babu:1997st,Langacker:2008yv,Erler:2009jh,Erler:2011ud,Diener:2009vq,Accomando:2016eom,Rizzo:2006nw,Batley:2015lha,Merkel:2014avp,Anastasi:2016ktq,Archilli:2011zc,Babusci:2014sta,Aaij:2017rft,Sirunyan:2018dsf,Batell:2014mga,Marsicano:2018krp,Gninenko:2012eq,Kou:2018nap,Ilten:2016tkc,Ilten:2015hya,Bediaga:2018lhg,Feng:2017uoz,Adrian:2018scb,Berlin:2018pwi,Karliner:2015tga,Curtin:2014cca,DOnofrio:2019dcp,Bilmis:2015lja}, or with \cite{Demir:2005ti,Anoka:2004vf,Corcella:2014lha,Corcella:2012dw,Corcella:2015zea,Corcella:2013cma,Gherghetta:1996yr,Chang:2011be,Baumgart:2006pa,Kang:2004bz,Demir:1998dm,Langacker:2008ip,Bhattacherjee:2018xzf}. 
The additional neutral gauge bosons have received significant attention from the experimentalists, and  have been searched for extensively at the LHC,   in dilepton channels  \cite{Aad:2019fac}, dijet channels \cite{Aaboud:2017yvp,Sirunyan:2018xlo}, di-tau \cite{Aaboud:2017sjh},  or $t {\bar t}$ decays \cite{Aaboud:2019roo}. 
Mass limits of 4 TeV or above, depending of the particular $U(1)^\prime$ scenario chosen, have hindered extensive analyses of their implications for phenomenology, as the prospects of observing them at the LHC do not appear to be  promising.

In supersymmetric  $U(1)^{\prime}$ models, an additional problem arises. The mass of the $Z^{\prime}$ boson is, as usual, proportional to the VEV of the singlet Higgs boson $S$. But  this parameter also determines the scale of the chargino/neutralino sector, thus a heavy $Z^{\prime}$ implies a heavy electroweakino sector, reducing further the interest in such models at the LHC. To avoid this link, we work in a {\it secluded} scenario \cite{Erler:2002pr,Frank:2012ne}, where the scalar sector of the $U(1)^{\prime}$ model is augmented by three additional singlet superfields\footnote{While only one additional singlet superfield would split the mass scales, three  are needed for anomaly cancellation.}, whose role is to decouple the mass of the $Z^{\prime}$ from the scale of chargino and neutralino masses.  The disadvantage is extending the particle spectrum of the SM more than the conventional  $U(1)^\prime$ model, while the gain is  being able to  preserve a large $Z^{\prime}$ mass while allowing light charginos/neutralinos, and in particular, a light dark matter candidate, which is the lightest supersymmetric particle (LSP)\footnote{The LSP can be the lightest neutralino or the lightest right-handed sneutrino.}.  In the secluded model, $Z^{\prime}$ decays into (light) chargino and neutralino pairs, or into sfermions can be significant, and affect the mass limits, albeit slightly. 

There have been studies of the $Z^{\prime}$ boson where the mass constraints were considerably reduced, assuming the model to be leptophobic \cite{Babu:1996vt,Chiang:2014yva,Frank:2020kvp,Araz:2017wbp} or quark-phobic \cite{Duan:2018akc}. For these, $Z^\prime$ couplings to the leptons or quarks are tuned (by assuming family non-universality \cite{Frank:2019nwk}, by using a specific value of the kinetic mixing \cite{Araz:2017wbp}, or by a choice of $U(1)^{\prime}$ charges \cite{Frank:2020kvp}). We propose here take a different point of view. We revisit the $U(1)^\prime$ model with a secluded sector, and allow for heavy $Z^{\prime}$ bosons, satisfying mass limit restrictions from LHC.  We concentrate on its production, decays and observability at high-luminosity (HL), and/or high-energy (HE) frontier of a future LHC, as well as the hadronic mode of a Future Circular Collider (FCC-hh).  

A heavy gauge boson, with mass $M_{Z^{\prime}} >4$ TeV cannot always be treated as a narrow width, as the ratio 
$\Gamma_{Z^{\prime}}/M_{Z^\prime}$ could be $ >10\%$. In this case, interference effects are important \cite{Accomando:2013sfa,Accomando:2019ahs,Frank:2020pui} and must be included. 
New physics contributions to the $Z^{\prime}$ width may 
significantly decrease the branching ratios into SM particles, and
therefore the mass limits quoted by the experiments may have to be
revisited. Furthermore, $Z^{\prime}$ decays into supersymmetric
particles  represent an excellent tool to investigate
the electroweak interactions at the LHC in a phase-space corner that
cannot be explored by employing the usual techniques. The secluded $U(1)^\prime$ model is ideally suited for this analysis, as it allows the electroweakinos to be light. Therefore, the
possible discovery of supersymmetry in $Z^{\prime}$ mediated processes would help to understand the role of $Z^{\prime}$ in
the SUSY breaking and open the path to additional investigations, since one would need to
formulate a consistent scenario accommodating both sparticles and heavy gauge
bosons.

The scope of this paper is indeed the investigation of the
phenomenology of $Z^{\prime}$ bosons at the LHC, assuming that they are heavy, and that they 
can decay into both SM and supersymmetric particles. We
will analyze the decay channels of the $Z^{\prime}$
boson, including decays to neutralinos, charginos, sleptons and 
Higgs bosons, which are normally neglected. 

In our  study, we will  allow the
$U(1)^{\prime}$ parameters to run within suitable ranges, taking
into account the recent experimental limits. Throughout this work,
we will focus especially on the decay of the $Z^{\prime}$
into slepton, chargino and neutralino pairs, eventually leading to multilepton 
final states $\twoleps$ or $\fourleps$ or $\sixleps$. To test the observability of such signals at the LHC, we devise benchmark scenarios and test their features at high integrated luminosity and at the center-of-mass $\sqrt{s}=14, \, 27$ and 100 TeV. We 
include calculations of he SM background from $VV$, $VVV$, and $VVVV$ processes
and we present a   simulation analysis for
the HL-LHC and future hadron colliders, indicating the significance of each scenario, and most promising observable for each signal.

Our paper is organized as follows. In Sec. \ref{sec:model} we describe the secluded $U(1)^\prime$ model with particular emphasis on its neutral gauge  (in \ref{subsec:Z'mass}) and neutralino  (in \ref{subsec:neutralinos}) sectors. We then proceed to analyze the implications of the model at colliders in Sec. \ref{sec:collider}, focusing first in choosing three benchmarks, which obey experimental constraints, and which  are able to reproduce the correct relic density, while maximizing $Z^\prime$ decays into supersymmetric particles, in \ref{subsec:BM}. Then we proceed with the analysis of $Z^\prime$ production and decays. We concentrate our analysis on multilepton signals \ref{subsec:analysis}, looking at $\twoleps$ (in \ref{subsubsec:2leptons}), $\fourleps$ (in \ref{subsubsec:4leptons}),  and $\sixleps$ (in \ref{subsubsec:6leptons}) signals. We summarize our findings and conclude in Sec. \ref{sec:conc}.

\section{The secluded $U(1)^{\prime}$ Model \label{sec:model}}

We present here the main ingredients of the secluded $U(1)^\prime$
model, with particular emphasis on the $Z^{\prime}$ boson. The model is based on the gauge group $SU(3)_c \otimes SU(2)_L \otimes U(1)_Y \otimes U(1)^\prime$, which breaks to the SM/MSSM $SU(3)_c \otimes SU(2)_L \otimes U(1)_Y$. The additional Abelian group introduces,  in addition to the MSSM superfields,  three right-handed neutrino superfields $\hat{N}_i^c$ (one for each generation), four singlet superfields  $\hat{S}$, $\hat{S}_1$, $\hat{S}_2$ and $\hat{S}_3$ and an additional neutral gauge boson and gaugino, corresponding to the gauge sector of $U(1)^\prime$. While only one scalar field $S$ is needed to break the symmetry,  three additional singlets $S_1$, $S_2$ and $S_3$ (the secluded sector) are introduced to split the mass scale of the additional gauge boson from that of electroweakinos.  

\begin{table}
\caption{Superfield (SF) configuration in the secluded $U(1)^\prime$ model, including notations for fermionic and bosonic states, number of generations, and charges under $U(1)_Y\otimes\, SU(2)_L \otimes\, SU(3)_C\otimes\, U(1)^\prime$.}
\label{tab:superfields}
		\begin{tabular*}{0.48\textwidth}{@{\extracolsep{\fill}} ccccc}
			\hline \hline
			SF & Spin 0 & Spin $\frac{1}{2}$ & Generations & Charges \\
			\hline
			$\hat{Q}$ & $\tilde{Q}$ & $Q$ & 3 & $(\frac{1}{6},{\bf 2},{\bf 3}, Q_Q^\prime) $ \\
			$\hat{L}$ & $\tilde{L}$ & $L$ & 3 & $(-\frac{1}{2},{\bf 2},{\bf 1}, Q_L^\prime) $ \\
			$\hat{H}_d$ & $H_d$ & $\tilde{H}_d$ & 1 & $(-\frac{1}{2},{\bf 2},{\bf 1}, Q_{H_d}^\prime) $ \\
			$\hat{H}_u$ & $H_u$ & $\tilde{H}_u$ & 1 & $(\frac{1}{2},{\bf 2},{\bf 1}, Q_{H_u}^\prime) $ \\
			$\hat{D}$ & $\tilde{D}_R^*$ & $D_R^*$ & 3 & $(\frac{1}{3},{\bf 1},{\bf \overline{3}}, Q_d^\prime $ \\
			$\hat{U}$ & $\tilde{U}_R^*$ & $U_R^*$ & 3 & $(-\frac{2}{3},{\bf 1},{\bf \overline{3}}, Q_u^\prime) $ \\
			$\hat{E}$ & $\tilde{E}_R^*$ & $E_R^*$ & 3 & $(1,{\bf 1},{\bf 1}, Q_e^\prime) $ \\
			$\hat{N}$ & $\tilde{N}^*$ & $N^*$ & 3 & $(0,{\bf 1},{\bf 1}, Q_v^\prime) $ \\
			$\hat{S}$ & $S$ & $\tilde{S}$ & 1 & $(0,{\bf 1},{\bf 1}, Q_s^\prime) $ \\
			$\hat{S}_1$ & $S_1$ & $\tilde{S}_1$ & 1 & $(0,{\bf 1},{\bf 1}, Q_{s_1}^\prime) $ \\
			$\hat{S}_2$ & $S_2$ & $\tilde{S}_2$ & 1 & $(0,{\bf 1},{\bf 1},Q_{s_2}^\prime) $ \\
			$\hat{S}_3$ & $S_3$ & $\tilde{S}_3$ & 1 & $(0,{\bf 1},{\bf 1}, Q_{s_3}^\prime) $ \\
			$\hat{\cal{Q}}$ & $\tilde{\cal{Q}}$ & $\cal{Q}$ & 3 & $(Y_{\cal{Q}},{\bf 1},{\bf 1}, Q_{\cal{Q}}^\prime) $ \\	
			$\hat{\overline{\cal{Q}}}$ & $\tilde{\overline{\cal{Q}}}$ & $\overline{\cal{Q}}$ & 3 & $(Y_{\overline{\cal{Q}}},{\bf 1},{\bf 1}, Q_{\overline{\cal{Q}}^\prime} )$ \\			
			$\hat{\cal{L}}$ & $\tilde{\cal{L}}$ & $\cal{L}$ & 2 & $(Y_{\cal{L}},{\bf 1},{\bf 1}, Q_{\cal{L}}^\prime) $ \\
			$\hat{\overline{\cal{L}}}$ & $\tilde{\overline{\cal{L}}}$ & $\overline{\cal{L}}$ & 2 & $(Y_{\overline{\cal{L}}},{\bf 1},{\bf 1}, Q_{\overline{\cal{L}}^\prime}) $ \\			\hline \hline
		\end{tabular*}
\end{table}

Unfortunately,  anomaly cancelation requires the presence of additional superfields (namely, the exotics $\hat{\cal{Q}}$ and  $\hat{\cal{L}}$),  with exotic quantum numbers, which are assumed to be heavy and decoupled form the rest of the spectrum. We list the superfields in the model, together with the number of generations and charge assignments under the $SU(3)_c \otimes SU(2)_L \otimes U(1)_Y \otimes U(1)^\prime$ gauge group in Table \ref{tab:superfields}.  The superpotential in this model including  the exotic fields is  given by
\begin{eqnarray}
	\label{eq:superpot}
	\widehat{W}&=&h_u\widehat{Q}\cdot \widehat{H}_u \widehat{U}+
	h_d\widehat{Q}\cdot \widehat{H}_d \widehat{D} + h_e\widehat{L}\cdot
	\widehat{H}_d \widehat{E} + h_s \widehat{S}\widehat{H}_u \cdot
	\widehat{H}_d \nonumber \\
	&+&  \frac{1}{M_R}  \widehat{S}_1 \widehat{L}\cdot
	\widehat{H}_u {\bf h_{\nu}} \widehat{N}+
	\bar{h}_s \widehat{S}_1 \widehat{S}_2 \widehat{S}_3 
	+ \sum_{i=1}^{n_{\cal{Q}}} {h}_Q^i \widehat{S} \widehat{\cal{Q}}_i
	\widehat{\cal{\overline{Q}}}_i\nonumber \\
	&+& \sum_{j=1}^{n_{\cal{L}}} {h}_L^j
	\widehat{S} \widehat{\cal{L}}_j \widehat{\cal{\overline{L}}}_j,
\end{eqnarray}
where the fields $\widehat{\cal Q},~\widehat{\cal L}$ are the exotics, $M_R$ is a large mass scale and
$h_{\nu}$ is the Yukawa coupling responsible for generating neutrino
masses. This non-renormailzable term is added in the original formulation of the model to account for Dirac neutrino masses (see, for example, \cite{PhysRevD.58.093017} for the origin of the term).  Thus, in this form, neutrinos are Dirac particles, whose masses imply, for the Yukawa coupling  \cite{Demir:2010is}, 
$\displaystyle
h_\nu \simeq 3 \times 10^{-13} \left (\frac {|m_\nu |^2}{2.8 \times 10^{-3}\, {\rm eV}^2}\right )^{1/2}$.

The effective $\mu$ term is generated dynamically as  $\mu=  h_s \langle S \rangle$. The scalar potential includes the $F$-term,  given  by
\begin{eqnarray}
	V_F&=&h_s^2 \Big( | H_u |^2  | H_d |^2 + |S|^2 |H_u|^2 + |S|^2 |H_d|^2\Big)\nonumber \\
	&+& 
	\bar{h}_s^2 \Big(|S_1|^2|S_2|^2 + |S_2|^2|S_3|^2 + |S_3|^2|S_1|^2\Big)\, ,
\end{eqnarray}
while the $D$-term scalar potential is 
\begin{eqnarray}
	V_D &=& \frac{g_1^2 + g_2^2}{8}  \Bigl( | H_d |^2 - | H_u |^2 \Big)^2  +
	\frac{1}{2} g_1^{\prime \,2} \Big( Q^\prime_S |S|^2 + Q^\prime_{H_u} |H_u|^2 \Bigr. \nonumber\\
	\Bigl.	&+& Q^\prime_{H_d} |H_d|^2 + \sum_{\mathclap{i=1}}^{3} Q^\prime_{S_i} |S_i|^2 \Big)^2,
\end{eqnarray}
where $g_1$, $g_2$ and $g_1^\prime$ are the coupling constants for the $U(1)_Y$, $SU(2)_L$ and $U(1)^\prime$ gauge groups while  $Q^\prime_{\phi}$ is the $U(1)^\prime$ charge of the field $\phi$. Finally, the potential includes the  SUSY-breaking soft terms, expressed in terms of soft-SUSY breaking mass parameters $M^2_i$ and triple scalar couplings $A_i$ as
\begin{eqnarray}
	V_{\rm soft} &=& M^2_{H_u} | H_u |^2 + M^2_{H_d} | H_d |^2 + M^2_{S} | S |^2  + \sum_{\mathclap{i=1}}^{3} M_{S_i}^2 |S_i|^2\nonumber \\
	&-&
	\big(A_s h_s S H_u H_d + A_{\bar{s}} \bar{h}_s S_1 S_2 S_3 + h.c.\big) \nonumber \\ &+&
	\big(M_{SS_1}^2 S S_1 + M_{SS_2}^2 S S_2 + M_{S_1 S_2}^2 S_1^\dagger S_2 + h.c.\big).
\end{eqnarray}
The symmetry-breaking sector of the model is very complex, and finding an acceptable
minimum of the Higgs potential,  even at the
tree level, is non-trivial \cite{Frank:2012ne}.  Once a
minimum is found, the mass of the lightest Higgs boson can be fine
tuned to 125 GeV by small variations in the parameter $\bar{h}_s$.
Setting masses for the additional scalars in the TeV range insures
that the mixing with the lightest Higgs boson is small, and  thus this Higgs will obey mass  \cite{Bechtle:2013wla} and signal bounds \cite{Bechtle:2013xfa} consistent with the SM-like Higgs found at the LHC.  Additional Higgs
states, in particular the lightest pseudoscalar, being heavy, will also 
satisfy constraints from $B_s \to \mu^+ \mu^-$ branching ratio
\cite{Aaij:2012nna}.

The $U(1)^{\prime}$ charges of the fields satisfy 
conditions arising  the requirement of cancellation of gauge and
gravitational anomalies. For instance, the charges for Higgs fields
in the model are chosen so that $$\displaystyle
Q^\prime_{S}=-Q^\prime_{S_1}=-Q^\prime_{S_2}=Q^\prime_{S_3}/2,
~~Q^\prime_{H_u}+Q^\prime_{H_d}+Q^\prime_{S}=0\nonumber.$$ The $U(1)^\prime$
charge of the quark doublet $\widehat{Q}$ is kept as a free
parameter after the  normalization  $$Q^\prime_{H_u}=-2,
Q^\prime_{H_d}=1, Q^\prime_{S}=1, Q^\prime_{S_1}=-1,
Q^\prime_{S_2}=-1, Q^\prime_{S_3}=2.$$ A  complete list of
conditions for anomaly cancellations in the model, and a choice of charge assignments of the SM and
exotic quarks and leptons in the model
can be found in \cite{Demir:2010is}. 

\subsection{Gauge boson masses and mixing}
\label{subsec:Z'mass}
Through spontaneous breakdown of the 
group $SU(2)_L\\ \otimes U(1)_Y \otimes U(1)^{\prime}$ to $U(1)_{\rm em}$ the Higgs acquire the 
VEVs
\begin{eqnarray}\displaystyle 
	\langle H_u \rangle = \left(\begin{array}{c} 0\\
		\frac{v_u}{\sqrt{2}}\end{array}\right), \langle H_d \rangle =
	\left(\begin{array}{c} \frac{v_d}{\sqrt{2}}\\
		0\end{array}\right), \langle S \rangle =
	\frac{v_s}{\sqrt{2}}, \langle S_i \rangle =
	\frac{v_{s_i}}{\sqrt{2}}
\end{eqnarray}
Here the first two VEVs are required to break the gauge symmetries of the SM, and the third to break $U(1)^\prime$. After symmetry breaking,  one massless state (the photon) and two massive states
(the $Z_0$ and $Z_0^\prime$ bosons which are not yet the physical eigenstates due to a non-zero mass mixing term, to be introduced below) arise as orthonormal combinations of
$W^{3}_{\mu}$, $Y_\mu$ and $Y^\prime_{\mu}$ gauge bosons. The
$W^{1}_{\mu}$ and $W^{2}_{\mu}$  combine to form
$W^{\pm}_{\mu}$, the charged vector bosons in the model. Unlike in the 
MSSM, the $Z_0$ boson is not a physical state by
itself but mixes with the $Z_0^\prime$ boson. This mass mixing term
arises from the fact that the Higgs doublets $H_{u,d}$ are charged
under each factor of $SU(2)_L\otimes U(1)_Y\otimes U(1)^{\prime}$,
and the associated mass-squared matrix is given by
\begin{eqnarray}
	\label{mzzp}
	M^{2}_{Z_0 Z_0^\prime} = \Bigg(\begin{array}{cc} M_{Z_0}^2 & \Delta^2 \\[0.5em]
		\Delta^2 & M_{Z_0^\prime}^2\end{array}\Bigg)\,,
\end{eqnarray}
in the $\left(Z_{0\mu}, Z_{0\mu}^\prime\right)$ basis, where the matrix elements
are
\begin{eqnarray}
	\label{eqn7}
	M_{Z_0}^2 &=& \frac{1}{4} g_Z^{ 2} \left(v_u^2 + v_d^2\right),\nonumber\\
	M_{Z_0^\prime}^2 &=& g_1^{\prime \, ^2} \left( Q^{\prime\ 2}_{H_u}
	v_u^2 + Q^{\prime\ 2}_{H_d} v_d^2 + Q^{\prime\ 2}_{S}
	v_s^2+\sum^3_{i=1}Q^{\prime\ 2}_{S_i} v_{s_i}^2
	\right)\,,\nonumber\\
	\Delta^2 &=& \frac{1}{2} g_Z g_1^{\prime} \left( Q^{\prime}_{H_u}
	v_u^2 - Q^{\prime}_{H_d} v_d^2\right)\,,
\end{eqnarray}
where $g_Z^2 = g_2^2 + g_1^2$. The physical neutral vector bosons,
$Z, \, Z^\prime$, are obtained by diagonalizing $M^{2}_{Z_0Z_0^\prime}$:
\begin{eqnarray}
	\label{mzzp-angle} \left(\begin{array}{c} Z\\Z^\prime
	\end{array}\right) =
	\left(\begin{array}{cc} \cos\theta_{Z_0Z_0^{\prime}} & \sin\theta_{Z_0Z_0^{\prime}} \\
		-\sin\theta_{Z_0Z_0^{\prime}} &
		\cos\theta_{Z_0Z_0^{\prime}}\end{array}\right) \left(\begin{array}{c} Z_0
		\\Z_0^\prime
	\end{array}\right)\,,
\end{eqnarray}
where
\begin{eqnarray}
	\theta_{Z_0Z_0^{\prime}} = - \frac{1}{2} \arctan \left( \frac{ 2
		\Delta^2}{M_{Z_0^\prime}^2 - M_{Z_0}^2}\right)
\end{eqnarray}
is their mass mixing angle, and
\begin{eqnarray}
	M^{2}_{Z,\, Z^\prime}= \frac{1}{2} \left[ M_{Z_0^\prime}^2 + M_{Z_0}^2  \mp
	\sqrt{\left(M_{Z_0^\prime}^2 - M_{Z_0}^2\right)^2 + 4
		\Delta^4}\right]\
\end{eqnarray}
are their squared masses of the corresponding mass eigenstates. The collider searches 
plus various indirect observations require the $Z_0$--$Z_0^{\prime}$ mixing
angle $\theta_{Z_0Z_0^{\prime}}$ to be at most a few times $10^{-3}$ \cite{Langacker:2008yv}, where
unavoidable model dependence arises from $Z^{\prime}$ couplings.
This bound requires either $M_{Z_0^\prime}$ to be large enough (well in the
${\rm TeV}$ range) or $\Delta^2$ to be sufficiently suppressed by
the vacuum configuration, that is, $\tan^2\beta\equiv v_u^2/v_d^2
\sim Q^{\prime}_{H_d}/Q^{\prime}_{H_u}$. Which of these options is
realized depends on the $U(1)^{\prime}$ charge assignments and the
soft-breaking mass parameters in the Higgs sector. Having large $M_{Z_0^\prime}$ term in Eq.~\ref{mzzp} insures a small mixing angle.\\[0.5em]

We expand more on the reason for introducing the extra scalars $S_1,S_2,$ and $S_3$. In their absence, $M_{Z_0^\prime}$ term  is equal to the one given in Eq.~\ref{eqn7} without the summation term  at the end.
  Hence, the mass of $Z^\prime$ boson will be determined by the charges under $U(1)^\prime$, and the vacuum expectation value of the singlet scalar $\langle S \rangle =v_s/\sqrt{2}$ 
and thus effectively proportional to the VEV $v_s$. At the same time, the masses of charginos and neutralinos in the model would be proportional to the effective $\mu=  h_s v_s/\sqrt{2}$, and the two scales, barring serious fine-tuning of the $h_s$ coupling, are connected. Introducing the extra scalars,  their VEVs will be added to the contribution to the  $M_{Z_0^\prime}$ term (eventually determining the mass of $Z^\prime$ boson), while $\mu$, determining the mass scale of charginos and neutralinos, remains unchanged. As strong constraints are imposed on the $Z^\prime$ boson mass at LHC, while the electroweakinos can remain relatively light, decoupling these two scales is desirable. 

 Thus, in our investigations, the mass of the $Z^\prime$ boson will be expected to be heavy, $M_{Z^\prime}>4$ TeV. In this case, the mixing angle between $Z_0$ and $Z_0^\prime$ becomes rather small\footnote{For our chosen benchmarks, the mixing angle $\theta_{Z_0Z_0^{\prime}}$ is of ${\cal O}(10^{-4})$.}, and the mass eigenstates $(Z,Z^\prime)$ are almost identical to the the original gauge states $(Z_0,Z_0^\prime)$.  However, despite the smallness of the mixing angle $\theta_{Z_0Z_0^{\prime}}$, we keep it in our numerical analysis.

\subsection{Neutralinos in the secluded $U(1)^\prime$ model}
\label{subsec:neutralinos}
While the chargino sector is the same as in MSSM, the neutralino content is significantly enlarged  by the additional $U(1)^\prime$ gaugino, and the four additional singlinos $\tilde{S}, \tilde{S}_1, \tilde{S}_2$ and $\tilde{S}_3$. In the basis 
$\{\tilde Y, \tilde W^3,\tilde H_{d}^0, \tilde H_{u}^0,
\tilde S, \tilde Y^\prime,\tilde S_{1}, \tilde S_{2},\tilde S_{3}\}$ where $\tilde{Y}, \tilde{Y}^\prime$ and $\widetilde{W}^3$ are the neutral gauge fermions of $U(1)_Y, U(1)^\prime$ and $SU(2)_L$, the neutralino $ 9 \times 9$ mass matrix is given as

\begin{widetext}
\begin{eqnarray}
{\cal
		M_N} &=& \left(
	\begin{array}{ccccccccc}
		M_{\tilde Y}&0&-M_{\tilde Y \tilde H_d}&M_{\tilde Y \tilde H_u}&0&M_{\tilde Y \tilde Y^\prime}&0&0&0 \\[1.ex]
		0&M_{\tilde W}&M_{\tilde W \tilde H_d}&-M_{\tilde W \tilde H_u}&0&0&0&0&0\\[1.ex]
		-M_{\tilde Y \tilde H_d}&M_{\tilde W \tilde H_d}&0&-\mu_{\rm eff}&-\mu_{H_u}&\mu^\prime_{H_d}&0&0&0\\[1.ex]
		M_{\tilde Y \tilde H_u}&-M_{\tilde W \tilde H_d}&-\mu_{\rm eff}&0&-\mu_{H_d}&\mu^\prime_{H_u}&0&0&0\\[1.ex]
		0&0&-\mu_{H_u}&-\mu_{H_d}&0&\mu'_S&0&0&0\\[1.ex]
		M_{\tilde Y \tilde Y^\prime}&0&\mu'_{H_d}&\mu'_{H_u}&\mu^\prime_S&M_{\tilde Y^\prime}&\mu^\prime_{S_1}&\mu^\prime_{S_2}&\mu'_{S_3}\\[1.ex]
		0&0&0&0&0&\mu^\prime_{S_1}&0&-\frac{\bar{h}_s v_{s_3}}{\sqrt{2}}&-\frac{\bar{h}_s v_{s_2}}{\sqrt{2}}\\[1.ex]
		0&0&0&0&0&\mu^\prime_{S_2}&-\frac{\bar{h}_s v_{s_3}}{\sqrt{2}}&0&-\frac{\bar{h}_s v_{s_1}}{\sqrt{2}}\\[1.ex]
		0&0&0&0&0&\mu^\prime_{S_3}&-\frac{\bar{h}_s v_{s_2}}{\sqrt{2}}&-\frac{\bar{h}_s v_{s_1}}{\sqrt{2}}&0\\[1.ex]
	\end{array}
	\right) \, 
	\label{eq:neutralmatrix}
\end{eqnarray}
\end{widetext}
and is diagonalized by
$N {\cal
	M_N}N^\dag$=diag$(m_{\tilde\chi_1^0},.....,m_{\tilde\chi_9^0})$,\hfill\\  $0\leq m_{\tilde\chi_1^0}\leq.......\leq
m_{\tilde\chi_9^0}.$ The parameters introduced in the neutralino mass matrix elements in Eq. \ref{eq:neutralmatrix} are defined as
\begin{widetext}
\[\arraycolsep=1.2pt\def\arraystretch{1.6}
\begin{array}{llll}
	M_{\tilde Y \tilde H_d}=M_{Z_0} \sin\theta_W\cos\beta, &\hspace{0.5cm}
	M_{\tilde Y \tilde H_u}=M_{Z_0}\sin\theta_W\sin\beta,&\hspace{0.5cm}
	M_{\tilde W \tilde H_d}=M_{Z_0}\cos\theta_W\cos\beta,&\hspace{0.5cm}
	M_{\tilde W \tilde H_u}=M_{Z_0}\cos\theta_W\sin\beta, \\
	\mu^\prime_{H_d}=g^\prime_{1}Q^\prime_{H_d}v_d, &\hspace{0.5cm}
	\mu^\prime_{H_u}=g^\prime_{1}Q'_{H_u}v_u, &\hspace{0.5cm}
	\mu^\prime_{S}=g^\prime_{1}Q^\prime_{S}v_s, &\hspace{0.5cm}
	\mu^\prime_{S_i}=g^\prime_{1}Q'_{S_i}v_{s_i}, \\
	\mu_{\rm eff}=\frac{h_s v_s}{\sqrt{2}}, &\hspace{0.5cm}
	\mu_{H_d}=\frac{h_sv_d}{\sqrt{2}},&\hspace{0.5cm}
	\mu_{H_u}=\frac{h_sv_u}{\sqrt{2}}. &
\end{array}
\]
\end{widetext}
The gaugino mass parameters $M_{\tilde Y},\ M_{\tilde Y^\prime},$ and $M_{\tilde Y \tilde Y^\prime}$ are free parameters of the model, and we introduce the ratios
\begin{equation}
	R_{Y Y^\prime}=\frac{M_{\tilde Y \tilde Y^\prime}}{M_{\tilde Y}}\,,  \hspace{0.5cm} R_{Y^\prime}=\frac{M_{\tilde Y^\prime}}{M_{\tilde Y}}\, .
\end{equation}
These parameters, representing mixing of  $U(1)_Y$ and $U(1)^\prime$ gauginos, and mass parameter of the $U(1)^\prime$ gaugino, measured relative to the $U(1)_Y$ gaugino mass parameter, will be seen to be important in scanning over the parameter space, as the underlying physics will be sensitive to their variation. 
\begin{table*}[htbp]
	\begin{center}
		\caption{The parameters characterizing benchmarks BP1, BP2 and BP3  for the
			secluded $U(1)^{\prime}$ model. The values of dimensionful parameters are given in GeV.}
		\label{tab:inputs}
		\setlength{\extrarowheight}{2.5pt}	
		\begin{tabular*}{0.99\textwidth}{@{\extracolsep{\fill}} cccc}
			\hline\hline
			$\rm Parameters$ & $\rm BP1$  &$\rm BP2$ &$\rm BP3$ \\ 
			\hline
			$g_1^{\prime}$&$0.2$&0.12&0.15\\
			$\tan\beta$&$1.345$&1.198&1.175\\
			$Q^{\prime}_Q$&$0.6$&0.1&-0.81\\
			$\mu_{\rm eff}$&260&280&250\\
			$(h_{\nu},\ h_s,\ \bar{h}_s)$&(1.0,\ 0.739,\ 0.1)&(1.0,\ 0.7235,\ 0.1)&(1.0,\ 0.724,\ 0.1)\\
			$(A_s,\ A_{\bar{s}}) $&557.7& (557.7,\ 2200)&(557.7,\ 1200)\\
			$(v_{s_1},\ v_{s_2},\ v_{s_3})$&(8675,\ 8650,\ 8675)&(6675,\ 15600,\ 14675)&(12100,\ 14550,\ 14500)\\
			$(M_{\tilde Y},\ M_{\tilde W},\ M_{\tilde g}) $&(-200,\ 2000,\ 2500)&(-760,\ 750,\ 2500)&(-260,\ 300,\ 2500)\\
			$(R_{Y^\prime},\ R_{Y Y^\prime})$&$(5.0,\ 4.8)$&(1.0,\ 0.01)&(1.0,\ 0.01)\\
			$(M_{\tilde \nu{_e{_R}}},\ M_{\tilde \nu{_\mu{_R}}},\ M_{\tilde \nu{_\tau{_R}}} ) $ & 500 &3000&500\\
			$(M_{L_1},\ M_{L_2},\ M_{L_3})$&520&450 & 200\\
			$(M_{E_1},\ M_{E_2},\ M_{E_3})$&450&2125 &1700\\
			$(M_{Q_1},\ M_{Q_2},\ M_{Q_3})$&(2200,\ 2200,\ 2400)&(2200,\ 2200,\ 2400) &(2200,\ 2200,\ 2400)\\
			$(M_{U_1},\ M_{U_2},\ M_{U_3})$&(2200,\ 2200,\ 2500)&(2200,\ 2200,\ 2500) &(2200,\ 2200,\ 2500)\\
			$(M_{D_1},\ M_{D_2},\ M_{D_3})$&(2300,\ 2300,\ 2500)&(2300,\ 2300,\ 2500) &(2300,\ 2300,\ 2500)\\
			$(M^2_{SS_{1}},\ M^2_{SS_{2}},\ M^2_{S_1 S_{2}})$ & 
			$(-9\times10^{6},\ -9\times10^{6},\ 0)$ & 
			$(-9\times10^{6},\ -9\times10^{6},\ 0)$ & 
			$(-9\times10^{6},\ -9\times10^{6},\ 0)$ \\
			$(A_t,\ A_b) $ &(-697.75,\ -959.66)&(-697.75,\ -959.66)&(-697.75,\ -959.66)\\
			\hline\hline
		\end{tabular*}
	\end{center}
\end{table*}
\section{$Z^{\prime}$ Boson in the $U(1)^{\prime}$ model at the current and future hadron colliders  }
\label{sec:collider}

We now proceed to the main analysis in this work, looking at the consequences of a heavy neutral gauge boson at the collider. As we shall see, and as found before, $Z^\prime$ bosons satisfying all collider, cosmological and low energy constraints, do not offer promising prospects for observability at the present LHC, even operating at 3 ab$^{-1}$. Thus, we will also analyze the prospects of observing a signal at the HE-LHC  operating at 27  TeV as well as at the FCC-hh.  As the parameter space is large, choosing realistic benchmarks is a more transparent method to show  physics results than a scan.  
Some previous analyses of $Z^\prime$ at present and future colliders exist, e.g., in \cite{CidVidal:2018eel}, but in that case, the authors have considered an $E_6$-inspired  leptophobic model, constrained to yield light $Z^\prime$ masses. While we share with that analysis a light chargino-neutralino sector (insured in our case, by the existence of a secluded model), ours is an analysis for signals from a heavy $Z^\prime$ boson scenario.
\subsection{$U(1)^\prime$ Benchmark Points and Relic Density}
\label{subsec:BM}

In order to give definite predictions for the production and decay rates of the $Z^\prime$ boson, we scan the parameter space for benchmark scenarios to showcase the salient points of the model.

The benchmark points  chosen must obey five important conditions:
\begin{itemize}
	\item The parameters chosen had to insure the stability of the vacuum;
	\item The points had to satisfy relic density constraints from WMAP of cold  dark matter \cite{Spergel:2006hy} 
	for the LSP, assumed here to be the lightest neutralino;  
	\item The mass of the $Z^\prime$ boson has to satisfy mass constraints from ATLAS and CMS, as discussed in the next subsection;
	\item Of the parameter points satisfying the above two conditions,  benchmarks were chosen to enhance
	the supersymmetric decay signals of the  $Z^{\prime}$ boson; and 
	\item In each scenario, the  lightest Higgs boson is SM-like and has $m_{H_1^0}=125$ GeV. 
\end{itemize}
\begin{table*}[htbp]
	\begin{center}
		\caption{The  mass spectra (in GeV) for
			the supersymmetric sector and the relic density $\Omega_{DM}h^2$ values of the benchmark points given in
			Table~\ref{tab:inputs} for the Secluded $U(1)^\prime$.}
		\label{tab:spec}
		\setlength{\extrarowheight}{2.5pt}	
		\small
		\begin{tabular*}{1.0\textwidth}{@{\extracolsep{\fill}} p{0.9in}p{1.4in}p{1.4in}p{1.4in}p{1.0in}}
			\hline\hline
			$\rm Masses$ & $\rm BP1$  &$\rm BP2$ &$\rm BP3$ & Bounds \\
			\hline
			$m_{Z^\prime}$& 4250&4069&5195 & 3900 \cite{Tanabashi:2018oca}\\
			$m_{H^0_i,\ i=1,...,6}$&
			(125.9,\,543,\,671,\,1077, \hspace*{0.05cm} 4237,\,17719) & 
			(125.0,\,557,\,1148,\,2418, \hspace*{0.05cm} 4171,\,19151) & 
			(125.3,\,524,\,1045,\,1611, \hspace*{0.05cm} 5210,\,22171) &  $m_{H^0_1} = 125.2$ \cite{Tanabashi:2018oca} 
			\\
			$m_{A^0_i,\ i=1,...,4}$&
			(550,\,719,\,1012,\,17718) & 
			(563,\,592,\,20769,\,19151) & 
			(531,\,572,\,1882,\,22170)  & $m_{A^0_1}>93.4$ \cite{Tanabashi:2018oca}
			\\
			$m_{\tilde\chi^0_i,\ i=1,...,5}$&
			(51,\,167,\,262,\,312,\,613) & 
			(48,\,269,\,328,\,762,\,763) & 
			(52,\,195,\,264,\,303,\,360) &
			$m_{\tilde\chi^0_1} > 50$ \cite{Tanabashi:2018oca}
			\\
			$m_{\tilde\chi^0_i,\ i=6,...,9}$&
			(1226,\,2004,\,4222,\,4638) & 
			(1170,\,1740,\,4047,\,4237) & 
			(1036,\,1939,\,4908,\,5551) & -
			\\
			$(m_{\tilde\chi^\pm_1},\ m_{\tilde\chi^\pm_2})$&
			(256,\,2004)& (267,\,763)& (192,\,359) &
			$m_{\tilde\chi^\pm_1}> 103.5 $ \cite{Tanabashi:2018oca} 
			\\
			$m_{H^{\pm}}$&540.9&554.9&522.4 & $m_{H^{\pm}} > 80$ \cite{Tanabashi:2018oca}
			\\
			$(m_{\tilde e_L},\,m_{\tilde \mu_L},\,m_{\tilde \tau_1})$&
			(503,\,503,\,457) &503&(1412,\,1412,\,473) 
			&  $m_{\tilde \ell} > (82 - 107)$ \cite{Tanabashi:2018oca}
			\\
			$(m_{\tilde e_R},\,m_{\tilde \mu_R},\,m_{\tilde \tau_2})$&
			(457,\,457,\,503) & 1850 & (473,\,473,\,1412) & $m_{\tilde \ell} > (82 - 107)$ \cite{Tanabashi:2018oca}
			\\
			$(m_{\tilde\nu_e},\, m_{\tilde\nu_\mu},\,m_{\tilde\nu_\tau})$&
			501&501&1412 & $m_{\tilde\nu_\ell}> 41$ \cite{Tanabashi:2018oca}
			\\
			$(m_{\tilde\nu_{eR}},\, m_{\tilde\nu_{\mu R}},\,m_{\tilde\nu_{\tau R}})$&
			553&3472&645 & -
			\\[0.1em]
			\hline
			$\Omega_{DM}h^2$&0.117&0.121&0.119  & 0.111 \cite{Spergel:2006hy} \\
			\hline\hline
		\end{tabular*}
	\end{center}
\end{table*}

To analyze the model,  we used   {\tt CalcHEP}  \cite{Belyaev:2012qa},  \texttt{micrOMEGAs} \cite{Belanger:2018ccd}, {\tt PYTHIA8} \cite{Sjostrand:2006za,Sjostrand:2014zea}, {\tt Delphes} \cite{deFavereau:2013fsa,Selvaggi:2014mya,Mertens:2015kba}, and {\tt MadAnalysis} \cite{Conte:2012fm}
to prepare the model, calculate the mass spectrum and branching ratios, calculate the relic density, generate events and eventually carry out the simulation.  Our goal  was to find benchmarks that satisfy cosmological constraints on dark matter  and satisfy collider constraints  (the invisible width of the $Z$ boson, limits on charged sparticle masses, charginos mass, first and second-generation squark  masses, lightest Higgs boson mass, Br$(B_{s,d} \to \mu^+ \mu^-$), Br$(B \to X_s \gamma$), $\Delta M_{B_{s,d}}$ and various others), as outlined in \cite{Aad:2015baa,Caron:2016hib}, and choose those exhibiting distinct decay features, while offering some promise for  collider observability. 

\begin{table*}[htbp]
	\begin{center}
		\caption{Leptonic anomalous moments corrections and flavor observables for each benchmark scenario considered in this study.}
		\label{tab:low}
		\setlength{\extrarowheight}{5pt}	
		\begin{tabular*}{0.99\textwidth}{@{\extracolsep{\fill}} ccccc}
			\hline\hline
			$\rm Observable$ & $\rm BP1$  &$\rm BP2$ &$\rm BP3$ &Bounds \\
			\hline
			$\Delta a_e$& $5.68\times 10^{-16}$&$1.14\times 10^{-15}$&$3.37\times 10^{-16}$ &$-(8.7\pm 3.6)\times 10^{-13}$ \cite{Laporta:2017okg,Aoyama:2017uqe,Aoyama:2012wj} \\
			$\Delta a_\mu$& $2.43\times 10^{-11}$&$4.86\times 10^{-11}$&$1.44\times 10^{-11}$ & $(2.7\pm 0.9)\times 10^{-9}$ \cite{Bennett:2006fi,Aoyama:2012wk}\\
			$\Delta a_\tau$& $7.82\times 10^{-9}$&$1.12\times 10^{-8}$&$-4.78\times 10^{-10}$ & $|\Delta a_\tau|<1.75\times 10^{-5}$ \cite{Chen:2018cxt}\\
			$ \frac{{\rm Br}(B\to X_s \gamma)}{{\rm Br}(B\to X_s \gamma)_{\rm SM}}$&
			1.18 & 1.17 & 1.15 & 1.05$\pm$ 0.11 \cite{Tanabashi:2018oca} \\[0.4em]
			$\frac{{\rm Br}(B^0_{s,d}\to \mu^+\mu^-)}{{\rm Br}(B^0_{s,d}\to \mu^+ \mu^-)_{\rm SM}}$&
			1.09 & 1.11 & 1.10  & 0.83$\pm$ 0.25 \cite{Tanabashi:2018oca}\\
			$\frac{{\rm Br}(B^+\to \tau^+\nu_\tau)}{{\rm Br}(B^+\to \tau^+\nu_\tau)_{\rm SM}}$&
			0.991 & 0.991 & 0.991  & 1.04$\pm$ 0.34\, \cite{Tanabashi:2018oca}\\
			$\Delta M_{B_{(s,d)}}/\Delta M_{B_{(s,d)}}^{\rm SM}$&
			(1.10,\ 1.04) & (1.12,\ 1.04) & (1.12,\ 1.04) & ($1.00\pm 0.15,0.86\pm 0.28$) \cite{Tanabashi:2018oca} \\
			$R_K/R_K^{\rm SM}$&
			1.00 & 1.00 & 1.00  & $1.00 \pm 0.17$ \cite{Tanabashi:2018oca} \\
			$\epsilon_K/\epsilon_K^{\rm SM}$&
			1.00 & 1.00 & 1.00 & $0.99\pm 0.18$ \cite{Tanabashi:2018oca,Sala:2016evk}\\
			$\Delta M_K/\Delta M_K^{\rm SM}$&
			1.00 & 1.00 & 1.00 & $1.24 \pm 0.16$ \cite{Tanabashi:2018oca,Buras:2014maa}\,\footnotemark \\[0.4em]			\hline\hline
		\end{tabular*}
	\end{center}
\end{table*}

In general, the $Z^{\prime}$ boson in this model can decay into all SM fermions,  into supersymmetric particles: 
squark, slepton, sneutrino, neutralino, chargino, in addition to Higgs-boson
pairs, $W$-boson pairs and $ZH$.

The three benchmark points, and all the parameters associated with them,  are given in Table~\ref{tab:inputs}. We
give VEVs, Yukawa couplings, trilinear couplings, mass ratios and
mixings for the gauginos and soft scalar fermion mass parameters. The 
low value of $\tan \beta \approx 1$ is consistent with  constraints
from $B_s \to \mu^+ \mu^-$ branching ratio \cite{Aaij:2012nna}. For each
benchmark scenario, the mass spectra for the supersymmetric partners
obtained are given in Table~\ref{tab:spec}. The mass of the
additional $Z^\prime$ boson is  $\gsim 4$ TeV, and consistent with the  ATLAS \cite{Aad:2019fac} analyses on $Z^\prime$ dilepton decays.  As seen in
Table~\ref{tab:inputs}, the VEVs of the additional scalars
($S_1,S_2$ and $S_3$) $v_{s_i}, i=1,2,3$  are mostly taken above the
TeV scale so that the $Z^\prime$ mass bound is satisfied independent of the value
of the chosen VEV of the scalar field $S$. For convenience, the
parameters $\mu_{\rm eff}$ and $h_s$ are taken as free parameters and
the VEV of $S$ is determined using the relation
\begin{eqnarray}
	\hspace{0.5cm} \mu_{\rm eff} =\frac{h_{s} v_s}{\sqrt{2}}.
	\hspace{0.5cm} \label{eq:rho}
\end{eqnarray}
	\footnotetext[4]{In the theoretical calculation of the quoted value for $\Delta M_K$ ratio assumes contributions from both the so-called short distance and long distance physics. However, the latter part is not very reliable and needs improvement which might drive the value in either direction. The ratio becomes  $1.12\pm 0.44$ if only the short distance contribution is kept.}

The differences between the benchmarks are the following. In BP1, the gaugino mass parameters $M_{\tilde Y} (200~ {\rm GeV}) \ll M_{\tilde W} (2000~ {\rm GeV})$, $R_{Y  Y^\prime}=4.8 $ is large, while the light (left-handed)  sneutrinos and sleptons have mass $\sim 500$ GeV and are approximately degenerate. The right-handed sneutrinos are slightly heavier. In BP2, the gauginos have intermediate mass parameters $M_{\tilde Y} (760 ~{\rm GeV})\simeq  M_{\tilde W} $, $R_{ Y Y^\prime}=0.01$ is very small, while the masses of the light (left-handed) sneutrinos  and light sleptons are degenerate and around 500 GeV. The heavy ones split from the light sector significantly and can have much larger masses (up to $\sim 3500$ GeV).
 In BP3, the gaugino mass parameters are both light $M_{\tilde Y} (260~ {\rm GeV})\simeq M_{\tilde W}$, $R_{ Y  Y^\prime}=0.01$ is very small, while, unlike the other two scenarios, the  right-handed sneutrino masses are much lighter than those of the left-handed sneutrinos. The masses of the sleptons run in this range.
The neutralino parameters affect the LSP  and its composition, while the slepton and sneutrino masses affect branching ratios of $Z^\prime$ into sfermions.

\begin{table*}[htbp]
	\setlength{\extrarowheight}{2pt}
	\begin{center}
		\caption{ Decay width (in GeV),  width over mass ratios,  and dominant branching ratios (in \%) of
			$Z^{\prime}$ boson decay channels for the three scenarios
			considered.  The total branching ratios for decay modes  with ${\rm BR}_i< 1\% $ are also shown separately.}
		\label{tab:crsrlc}
		\begin{tabular*}{0.99\textwidth}{@{\extracolsep{\fill}} lccc}
			\hline\hline
			$\rm Width~[GeV]~and~Branching~ Ratios ~ [\%]$  &$\rm BP1 $
			\\ \hline 
			$\Gamma_{Z^\prime}$& 386  &&\\ 
			$\Gamma_{Z^\prime}/M_{Z^\prime}[\%]$&9.0 &&\\  
			\hline
			${\rm BR} (Z^{\prime}\rightarrow \sum_\ell \tilde{\nu}_{\ell_R}
			\tilde{\nu}_{\ell_R}
			)$ &15.69 &&\\
			${\rm BR} (Z^{\prime}\rightarrow \tilde\chi^{\pm}_1\tilde\chi^{\mp}_1
			)$ &2.93  &&\\
			${\rm BR}(Z^{\prime}\rightarrow \tilde\chi^0_3\tilde\chi^0_4
			)$ &2.09  &&\\
			${\rm BR}(Z^{\prime}\rightarrow \sum_\ell \nu_{\ell} \bar{\nu}_{\ell})$ &38.70  \\
			${\rm BR}(Z^{\prime}\rightarrow \sum_q q_d \bar{q}_d
			)$&15.39  &&\\
			${\rm BR}(Z^{\prime}\rightarrow \sum_q q_u \bar{q}_u)$ &12.33 &&\\
			${\rm BR} (Z^{\prime}\rightarrow \sum_\ell \ell \bar{\ell}
			)$ &4.08  &&\\
			$\sum_i \big[{\rm BR_i}(Z^{\prime}\rightarrow {\rm others}) <1\%\big]$  &8.79  &&\\			
			\hline
			$\rm Width~[GeV]~and~Branching~ Ratios~ [\%] $ & &$\rm BP2 $ &
			\\ \hline
			$\Gamma_{Z^\prime}$& &70.8  &\\ 
			$\Gamma_{Z^\prime}/M_{Z^\prime}[\%]$& &1.7 & \\  
			\hline
			${\rm BR}(Z^{\prime}\rightarrow \tilde\chi^{\pm}_1\tilde\chi^{\mp}_1
			)$ & &5.27 & \\
			${\rm BR}(Z^{\prime}\rightarrow \tilde\chi^0_2\tilde\chi^0_3
			)$ & &4.09 & \\
			${\rm BR} (Z^{\prime}\rightarrow H^+H^-)$ & &1.00 &\\
			${\rm BR} (Z^{\prime}\rightarrow H^0_2A^0_1)$ & &1.26 & \\
			${\rm BR}(Z^{\prime}\rightarrow \sum_q q_u \bar{q}_u)$ & &36.60 &\\
			${\rm BR}(Z^{\prime}\rightarrow \sum_\ell \nu_{\ell} \bar{\nu}_{\ell})$ & & 28.98 & \\
			${\rm BR}(Z^{\prime}\rightarrow \sum_q q_d \bar{q}_d
			)$ &  & 12.24 & \\
			${\rm BR} (Z^{\prime}\rightarrow W^+W^-)$ & &1.42 & \\
			${\rm BR} (Z^{\prime}\rightarrow ZH^0_1)$ & &1.21 & \\
			${\rm BR} (Z^{\prime}\rightarrow \sum_\ell \ell \bar{\ell}
			)$ & & 3.72 &\\
			$\sum_i \big[{\rm BR_i}(Z^{\prime}\rightarrow {\rm others}) <1\% \big]$  & &4.21  &\\
			\hline 
			$\rm Width~[GeV]~and~Branching~ Ratios~ [\%] $  && &$\rm BP3 $
			\\ \hline 
			$\Gamma_{Z^\prime}$& && 351  \\ 
			$\Gamma_{Z^\prime}/M_{Z^\prime}[\%]$& && 6.7 \\  
			\hline
			${\rm BR}(Z^{\prime}\rightarrow \sum_{\ell=1}^2 \tilde\ell_R\tilde\ell_R
			)$ &&& 6.02 \\
			${\rm BR}(Z^{\prime}\rightarrow \tilde\tau_1\tilde\tau_1
			)$ &&& 3.01 \\
			${\rm BR}(Z^{\prime}\rightarrow \tilde\chi^0_2\tilde\chi^0_4
			)$ &&& 1.02\\
			${\rm BR}(Z^{\prime}\rightarrow \sum_{\ell=1}^2 \tilde\ell_L\tilde\ell_L
			)$ &&& 2.02 \\
			${\rm BR}(Z^{\prime}\rightarrow \tilde\tau_2\tilde\tau_2
			)$ &&& 1.01 \\
			${\rm BR} (Z^{\prime}\rightarrow \sum_\ell \tilde{\nu}_{\ell_L}
			\tilde{\nu}_{\ell_L}
			)$ &&& 3.03 \\
			${\rm BR}(Z^{\prime}\rightarrow \sum_q q_u \bar{q}_u $ &&&34.50 \\
			${\rm BR}(Z^{\prime}\rightarrow \sum_\ell \ell \bar{\ell}
			)$ &&& 29.22 \\
			${\rm BR}(Z^{\prime}\rightarrow \sum_\ell \nu_{\ell} \bar{\nu}_{\ell})$ &&& 10.29  \\
			$\sum_i \big[{\rm BR_i}(Z^{\prime}\rightarrow {\rm others}) <1\% \big]$ & & &9.88  \\
			\hline\hline
		\end{tabular*}
	\end{center}
\end{table*}

\begin{figure*}[h]
	\begin{center}$
		\begin{array}{ccc}
			\hspace*{-0.6cm}
			\includegraphics[width=2.5in,height=2.5in]{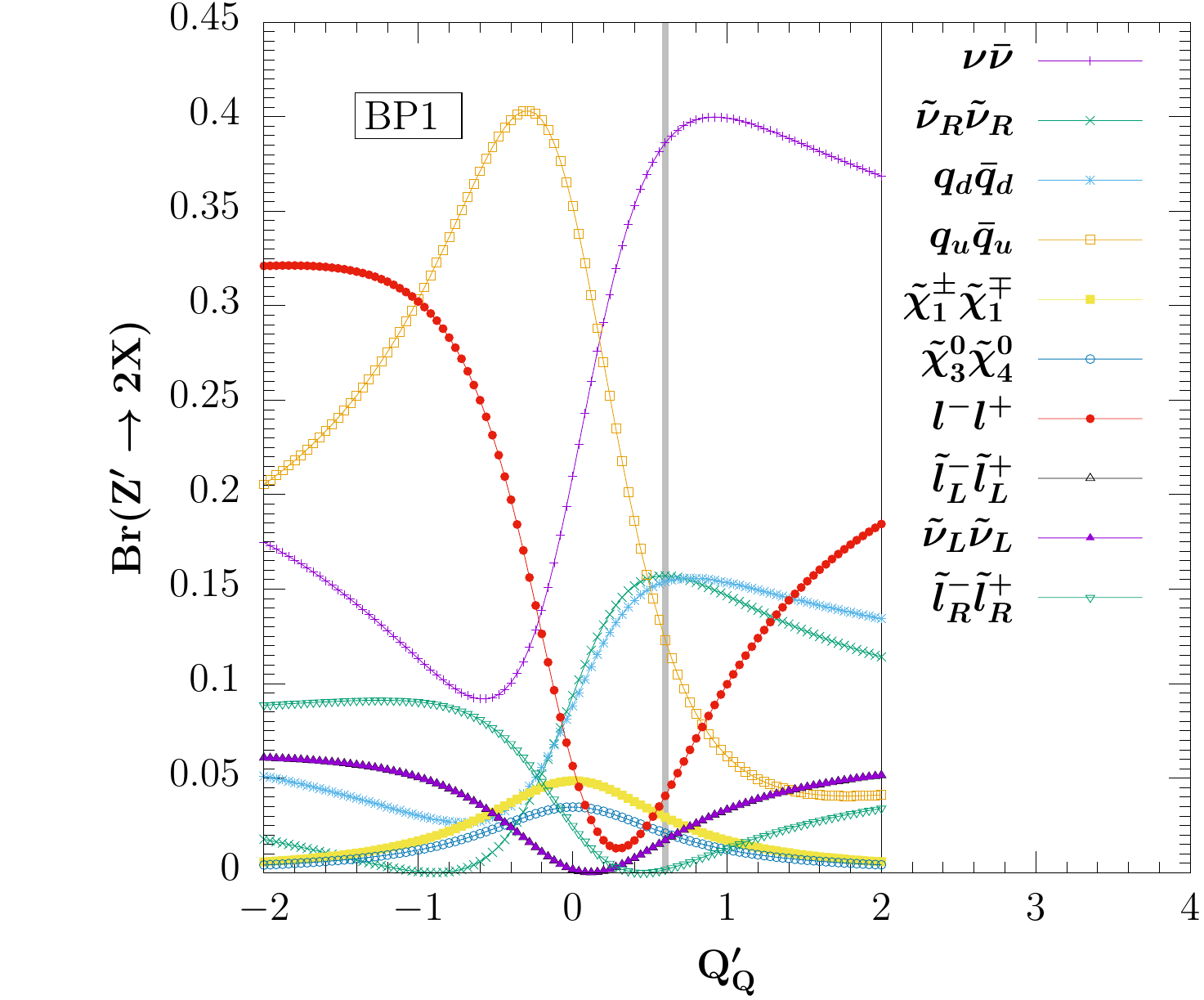}
			&\hspace*{-0.5cm}
			\includegraphics[width=2.5in,height=2.5in]{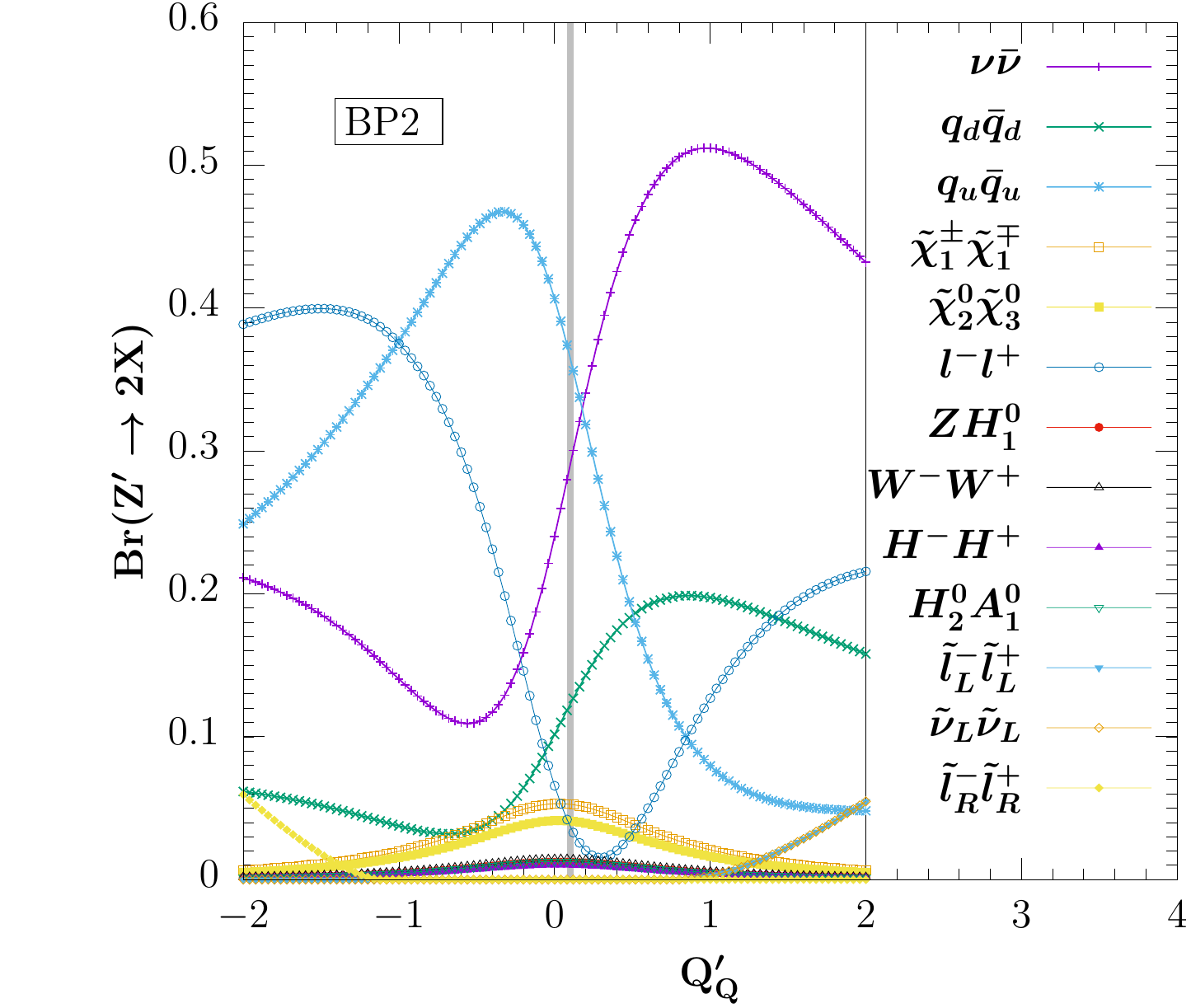}
			&\hspace*{-0.5cm}
			\includegraphics[width=2.5in,height=2.5in]{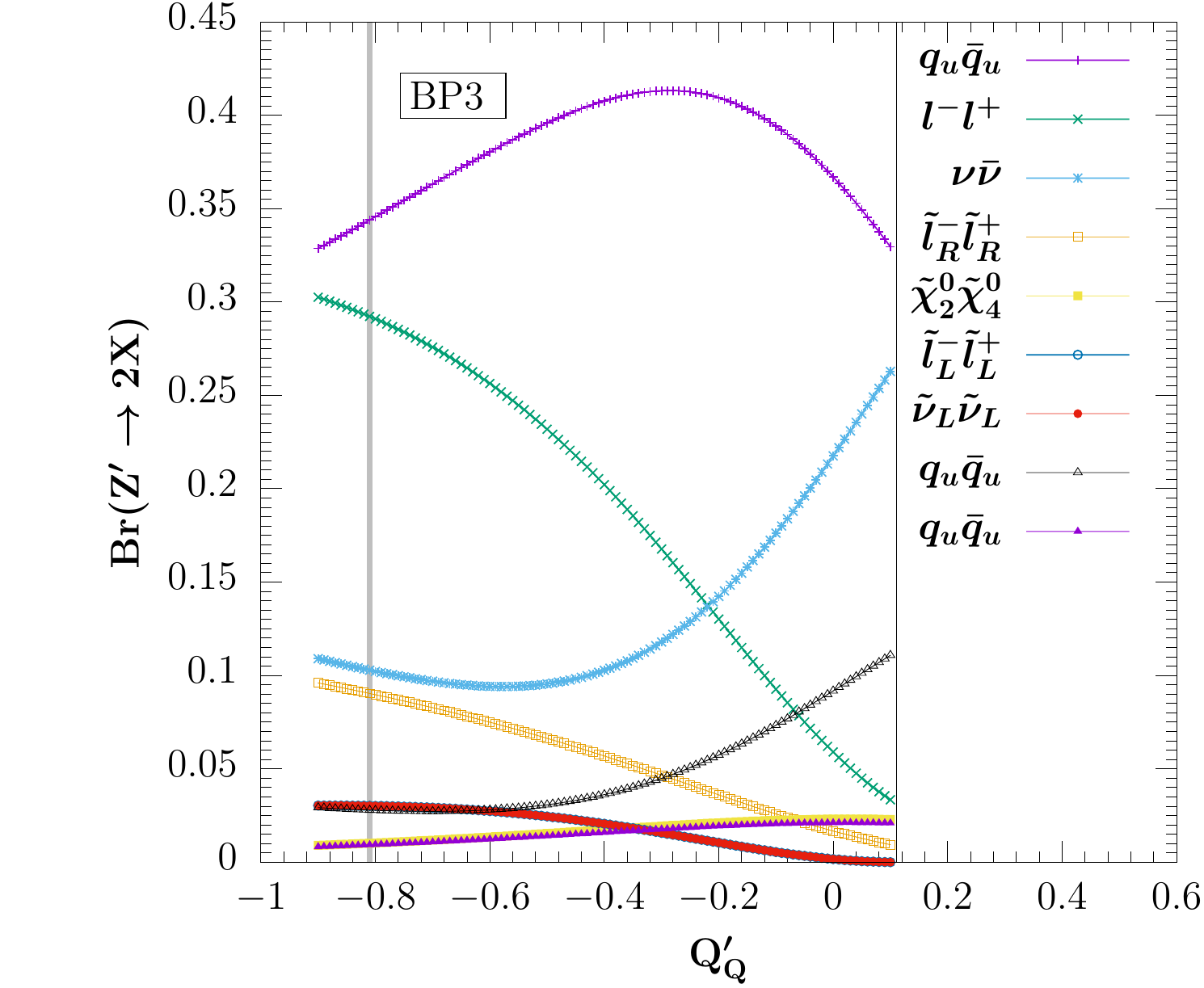}
		\end{array}$
	\end{center}
	\vskip -0.2in
	\caption{$Z^{\prime}$ decay branching ratios as a function of $Q^{\prime}_Q$ for BP1 (left), BP2  (middle) and BP3 (right).  For each benchmark we fix all other parameters as in Table~\ref{tab:inputs} except for $Q^{\prime}_Q$, which is allowed to vary. The choice of $Q^\prime_Q$ for each benchmark, chosen to maximize decays into supersymmetric particles, is indicated in each panel as a vertical grey line. Note that the branching ratios of $Z^\prime$ into the SM fermions and their superpartners as well as into the right-handed scalar neutrinos are all depicted after summing over three generations.} 
	\label{fig:ZprimeBR}
\end{figure*}
\begin{figure*}[htbp]
	\begin{center}$
		\begin{array}{ccc}
			\hspace*{-0.1cm}
			\fbox{\includegraphics[scale=0.32]{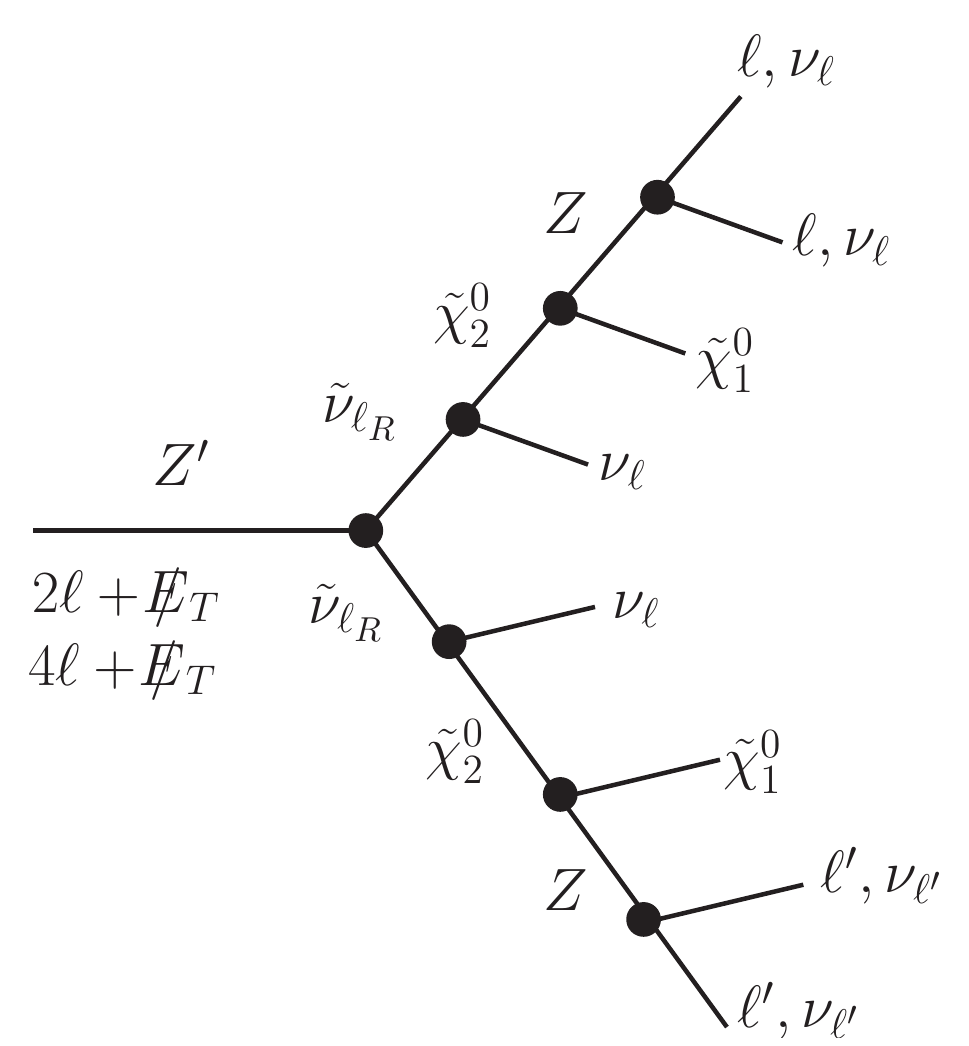}} & \fbox{\includegraphics[scale=0.4]{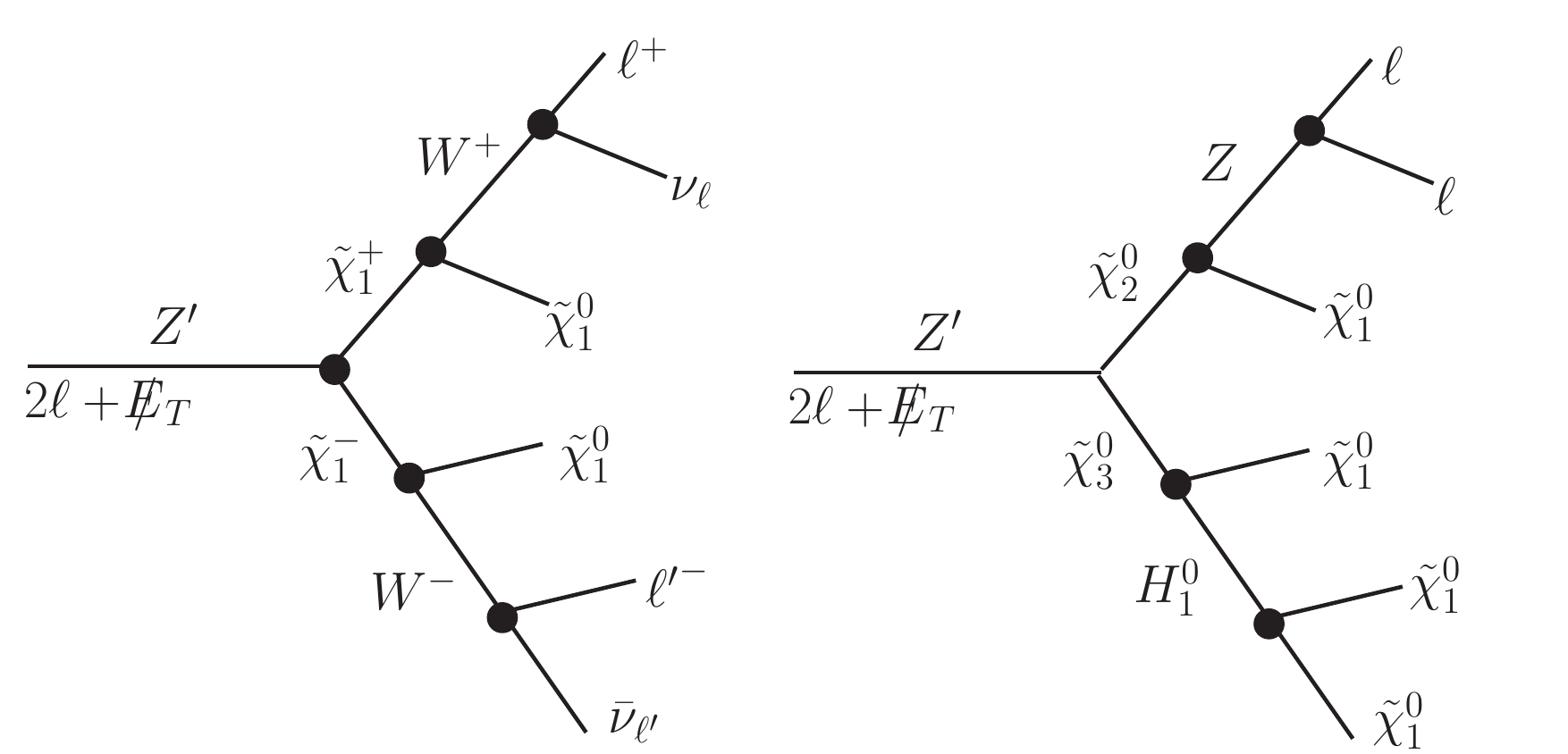}} &
			\fbox{\includegraphics[scale=0.33]{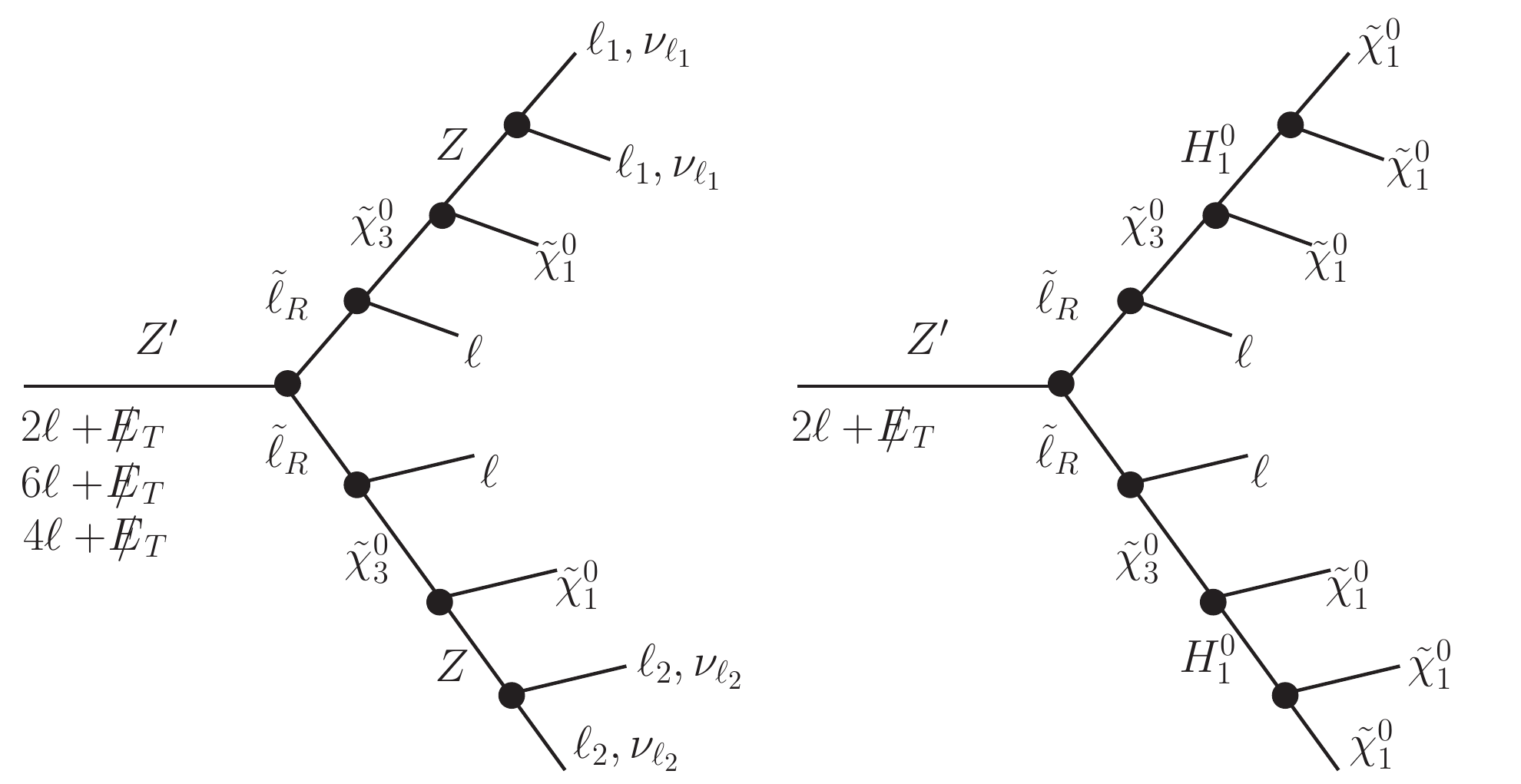}} \\[1em]    
			(a) &  (b) & (c) 
		\end{array}$
	\end{center}
	\vskip -0.2in
	\caption{The generic Feynman diagrams for the decay channels of the
		$Z^{\prime}$ in the secluded $U(1)^\prime$ model for BP1, BP2, and BP3 in the panels~(a), (b), and (c), respectively.}
	\label{fig:FeyndiagramsBPs}
\end{figure*}

The calculation of the relic density is performed importing the
model files from  {\tt CalcHEP}  \cite{Belyaev:2012qa}
into  the {\tt MicrOmegas} package \cite{Belanger:2018ccd}. All the numbers obtained are within the $1\sigma$ range of the WMAP
result  obtained from the Sloan
Digital Sky Survey \cite{Spergel:2006hy}
\begin{eqnarray}
	\Omega_{DM} h^2 = 0.111^{+0.011}_{-0.015}\,.
\end{eqnarray}
The relic density of dark matter $\Omega_{\rm DM} h^2$ is very
sensitive to the parameter $\displaystyle R_{Y^\prime}$. The value of the relic density  is shown in Table~\ref{tab:spec}, where we also give explicit values for masses of the physical eigenstates in the Higgs and sparticle sectors. In addition, we checked that the benchmarks satisfy low energy data. For this, various flavor observables are calculated with the help of the packages {\tt SARAH} \cite{Staub:2013tta,Staub:2015kfa} and {\tt SPheno} version 4.0.4 \cite{Porod:2003um,Porod:2011nf}. The results, normalized with  the corresponding SM values, are listed in Table~\ref{tab:low}. The values are all consistent with the current available data.  In the same table, we give the corrections of the secluded $U(1)^\prime$ to the SM values for the anomalous magnetic moments of electron, muon and tau, $\Delta a_e, \Delta a_\mu, \Delta a_\tau$. The measured values for the first two indicate a departure from the SM, in opposite directions for the electron \cite{Laporta:2017okg,Aoyama:2017uqe,Aoyama:2012wj} and muon \cite{Bennett:2006fi,Aoyama:2012wk}:
\begin{eqnarray}
	\Delta a_e&=&a_e^{\rm exp}-a_e^{\rm SM}=-(8.7\pm 3.6)\times 10^{-13}\, , \nonumber \\
	\Delta a_\mu&=&a_\mu^{\rm exp}-a_\mu^{\rm SM}=(2.7\pm 0.9)\times 10^{-9}\, .
\end{eqnarray}
In our benchmarks, the  contributions are too small to saturate these differences, as these were chosen to instead yield interesting $Z^\prime$ phenomenology. If we would  aim to satisfy constraints on $(g-2)_\mu$, we would choose values for the chargino, neutralino, slepton and sneutrino masses consistent with anomalous magnetic moment constraints.  Since the chargino and neutralino sector masses are not directly connected to the $Z^\prime$ boson mass, they would affect the $Z^\prime$ phenomenology through decays.
As the chargino-sneutrino loop is expected to be similar to the one in MSSM, the difference will arise from the lightest neutralino contribution. Its mass and composition are important, and so is the mass of the slepton. In addition, in both MSSM and $U(1)^\prime$ models, the anomalous magnetic moment depends almost linearly on $\tan \beta$ \cite{Czarnecki:1998nd,Barger:2004mr}. A larger value of $\tan \beta$ may increase the contributions to $(g-2)_\mu$, but for our benchmarks, this is in conflict with flavor constraints.

Both ATLAS and CMS collaborations have searched for $Z^\prime$ bosons. The assumption is that they  are produced in $pp$, then decay into SM particles.  The decay channels explored  are  $jj$ \cite{Aad:2019hjw,CMS:2019oju,Aaboud:2018fzt}, $b\bar{b}$ \cite{Aaboud:2018tqo}, $t\bar{t}$ \cite{Sirunyan:2018ryr,Aaboud:2018mjh}, $e^{+}e^{-}$ \cite{CMS:2019tbu,Aad:2019fac}, $\mu^{+}\mu^{-}$ \cite{CMS:2019tbu,Aad:2019fac}, $\tau^{+}\tau^{-}$ \cite{Aaboud:2017sjh,Khachatryan:2016qkc}, $W^{+}W^{-}$ \cite{Aaboud:2017fgj,Sirunyan:2019vgt}, and $ZH_1^0$ \cite{Sirunyan:2019vgt}, within a variety of models with extended $U(1)^\prime$ and $SU(2)$ gauge groups.  Of these channels, the leptonic decays $e^{+}e^{-}$ and $\mu^{+}\mu^{-}$ impose the most  stringent constraints on the $Z^\prime$ mass, normally $M_{Z^\prime} \gsim 4.3$ TeV. However, all these analyses assumed non-supersymmetric scenarios. It has been shown that, including supersymmetry, these bounds can be reduced by $\sim 300$ GeV \cite{Araz:2017wbp}. Hence, as we wish to explore the largest parameter space possible, we shall assume that $M_{Z^\prime} \ge 4.0$ TeV.

We calculated the branching ratios of the $Z^{\prime}$ decaying
into various final states for the three selected benchmark points, and show the results for the dominant ones in Table \ref{tab:crsrlc}. As expected, branching ratios for decays into quarks (BP2 and BP3) or neutrinos (BP1) dominate over those into supersymmetric particles.  The benchmarks were chosen for non-negligible decays into SUSY particles pairs, and are dominated by decays into sneutrinos and chargino pairs (BP1),  chargino and neutralino pairs (BP2), and into slepton pairs (BP3). In Table~\ref{tab:crsrlc} we also test the width/mass ratio. For all benchmarks considered, $\Gamma_{Z^\prime}/M_{Z^\prime}$ remains safely under 10\%, justifying treating $Z^\prime$ as a narrow resonance. 
The branching ratios of $Z^\prime$ are very sensitive to variations in   $Q^\prime_Q$, the $U(1)^\prime$ charge for the left-handed quark doublet. To highlight $Z^\prime$ decays into supersymmetric channels, we fixed all parameters for the chosen benchmarks BP1, BP2 and BP3 as in Table \ref{tab:inputs}, except $Q^\prime_Q$, which is allowed to vary freely.  In Fig.\ref{fig:ZprimeBR} we plot  the branching ratios of the
$Z^{\prime}$ as a function of $Q^{\prime}_Q$ for each benchmark scenario. The particular choice for $Q^\prime_Q$ for each benchmark as given in Table~\ref{tab:inputs}, is obtained by requiring that some branching ratios into supersymmetric particles be maximal. We indicated these choices in each panel of Fig. \ref{fig:ZprimeBR} as a vertical grey line.

Typically, the SUSY decay modes include (i) $Z^{\prime}\rightarrow
\tilde{\nu}_{\ell_R} \tilde{\nu}_{\ell_R}\\ \rightarrow \twoleps$ or
$\fourleps$, (ii) $Z^{\prime}\rightarrow
\tilde\chi^{\pm}_1\tilde\chi^{\mp}_1 \rightarrow \twoleps$,
(iii) $Z^{\prime}\rightarrow \tilde\chi^{0}_2\tilde\chi^{0}_3
\rightarrow \twoleps$, (iv) $Z^{\prime}\rightarrow
\tilde\ell_R\tilde\ell_R\rightarrow \twoleps$, or $\fourleps$, or
$\sixleps$ etc. Such pure leptonic modes give rise to a signature
consisting of charged-leptons and large missing energies, which are particularly well suited for observation at the LHC.
To determine and classify all possible signals for the three
scenarios we look into the decay topology of these particles.
We classify signals according to the final number of leptons present
in the signal events. The generic Feynman diagrams contributing dominantly to channels leading to signals with leptons and missing energy are shown in Fig. \ref{fig:FeyndiagramsBPs} for BP1 (a), BP2 (b) and BP3 (c).

\begin{table*}[htbp]
	\begin{center}
		\caption{\label{tab:cuts} The set of kinematical cuts used to isolate signal events from background.}
		\setlength{\extrarowheight}{2pt}
		\begin{tabular*}{0.99\textwidth}{@{\extracolsep{\fill}} cccc}
			\hline\hline
			& $\rm @14\,TeV$  &$\rm @27\,TeV$ &$\rm @100\,TeV$
			\\ \cline{1-4}\cline{1-3}
			$2\ell +\EmissT$ \\
			&$|\eta|<2.5$&$|\eta|<2.5$&$|\eta|<2.5$ \\
			&$\Delta R_{\ell \ell} \ge 0.5$&$\Delta R_{\ell \ell} \ge 0.5$&$\Delta R_{\ell \ell} \ge 0.5$ \\
			&$p_T(\ell_1)>475$ GeV&$p_T(\ell_1)>500$ GeV&$p_T(\ell_1)>2000$ GeV \\
			&$p_T(\ell_2)>50$ GeV&$p_T(\ell_2)>300$ GeV&$p_T(\ell_2)>1000$ GeV \\
			&$\EmissT> 50$ GeV&$\EmissT> 400$ GeV&$\EmissT> 2300$ GeV \\

			$4\ell +\EmissT$ \\
			&$|\eta|<2.5$&$|\eta|<2.5$&$|\eta|<2.5$ \\
			&$\Delta R_{\ell \ell} \ge 0.5$&$\Delta R_{\ell \ell} \ge 0.5$&$\Delta R_{\ell \ell} \ge 0.5$ \\
			&$p_T(\ell_1)>100$ GeV&$p_T(\ell_1)>100$ GeV &$p_T(\ell_1)>100$ GeV \\
			&$p_T(\ell_2)>50$ GeV&$p_T(\ell_2)>50$ GeV&$p_T(\ell_2)>50$ GeV \\
			&$p_T(\ell_3)>25$ GeV&$p_T(\ell_3)>25$ GeV&$p_T(\ell_3)>25$ GeV \\
			&$p_T(\ell_4)>15$ GeV&$p_T(\ell_4)>15$ GeV&$p_T(\ell_4)>15$ GeV \\
			&$\EmissT> $ 400 GeV &$\EmissT>350 $ GeV &$\EmissT>800 $ GeV \\

			$6\ell +\EmissT$& \\
			&$|\eta|<2.5$&$|\eta|<2.5$&$|\eta|<2.5$ \\
			&$\Delta R_{\ell \ell} \ge 0.2$&$\Delta R_{\ell \ell} \ge 0.2$&$\Delta R_{\ell \ell} \ge 0.2$ \\
			&$p_T(\ell_1)>50$ GeV&$p_T(\ell_1)>50$ GeV &$p_T(\ell_1)>100$ GeV \\
			&$p_T(\ell_2)>20$ GeV&$p_T(\ell_2)>20$ GeV &$p_T(\ell_2)>50$ GeV \\
			&$p_T(\ell_3)>20$ GeV&$p_T(\ell_3)>20$ GeV &$p_T(\ell_3)>20$ GeV \\
			&$p_T(\ell_4)>20$ GeV&$p_T(\ell_4)>20$ GeV &$p_T(\ell_4)>20$ GeV \\
			&$p_T(\ell_5)>10$ GeV&$p_T(\ell_5)>10$ GeV &$p_T(\ell_5)>15$ GeV \\
			&$p_T(\ell_6)>5$ GeV&$p_T(\ell_6)>5$ GeV &$p_T(\ell_6)>5$ GeV \\
			&$\EmissT>100 $ GeV &$\EmissT>100 $ GeV &$\EmissT>100 $ GeV \\
			
			\hline\hline
		\end{tabular*}
	\end{center}
\end{table*}

The events are generated at the partonic level with {\tt CalcHEP}
\cite{Belyaev:2012qa} and they are subsequently passed to  {\tt PYTHIA8} \cite{Sjostrand:2006za,Sjostrand:2014zea} for decay, showering and hadronization. Events which are saved in {\tt HepMC} format \cite{Dobbs:2001ck} are then passed to {\tt MadAnalysis} \cite{Conte:2012fm}
for applying cuts and  further data analysis. {\tt Delphes} \cite{deFavereau:2013fsa,Selvaggi:2014mya,Mertens:2015kba} is used for fast detector simulations. We simulated  events for
the  $\twoleps$, $\fourleps$ and $\sixleps$ signals at the LHC with  $14 \tev$ center-of-mass energy, the HE-LHC with $27 \tev $, and as well as the FCC-hh with $100 \tev$. We used different PDF sets to model parton distributions for the colliders at different center of mass energies.  While CTEQ6l1 PDF set \cite{Pumplin:2002vw} was used for the $14 \tev$ LHC case, the PDF set from the {\tt PDF4LHC15} collaboration \cite{Butterworth:2015oua} was used for both HE-LHC and FCC-hh. In the numerical study, for the calculation of signal significance,  we have taken the integrated luminosities  ${\cal L}=3$ ab$^{-1}$,\,15\ ab$^{-1}$, and 30 ab$^{-1}$  for the HL-LHC, HE-LHC, and FCC-hh, respectively. 

\begin{figure*}[htbp]
	\begin{center}$
		\begin{array}{ccc}
			\hspace*{-0.8cm}
			\includegraphics[height=6.5cm,width=6.5cm]{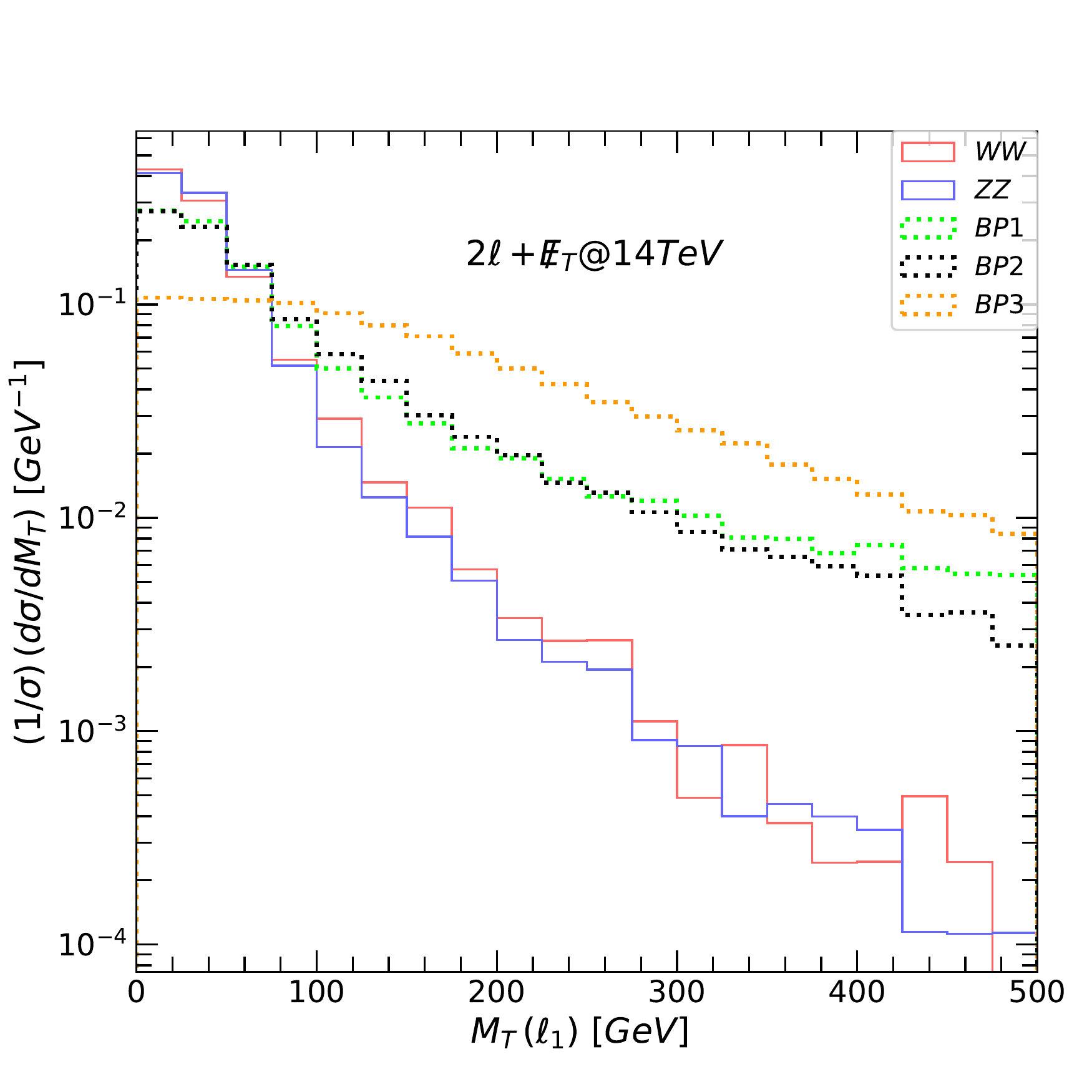} &\hspace*{-0.3cm}
			\includegraphics[height=6.5cm,width=6.5cm]{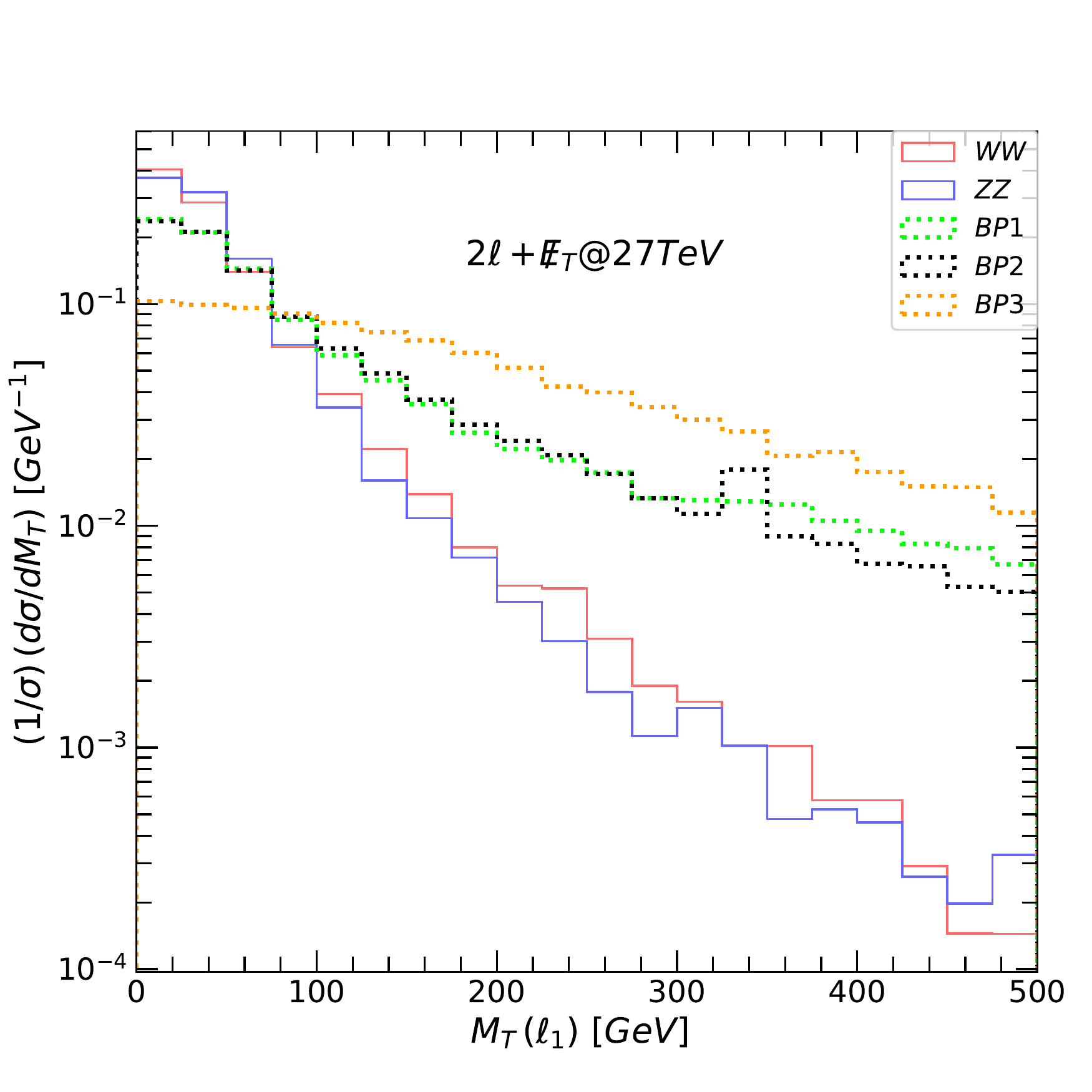} &\hspace*{-0.3cm}
			\includegraphics[height=6.5cm,width=6.5cm]{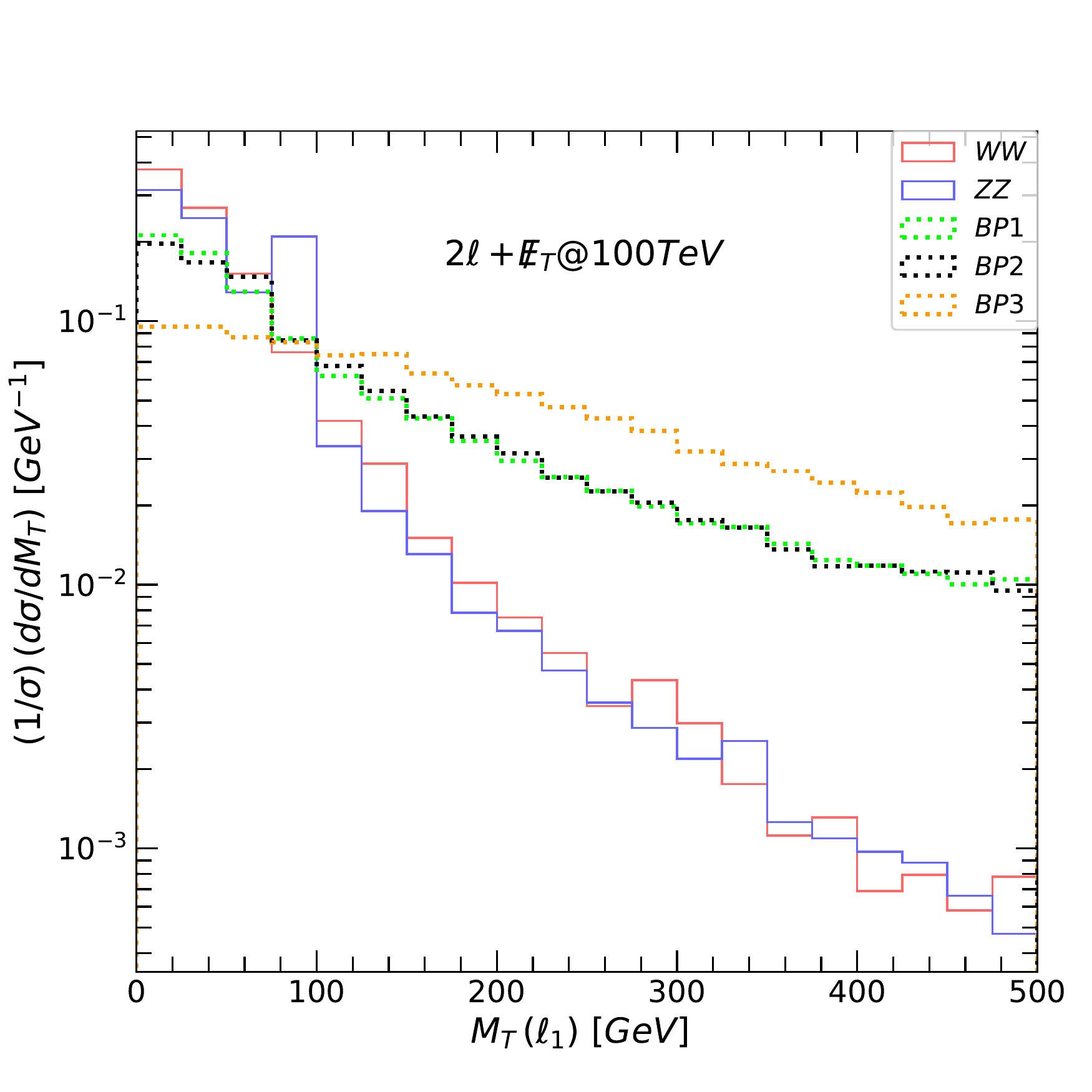} \\[-1em] \hspace*{-0.8cm}
			\includegraphics[height=6.5cm,width=6.5cm]{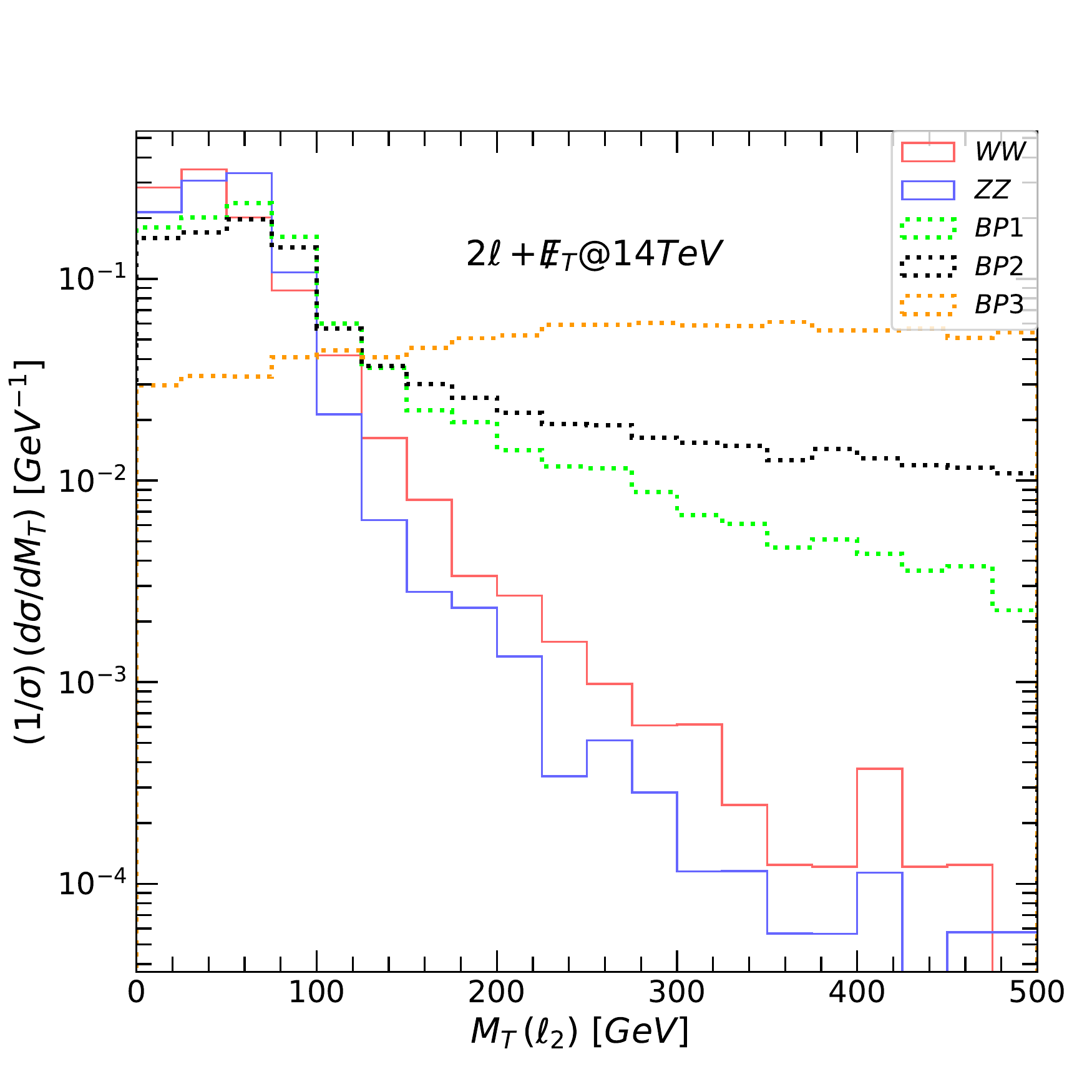} &\hspace*{-0.3cm}
			\includegraphics[height=6.5cm,width=6.5cm]{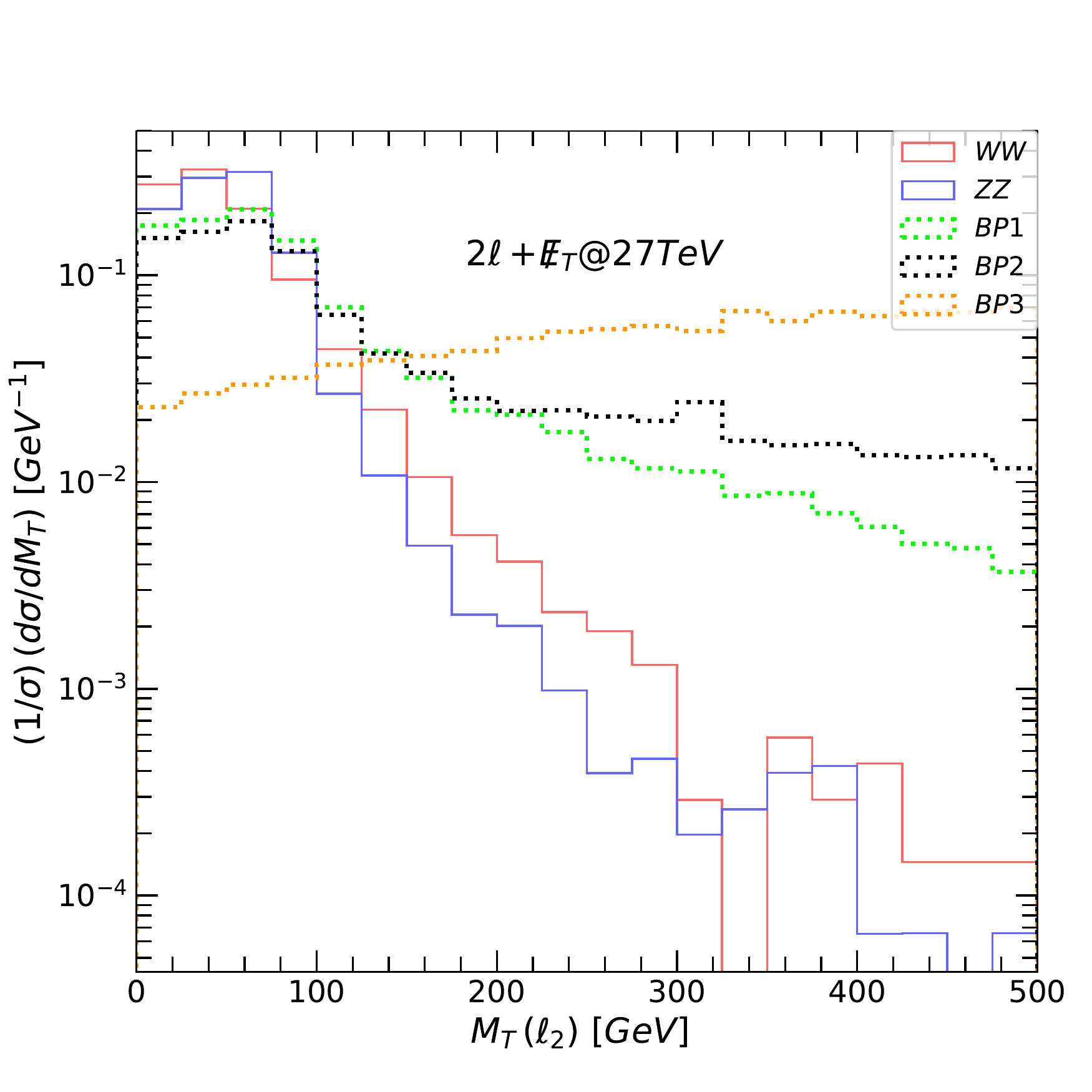} &\hspace*{-0.3cm}
			\includegraphics[height=6.5cm,width=6.5cm]{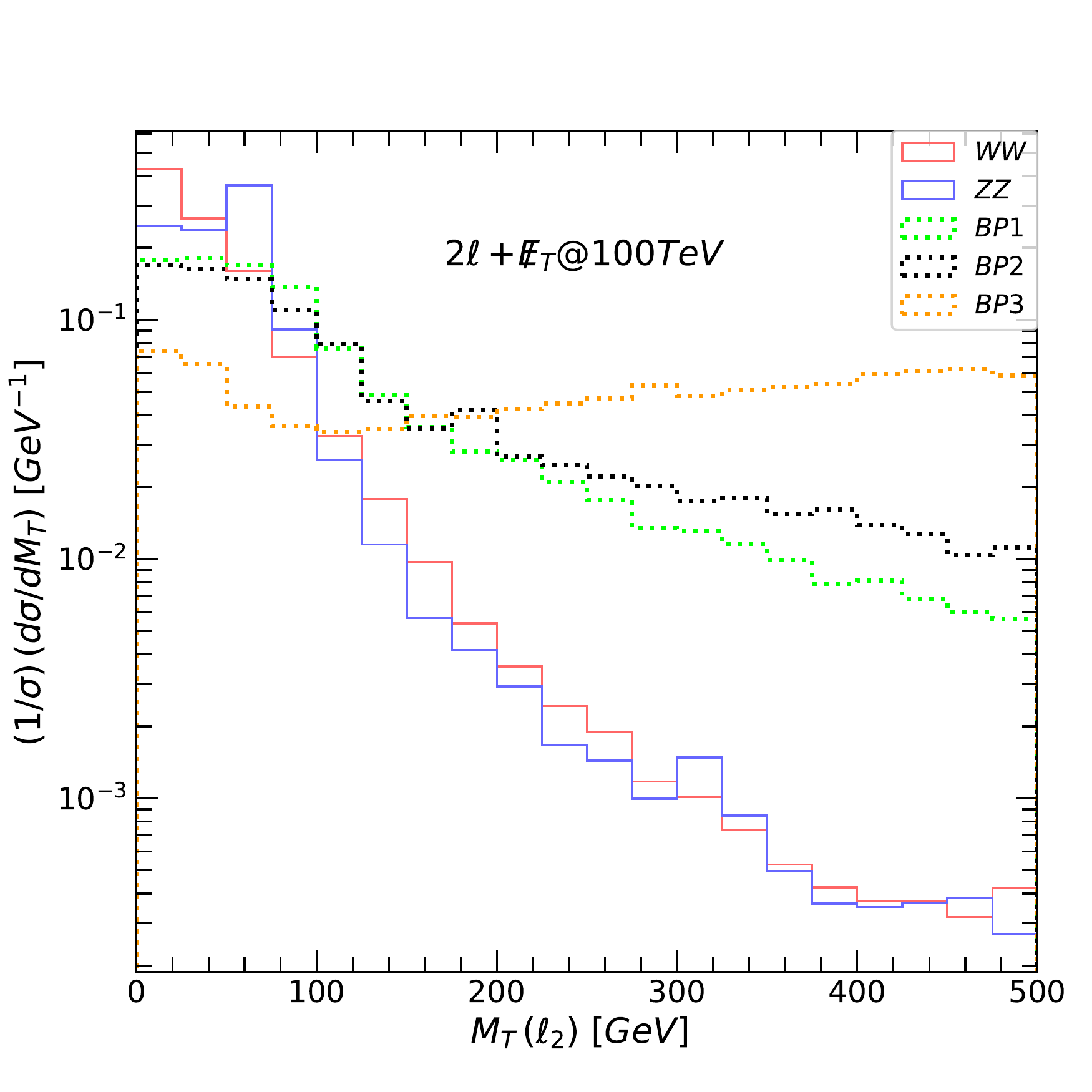}  
		\end{array}$
	\end{center}
	\vskip -0.3in
	\caption{The transverse mass $M_T$,  for the leading lepton $\ell_1$ (top panels) and next-to-leading lepton $\ell_2$ (bottom panels), for the signal and background in the $\twoleps$ scenario. (Left-hand)  signals and backgrounds at 14 TeV,  (middle)  at 27 TeV, and (right-hand) at 100 TeV. The main backgrounds (di-bosons) are indicated in solid lines while the signals are plotted in dotted lines: green for BP1, black for BP2 and orange for BP3.}
	\label{fig:2lmt}
\end{figure*}
\begin{figure*}[htbp]
	\begin{center}$
		\begin{array}{ccc}
			\hspace*{-0.8cm}
			\includegraphics[height=6.5cm,width=6.5cm]{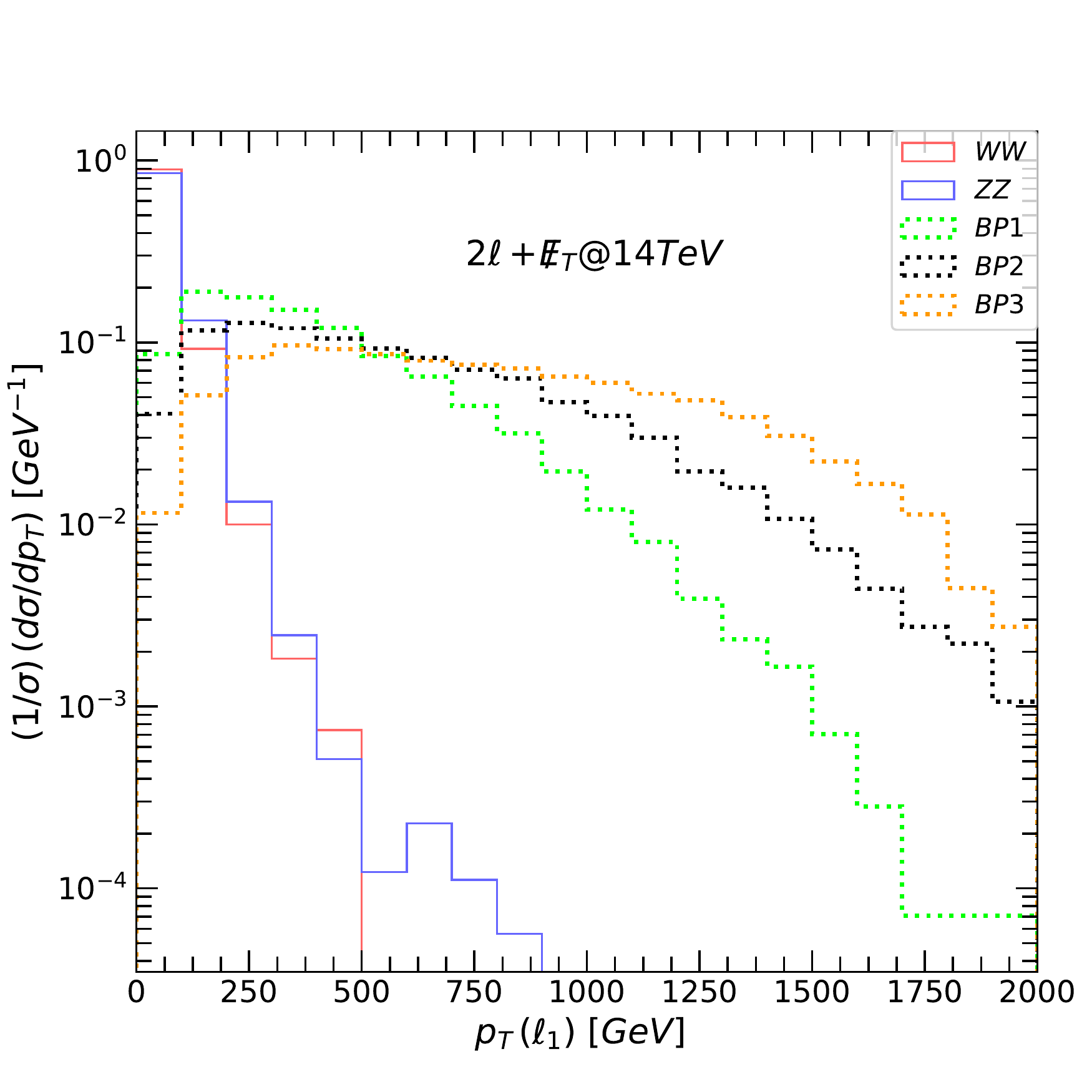} &\hspace*{-0.22cm}
			\includegraphics[height=6.5cm,width=6.5cm]{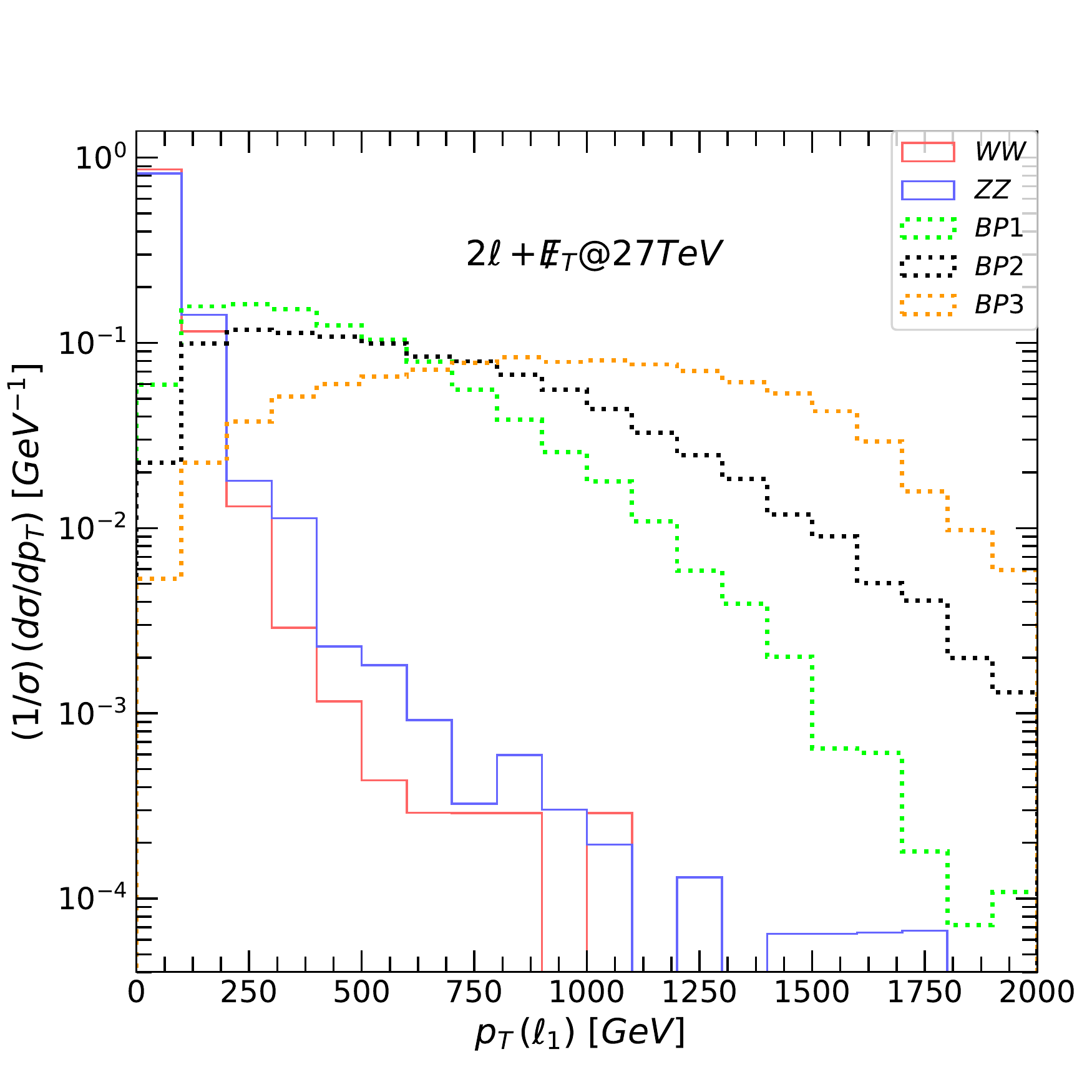} &\hspace*{-0.22cm}
			\includegraphics[height=6.5cm,width=6.5cm]{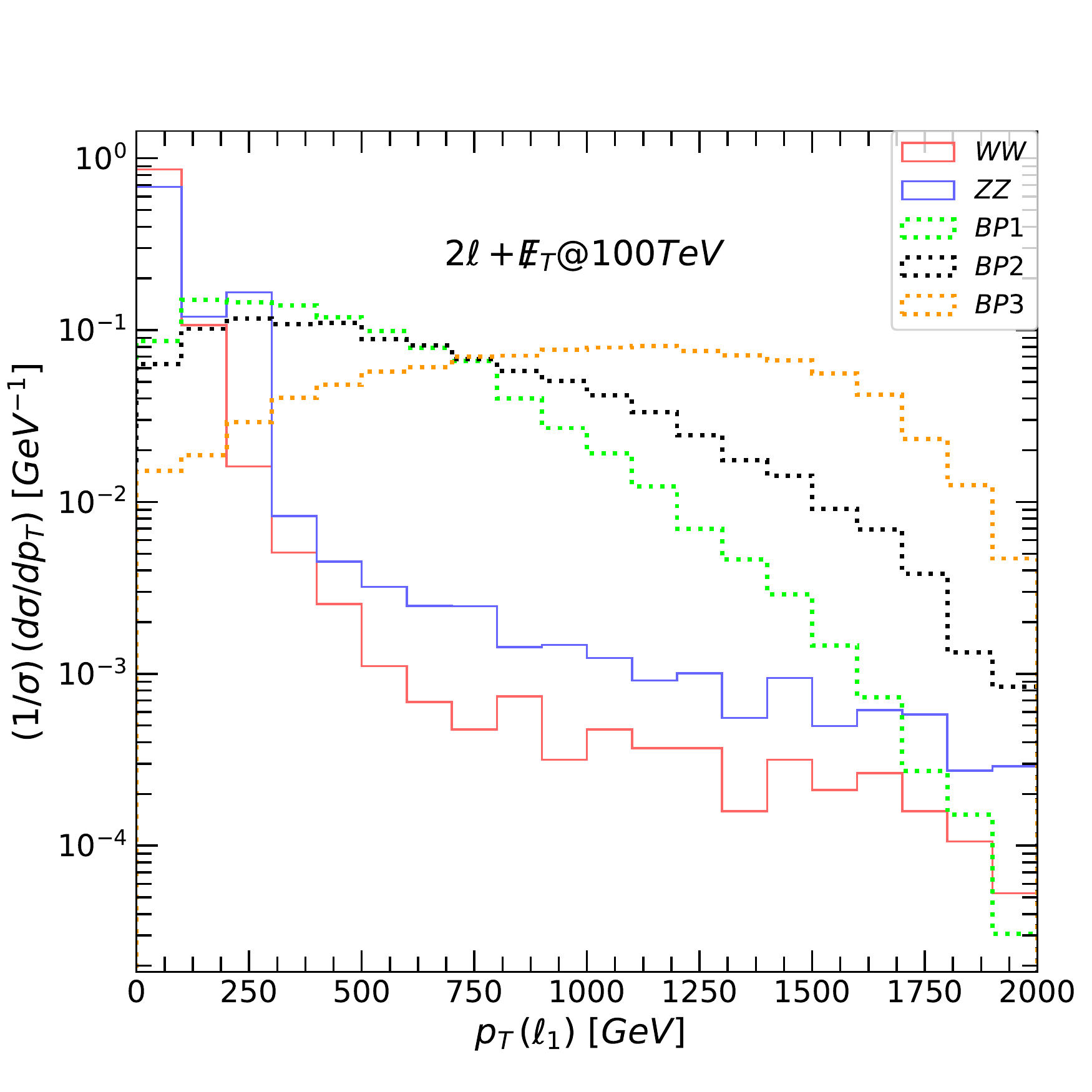} \\[-1em] \hspace*{-0.8cm}
			\includegraphics[height=6.5cm,width=6.5cm]{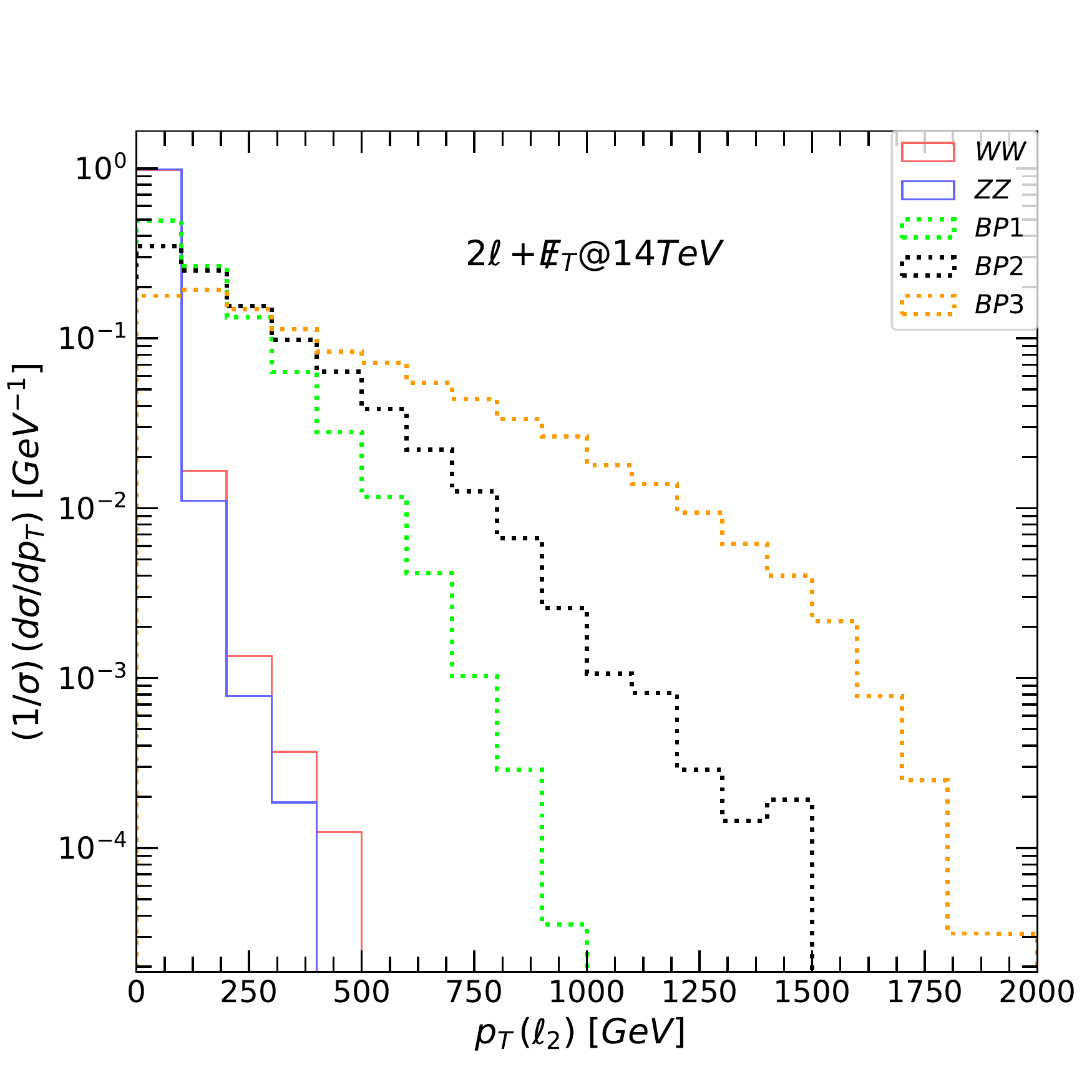} &\hspace*{-0.22cm}
			\includegraphics[height=6.5cm,width=6.5cm]{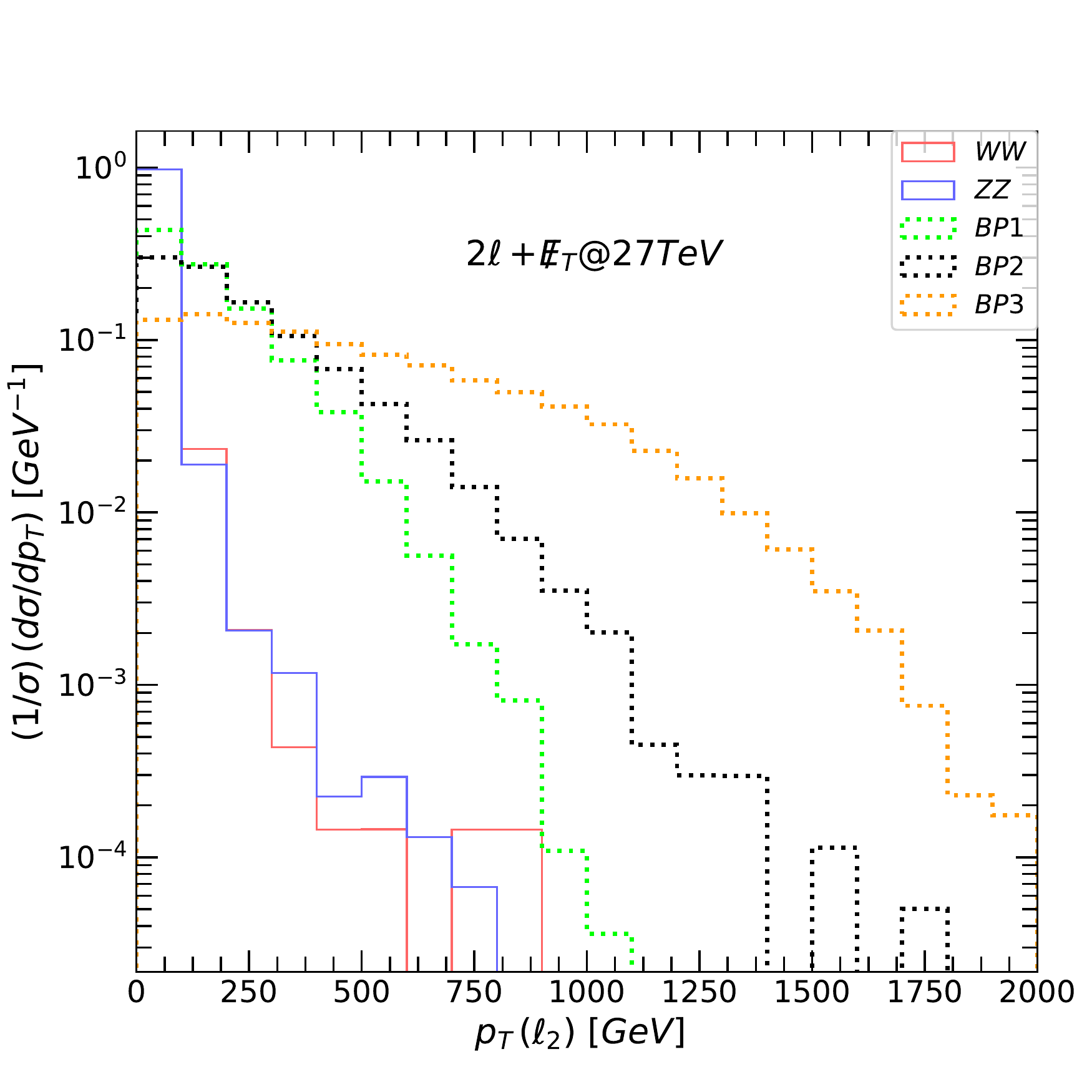} &\hspace*{-0.22cm}
			\includegraphics[height=6.5cm,width=6.5cm]{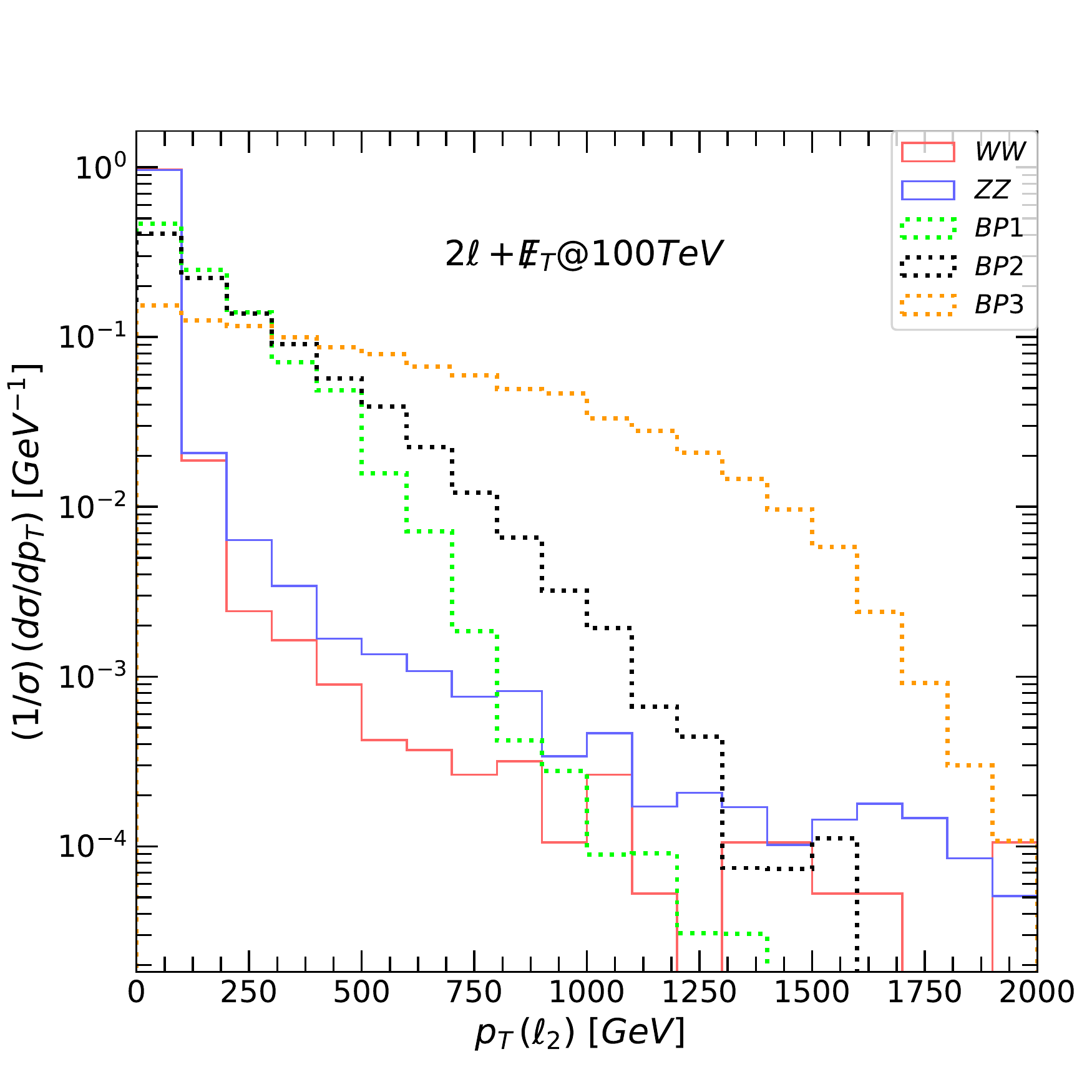}  
		\end{array}$
	\end{center}
	\vskip -0.3in	
	\caption{The transverse momentum $p_T$,  for the leading lepton $\ell_1$ (top panels) and next-to-leading lepton $\ell_2$ (bottom panels), for the signal and background in the $\twoleps$ scenario. (Left-hand)  signals and backgrounds at 14 TeV,  (middle)  at 27 TeV, and (right-hand) at 100 TeV. The main backgrounds (di-bosons) are indicated in solid lines while the signals are plotted in dotted lines: green for BP1, black for BP2 and orange for BP3.}
	\label{fig:2lpt}
\end{figure*}

For each benchmark, we use two different formulas for the significance of the signals \cite{Arganda:2018hdn,Abdallah:2015uba,Cowan:2010js}, where the first expression corresponds to the usual definition, and the second is more useful  for  smaller number of background events:
\begin{widetext}	
\begin{eqnarray}
	\sigma_{\cal A}&=&\frac{S}{\sqrt{S+B}}\,,\\
	\sigma_{\mathcal B}&=&\sqrt{2(S+B) \log \!\left[ \frac{(S+B)(B+\sigma_{B}^2)}{B^2+(S+B)\sigma_{B}^2}\right]-\frac{2 B^2}{\sigma_{B}^2}\log \! \left[1+ \frac{S\sigma_{B}^2}{B(B+\sigma_{B}^2)}\right]}\,,
\end{eqnarray}
\end{widetext}
where $S$ is the number of signal events, $B$ the number of background events, and $\sigma_{B}^2$ the standard deviation for background events.  We generated the events for the signal in each scenario, and we also simulated the SM background for the three benchmarks, separately for 14, 27 and 100  TeV. 

We concentrate on leptonic  final states, considered as  golden channels in experimental searches at LHC. To exploit these
features, this study will be focused on the decays of the
$Z^{\prime}$ boson into supersymmetric particles, leading to final
states with leptons and missing energy, due to the presence of
neutralinos or neutrinos. In the following, we present a study of
$Z^{\prime}$ decays into multileptonic final states for a given set of
the secluded sector $U(1)^{\prime}$ model parameters (BP1, BP2 and BP3), dividing our analysis into $\twoleps$, $\fourleps$ and $\sixleps$ signals.

\subsection{Multilepton analysis}
\label{subsec:analysis}
\begingroup
\setlength{\tabcolsep}{10pt} 
\renewcommand{\arraystretch}{0.95} 
\begin{table*}[htbp]
	\caption{ Signal selection strategy and cuts imposed in the  $\twoleps$ scenario at 14, 27 and 100 TeV. We give the cross-section for background and benchmark scenarios in fb. Statistical significances $\sigma_{\mathcal A}$ and $\sigma_{\mathcal B}$ of $\twoleps$ signal are given for each energy.}\label{tab:cuts-bckgr2l}
	\setlength{\extrarowheight}{2pt}
	\begin{tabular*}{\textwidth}{@{\extracolsep{\fill}} ccccc}
		\hline\hline
		$ 2\ell +\EmissT@14~{\rm TeV} $&\rm Background [fb]&$ \rm BP1~[fb]$&$\rm BP2~[fb]$&$\rm BP3~[fb]$\\
		\hline
		$\rm No~Cut$&$7.15\times 10^2$&$2.26\times 10^{-2}$&$8.25\times 10^{-3}$&$9.11\times 10^{-3}$\\ 
		$ |\eta_i|<2.5,\ \Delta R_{12} \ge 0.5$ &$6.99\times 10^2$&$1.14\times 10^{-2}$&$4.54\times 10^{-3}$&$9.06\times 10^{-3}$\\
		$ p_T(\ell_1)>475$ GeV&$7.13\times 10^{-2}$&$1.00\times 10^{-3}$&$1.61\times 10^{-3}$&$6.29\times 10^{-3}$\\
		$ p_T(\ell_2)>50$ GeV&$7.13\times 10^{-2}$&$1.52\times 10^{-4}$&$1.28\times 10^{-3}$&$6.01\times 10^{-3}$\\
		$ \EmissT>50~\rm GeV$ &$7.10\times 10^{-2}$&$1.5\times 10^{-4}$&$1.3\times 10^{-3}$&$5.9\times 10^{-3}$\\
		\hline
		$ Significance:\quad \sigma_{\mathcal A}$&${\cal L}=3\,{\rm ab}^{-1}$&$0.031\sigma$&$0.26\sigma$&$1.2\sigma$\\
		\hskip 2.1cm $\sigma_{\mathcal B}$ &$$&$0.022\sigma$&$0.18\sigma$&$0.85\sigma$\\
		\hline\hline
		$ 2\ell +\EmissT@27~\rm TeV $&&&&\\
		\hline
		$\rm No~Cut$&$1.59\times 10^3$&$3.54\times 10^{-1}$&$1.25\times 10^{-1}$&$1.94\times 10^{-1}$\\ 
		$ |\eta_i|<2.5,\ \Delta R_{12} \ge 0.5$ &$1.55\times 10^3$&$1.52\times 10^{-1}$&$6.43\times 10^{-2}$&$1.94\times 10^{-1}$\\
		$ p_T(\ell_1)>500~GeV$&$2.14$&$1.54\times 10^{-2}$&$1.18\times 10^{-2}$&$1.6\times 10^{-1}$\\
		$ p_T(\ell_2)>300$ GeV&$9.22\times 10^{-1}$&$3.35\times 10^{-4}$&$9.76\times 10^{-3}$&$1.12\times 10^{-1}$\\
		$ \EmissT>400~GeV$ &$1.8\times 10^{-1}$&$1.0\times 10^{-4}$&$4.4\times 10^{-3}$&$6.0\times 10^{-2}$\\
		\hline
		$\rm Significance:\quad \sigma_{\mathcal A}$&${\cal L}=15\,{\rm ab}^{-1}$&$0.029\sigma$&$1.2\sigma$&$14.9\sigma$\\
		\hskip 2cm $\sigma_{\mathcal B}$ &&$0.021\sigma$&$0.88\sigma$&$11.3\sigma$\\
		\hline\hline
		$ 2\ell +\EmissT@100~\rm TeV $&&&\\
		\hline
		$\rm No~Cut$&$1.89\times 10^4$&$9.11$&$3.23$&$6.52$\\ 
		$ |\eta_i|<2.5,\ \Delta R_{12} \ge 0.5$ &$8.41\times 10^3$&$3.36$&$6.62\times 10^{-1}$&$5.75$\\
		$ p_T(\ell_1)>2000$ GeV&$2.01\times 10^1$&$0.0$&$1.62\times 10^{-3}$&$5.15\times 10^{-2}$\\
		$\ p_T(\ell_2)>1000$ GeV&$ 8.94$&$0.0$&$1.2\times 10^{-4}$&$1.5\times 10^{-2}$\\
		$ \EmissT>2300~\rm GeV$ &$8.1\times 10^{-1}$&$0.0$&$0.0$&$3.5\times 10^{-4}$\\
		\hline
		${\rm Significance}:\quad \sigma_{\mathcal A}$&${\cal L}=30\,{\rm ab}^{-1}$&$0.0\sigma$&$0.0\sigma$&$0.067\sigma$\\
		\hskip 2cm $\sigma_{\mathcal B}$ 
		&&$0.0\sigma$&$0.0\sigma$&$0.048\sigma$\\
		\hline\hline	
	\end{tabular*}
\end{table*}
\endgroup

In this analysis, for each final state, we impose  cuts on the kinematical 
observables to suppress the SM background, as given in Table~\ref{tab:cuts}. Given the event topologies, stricter cuts on the leading lepton transverse momentum favor events with $\twoleps$ and $\fourleps$.  While the cuts on the angular variables and lepton separation remain the same, the kinematic cuts increase (in general), as expected going from 14 TeV to 27 TeV and eventually to 100 TeV.  Very stringent cuts are needed  for the case of $\twoleps$ signal,  and this is valid for the three center of mass energies but especially at 100 TeV. The final set of cuts are obtained by requiring  to maximize the signal significance. We proceed in turn to analyze each of the final states,  $\twoleps$, $\fourleps$ and $\sixleps$ signals and discuss their potential for observability.


\subsubsection{Two lepton signal: $\twoleps$}
\label{subsubsec:2leptons}

The main decay modes of $Z^\prime$ giving rise to dilepton final states are:
\begin{eqnarray}
	Z^{\prime}&\rightarrow& \tilde{\nu}_{\ell_R}
	\tilde{\nu}_{\ell_R}\rightarrow \twoleps,  \nonumber \\
	Z^{\prime}&\rightarrow& \tilde\chi^{\pm}_1\tilde\chi^{\mp}_1
	\rightarrow \twoleps, \nonumber \\
	Z^{\prime}&\rightarrow& \tilde\chi^{0}_2\tilde\chi^{0}_3 \rightarrow
	\twoleps, \nonumber \\
	Z^{\prime}&\rightarrow& \tilde\ell_R\tilde\ell_R\rightarrow
	\twoleps.
\end{eqnarray}
In the following figures, we first show the relevant kinematic variables for signals and background at 14, 27 and 100 TeV,  before any cuts were imposed. We plot the differential cross-section, normalized to unity, with individual bin contents  divided by the sum of all the data in the available bins. This way, the uncertainties are correlated across the bins, such that the uncertainties on the total integrated luminosity cancel. The resulting normalized differential fiducial cross-section is plotted as a function of various representative kinematic variables \cite{Chatrchyan:2012saa,PhysRevD.98.012003}. Let us define the variable transverse mass $M_T$ \cite{Sirunyan:2018mpc}
$$M_T(\ell)=
\sqrt{\Big(E_T(\ell)+\slashchar{E}_T\Big)^2- \Big(\vec{p}_T(\ell)+\slashchar{\vec{E}}_T\Big)^2}$$
for a system composed of a lepton $\ell$ and the invisible transverse momentum available in each event. Here $\slashchar{E}_T = |\slashchar{\vec{E}}_T|$. We show, in Fig. \ref{fig:2lmt}, $M_T$ of the leading lepton (top panels) and next-to-leading lepton (bottom panels), and in Fig. \ref{fig:2lpt}, the leading lepton transverse momentum  (top panels) and next-to-leading lepton transverse momentum (bottom panels). The left-hand side panels in both figures correspond to signals and backgrounds at 14 TeV, the middle at 27 TeV, and the right-hand side for 100 TeV. 


The main backgrounds (di-bosons for $\twoleps$) are indicated in solid lines while the signals are plotted in dotted lines, color coded: green for BP1, black for BP2 and orange for BP3. For our analysis we have checked  cross sections of other potential backgrounds (tri-bosons) for $\twoleps$,  such as $WWZ$ and $ZZZ$ and found that their cross sections at 14 TeV are about three orders of magnitude smaller than those of the di-bosons. Such tri-boson background suppressions hold also at 27 TeV and 100 TeV center of mass energies. Thus they are  subdominant and ignored in our analysis. We repeated the background calculations at each energy for the other signals (namely, $\fourleps$ and $\sixleps$) and only the dominant backgrounds, tri-bosons for $\fourleps$ and $ZZZZ$ for $\sixleps$, are kept in this analysis. As our signal is leptonic only, we selected certain number of isolated leptons + MET and vetoed any jets to avoid consistently signals and backgrounds due to QCD interactions.   As an extensive background calculation is beyond the scope of our work, we restricted ourselves with backgrounds without jets, thus neglecting any contribution from soft jets, or jets which will not survive cuts.

Both of these graphs show clearly that at large $M_T(\ell)$ and $p_T(\ell)$, the signal dominates the background, and the graphs justify our choice of kinematic cuts.  Table~\ref{tab:cuts-bckgr2l} gives values for signal and background cross-sections after each cut. We also show the signal significance, for both $\sigma_{\cal A}$ and $\sigma_{\cal B}$, for each benchmark, at proposed total integrated luminosity: for  14 TeV at ${\cal L}= 3$ ab$^{-1}$, for  27 TeV at ${\cal L}= 15$ ab$^{-1}$ and for  100 TeV at ${\cal L}= 30$ ab$^{-1}$.  While BP3 appears to be most promising, the significance for all benchmarks at 14 TeV is very low, dispelling any hope  for observing the $Z^\prime$ boson in the $\twoleps$ final state. This is not surprising, and in complete agreement with other findings \cite{Frank:2020kvp,Araz:2017qcs}. However, at 27 TeV the cuts imposed are especially effective for BP3, and we obtain large significances for both $\sigma_{\cal A}$ and $\sigma_{\cal B}$. At 100 TeV, the cuts imposed to reduce the background wiped out the signal, and we were not able to gain any predictable features. We had here two available options: either loose all the signal events  due to the stringent cuts applied, or allow the signals to be overwhelmed by large background events.

\begin{figure*}[htbp]
	\begin{center}$
		\begin{array}{ccc}
			\hspace*{-0.8cm}
			\includegraphics[height=6.5cm,width=6.5cm]{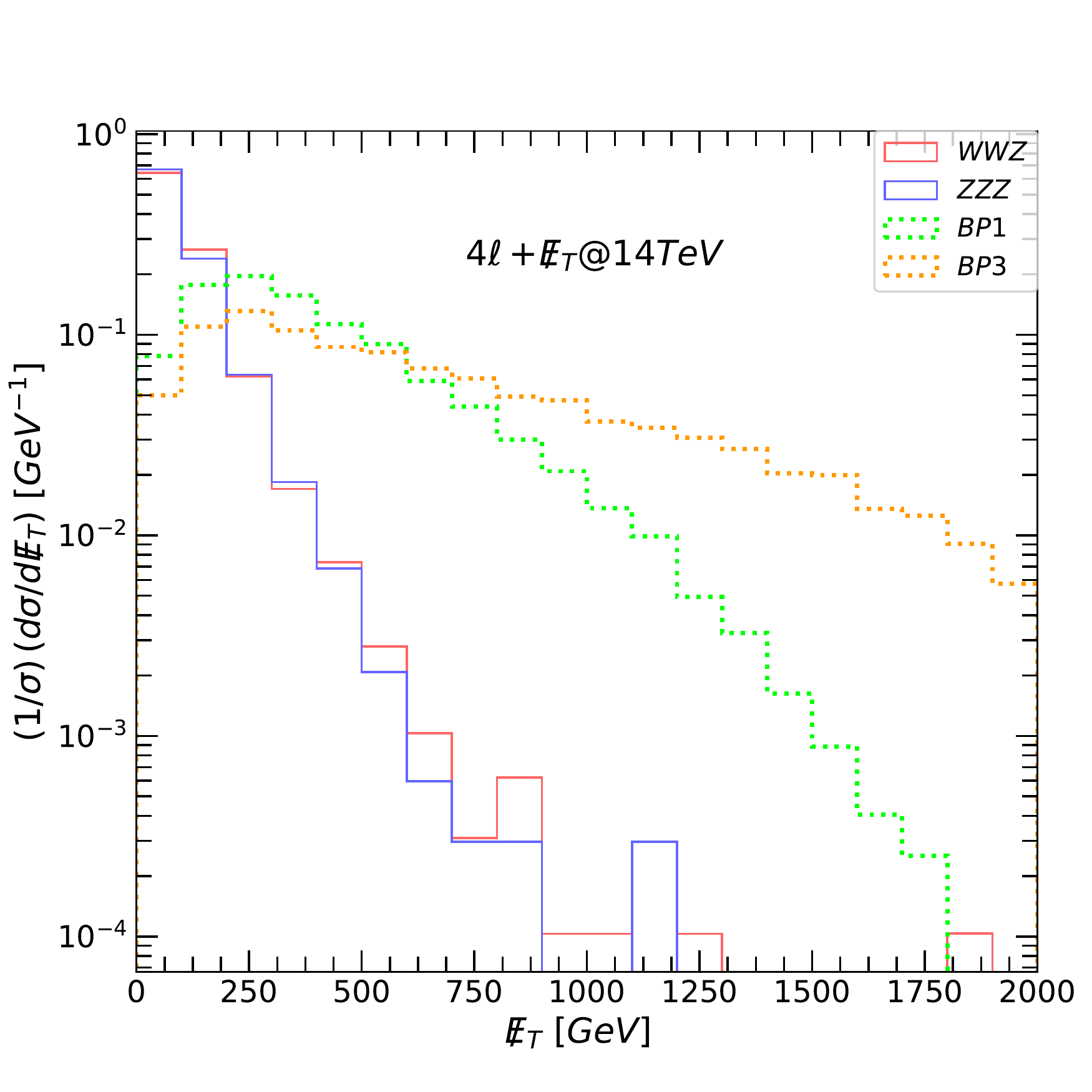} &\hspace*{-0.22cm}
			\includegraphics[height=6.5cm,width=6.5cm]{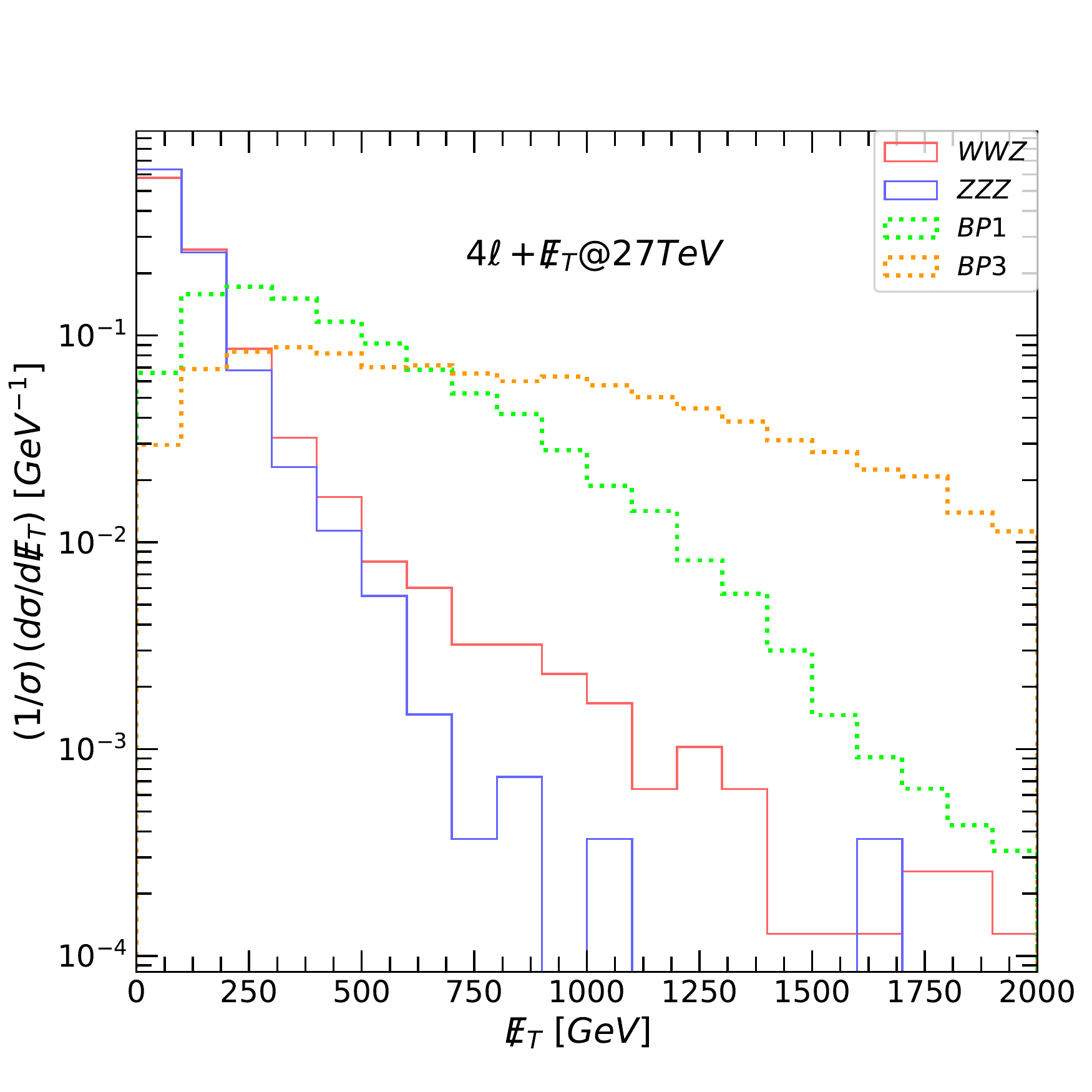} &\hspace*{-0.22cm}	\includegraphics[height=6.5cm,width=6.5cm]{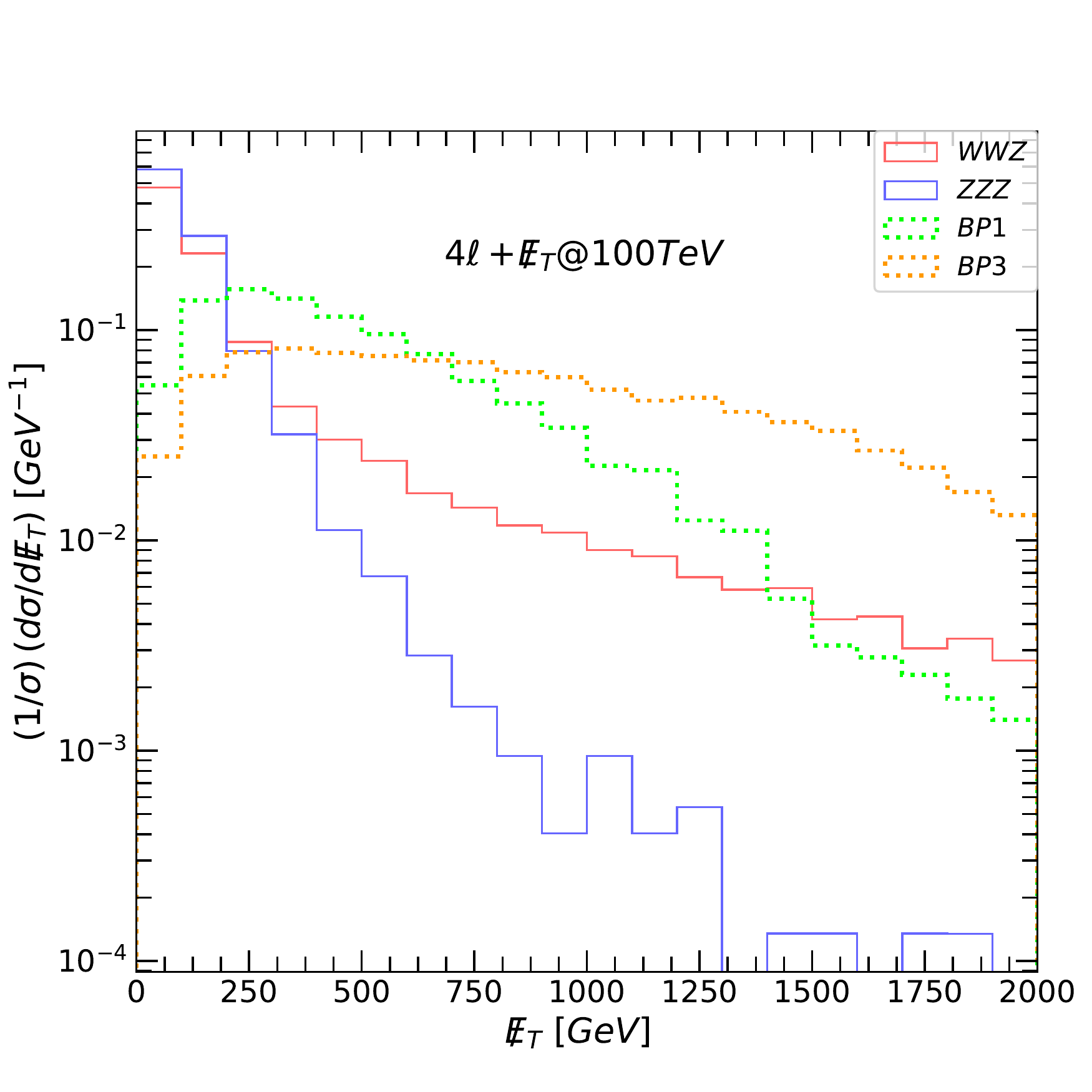}  
		\end{array}$
	\end{center}
	\vskip -0.3in	
	\caption{The total missing energy $\EmissT$  for the signal and background for the $\fourleps$ signal. (Left-hand)  signals and backgrounds at 14 TeV,  (middle)  at 27 TeV, and (right-hand) at 100 TeV. The main backgrounds (three-bosons) are indicated in solid lines while the signals are plotted in dotted lines: green for BP1,  and orange for BP3.}
	\label{fig:4lht}
\end{figure*}


\subsubsection{Four lepton signal: $\fourleps$}
\label{subsubsec:4leptons}
\begin{figure*}[htbp]
	\begin{center}$
		\begin{array}{ccc}
			\hspace*{-0.8cm}
			\includegraphics[height=6.5cm,width=6.5cm]{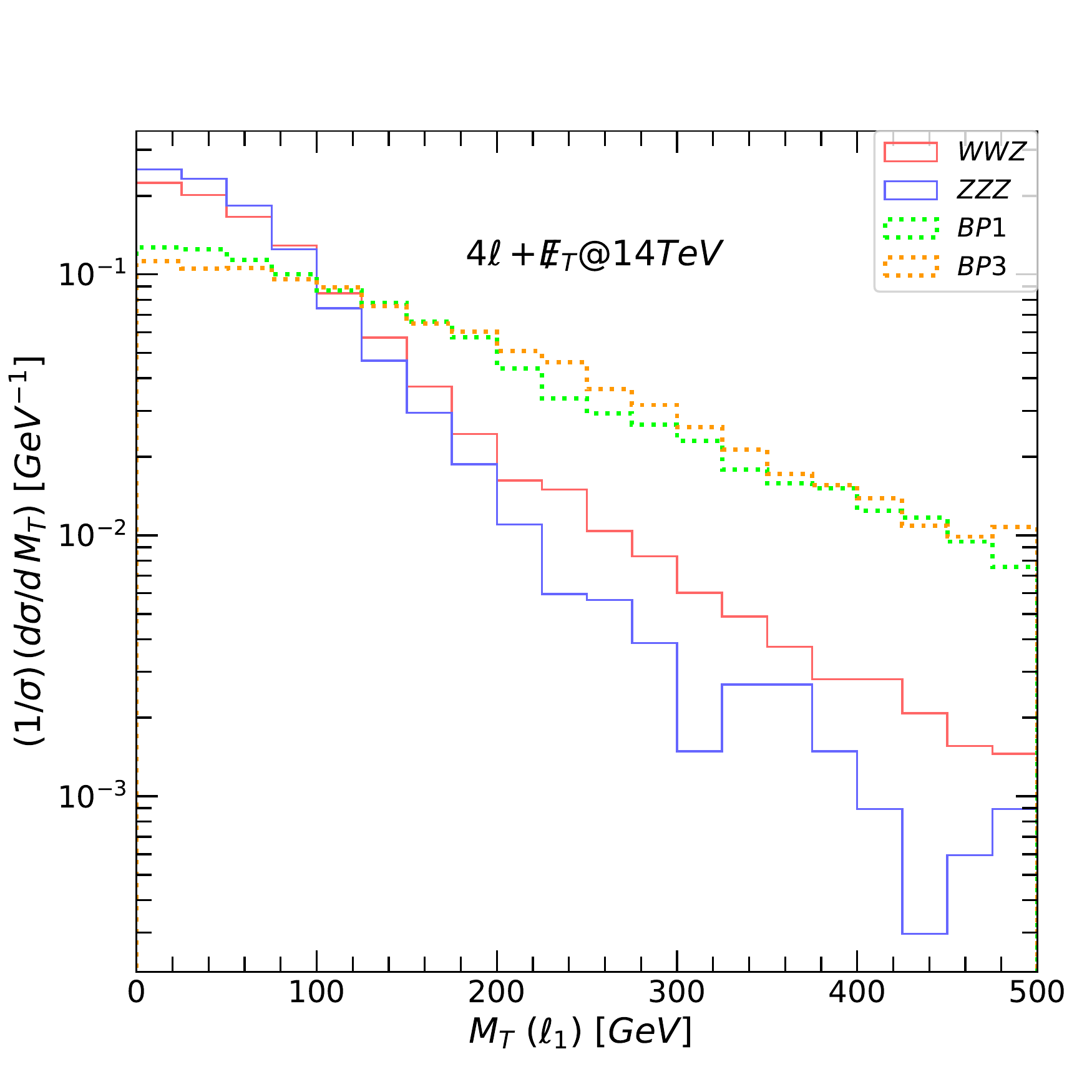} &\hspace*{-0.27cm}
			\includegraphics[height=6.5cm,width=6.5cm]{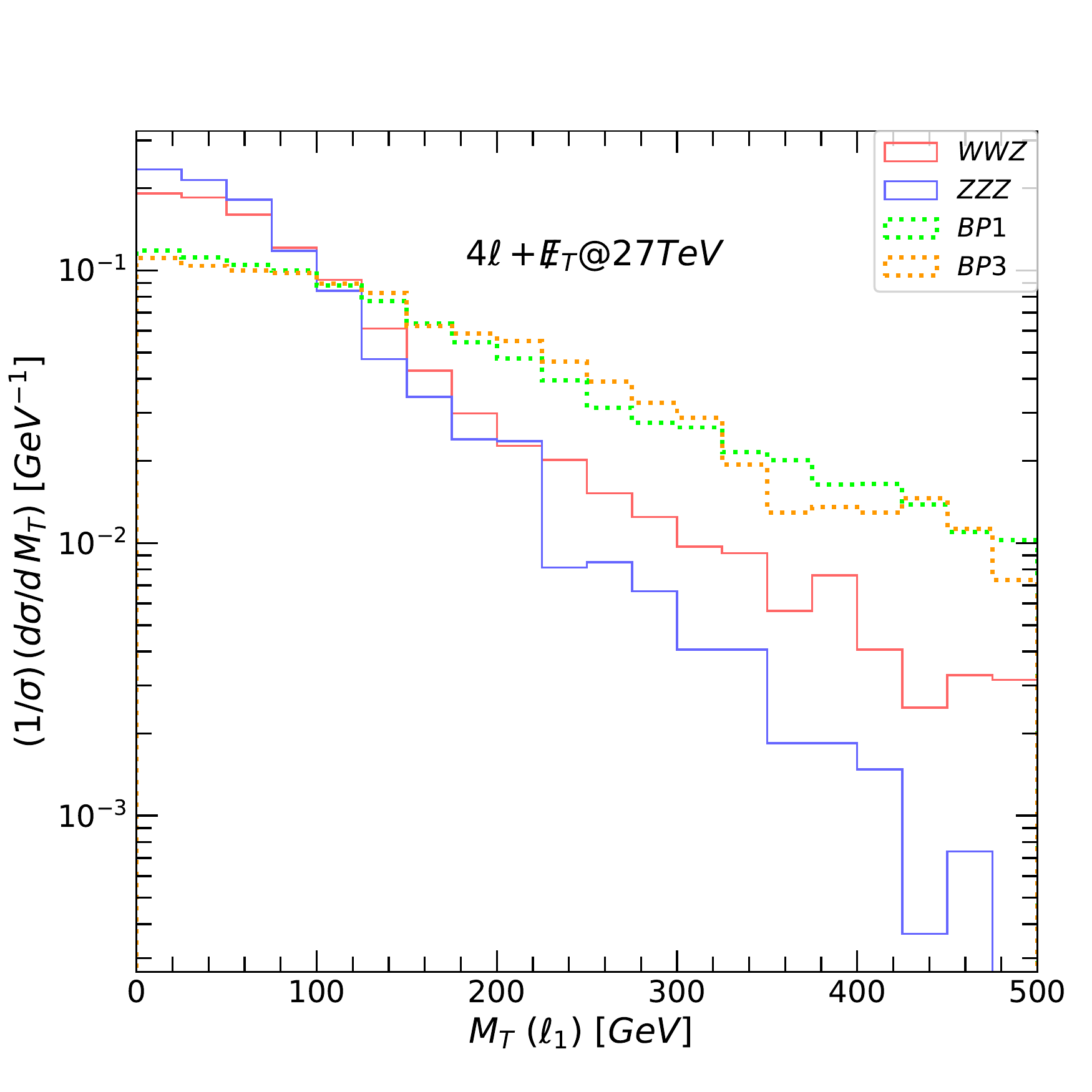} &\hspace*{-0.27cm}
			\includegraphics[height=6.5cm,width=6.5cm]{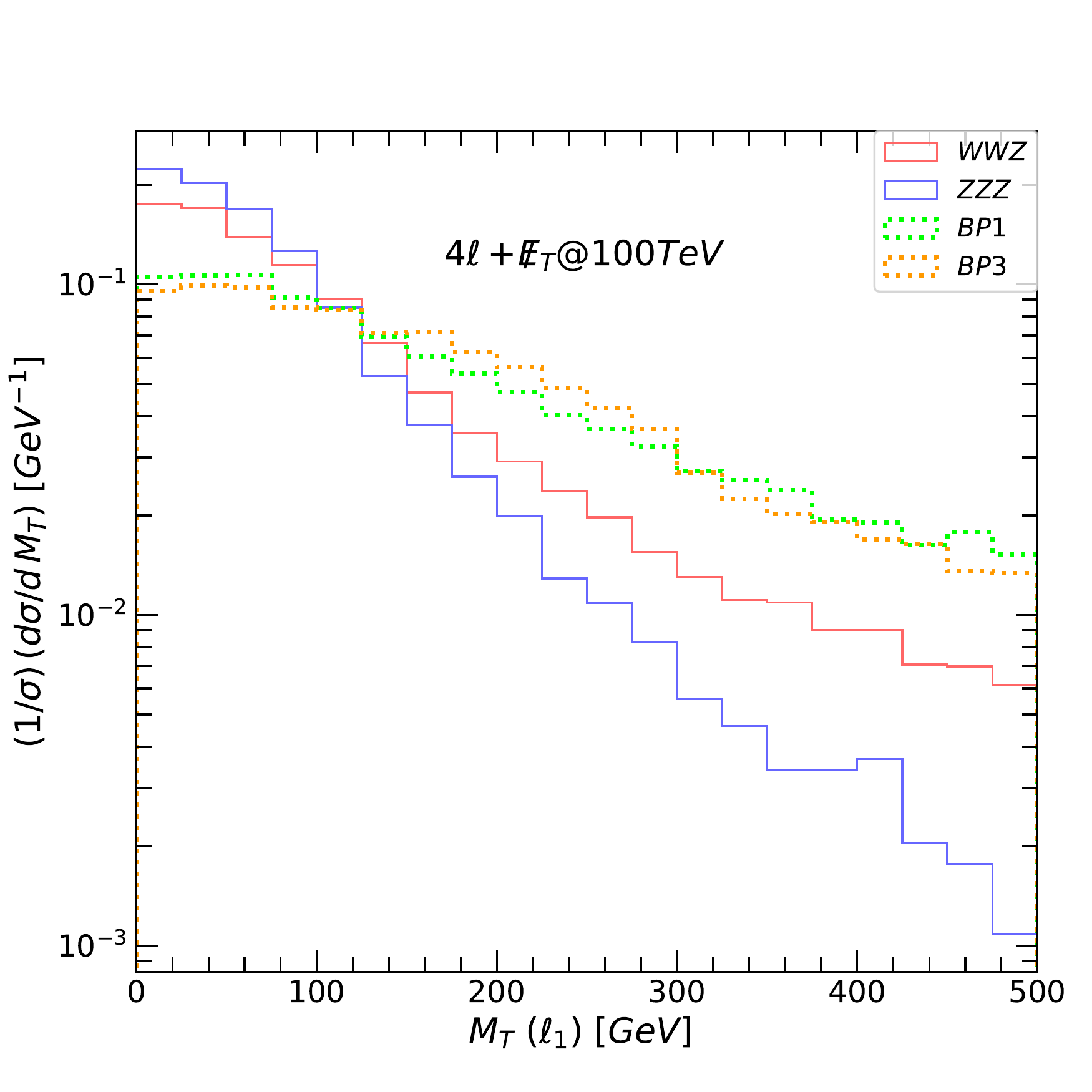} \\[-1em] \hspace*{-0.8cm}
			\includegraphics[height=6.5cm,width=6.5cm]{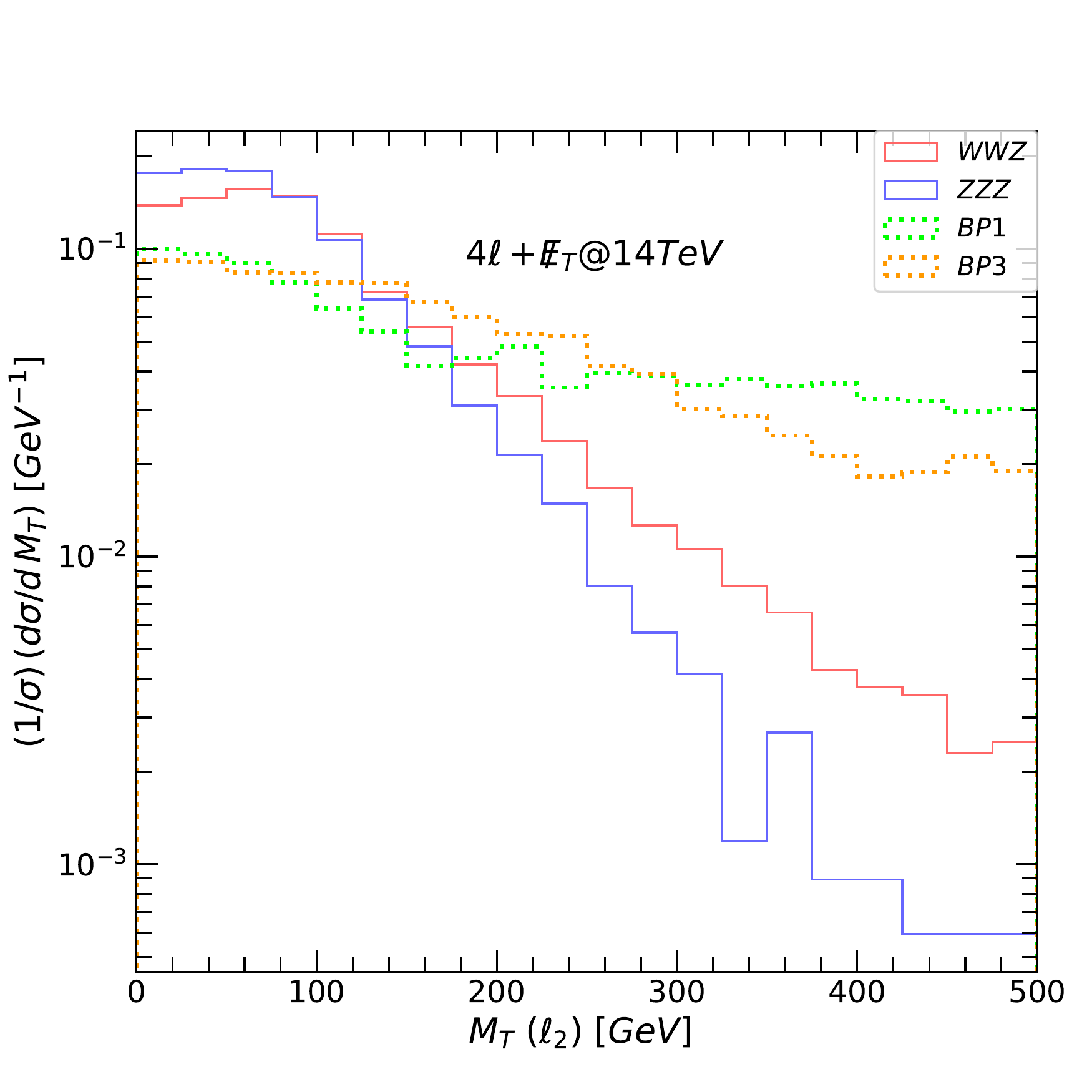} &\hspace*{-0.27cm}
			\includegraphics[height=6.5cm,width=6.5cm]{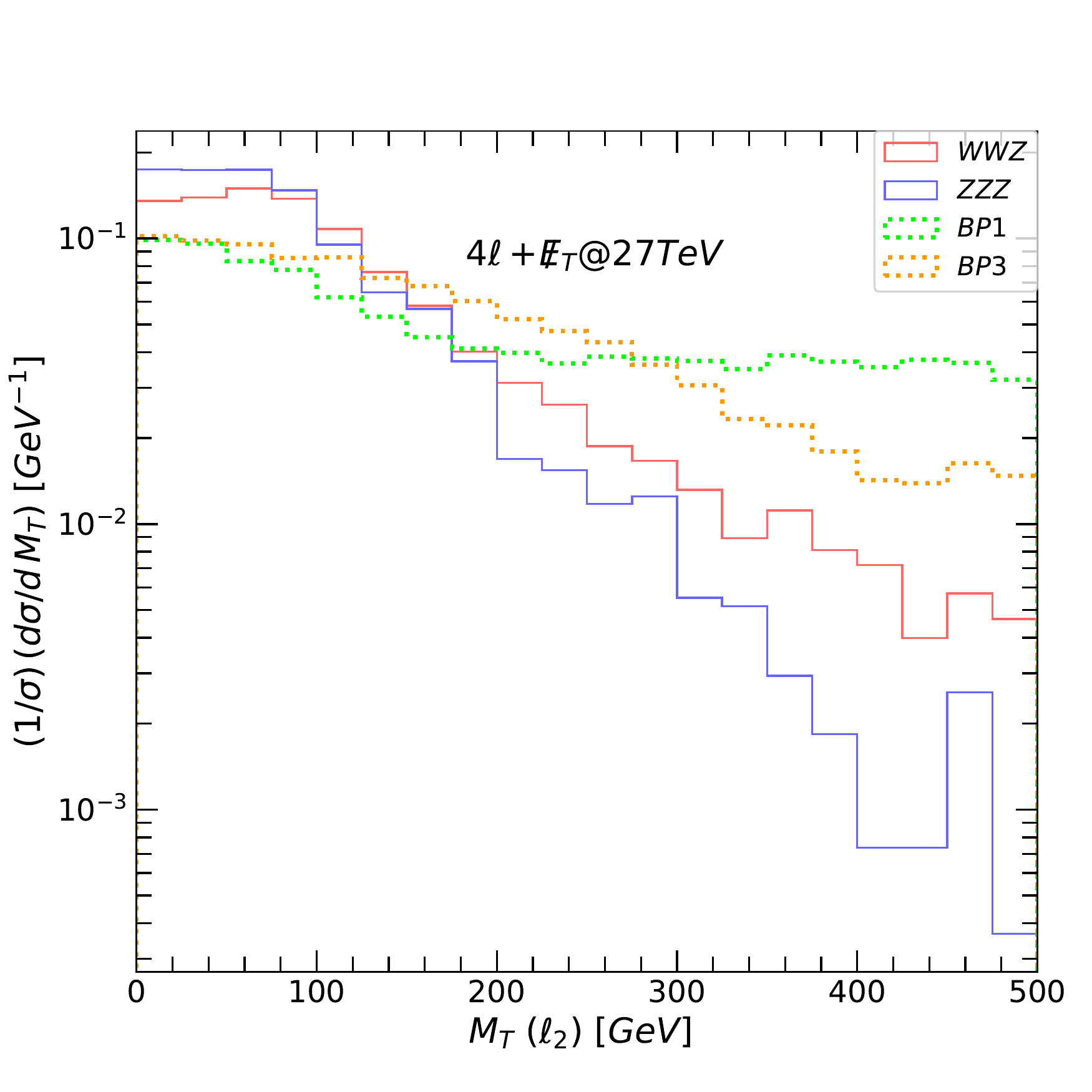} &\hspace*{-0.27cm}
			\includegraphics[height=6.5cm,width=6.5cm]{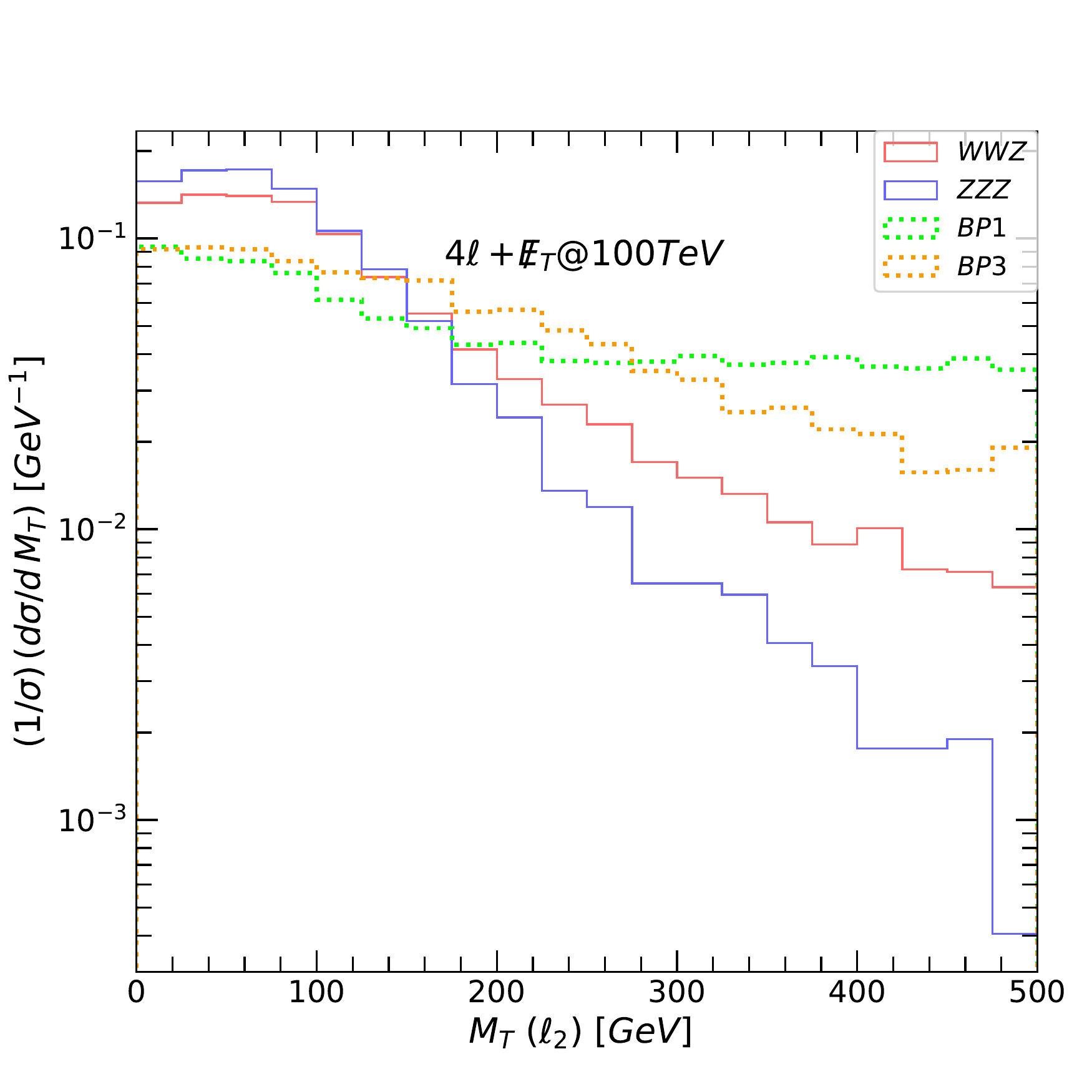}  
		\end{array}$
	\end{center}
	\vskip -0.3in	
	\caption{The transverse mass $M_T$,  for the leading lepton $\ell_1$ (top row) and next-to-leading lepton $\ell_2$ (bottom row),  for the signal and background for the $\fourleps$ signal. (Left-hand)  signals and backgrounds at 14 TeV,  (middle)  at 27 TeV, and (right-hand) at 100 TeV. The main backgrounds (three-bosons) are indicated in solid lines while the signals are plotted in dotted lines: green for BP1, and orange for BP3.}
	\label{fig:4lmt}
\end{figure*}

\begin{figure*}[htbp]
	\begin{center}$
		\begin{array}{ccc}
			\hspace*{-0.8cm}
			\includegraphics[height=6.5cm,width=6.5cm]{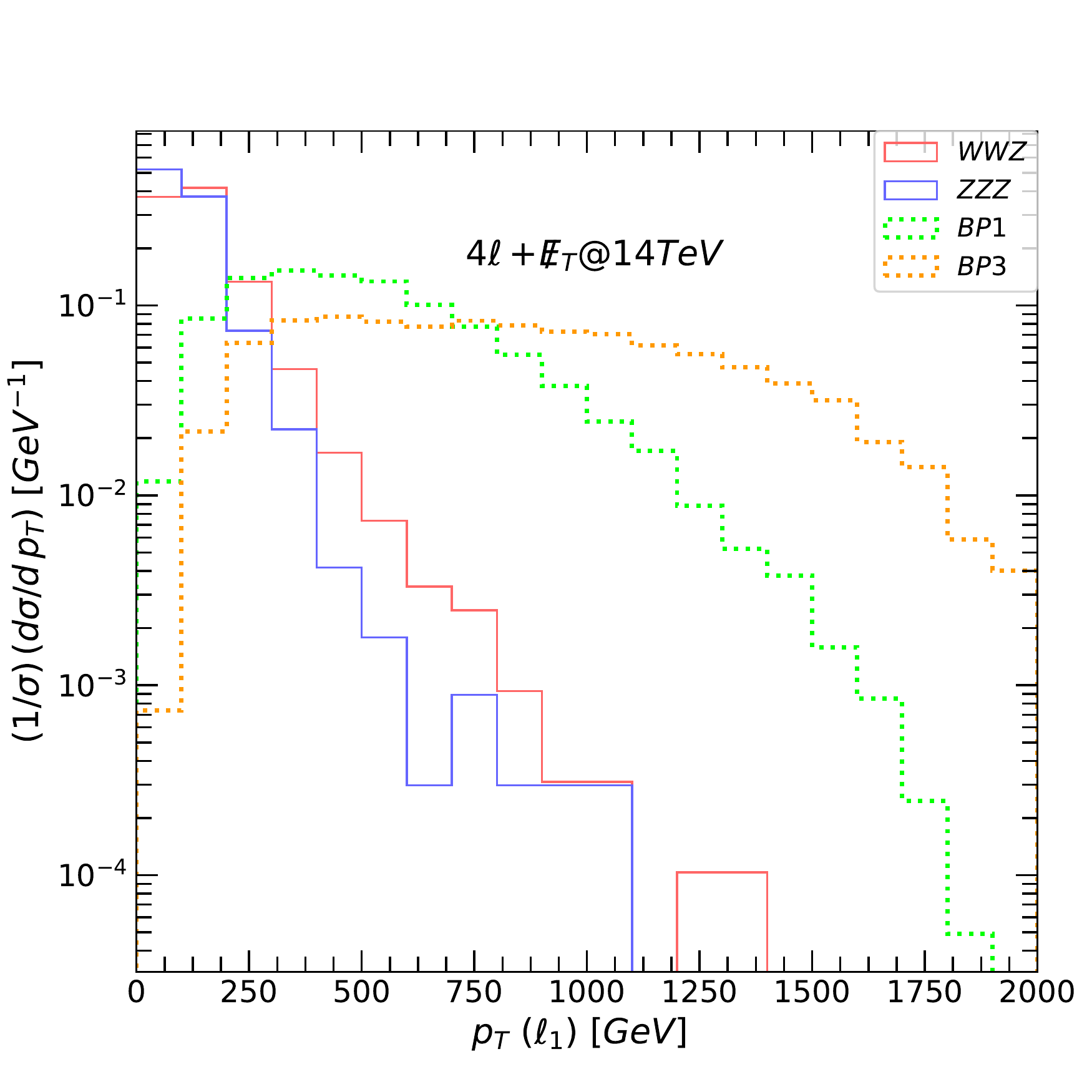} &\hspace*{-0.22cm}
			\includegraphics[height=6.5cm,width=6.5cm]{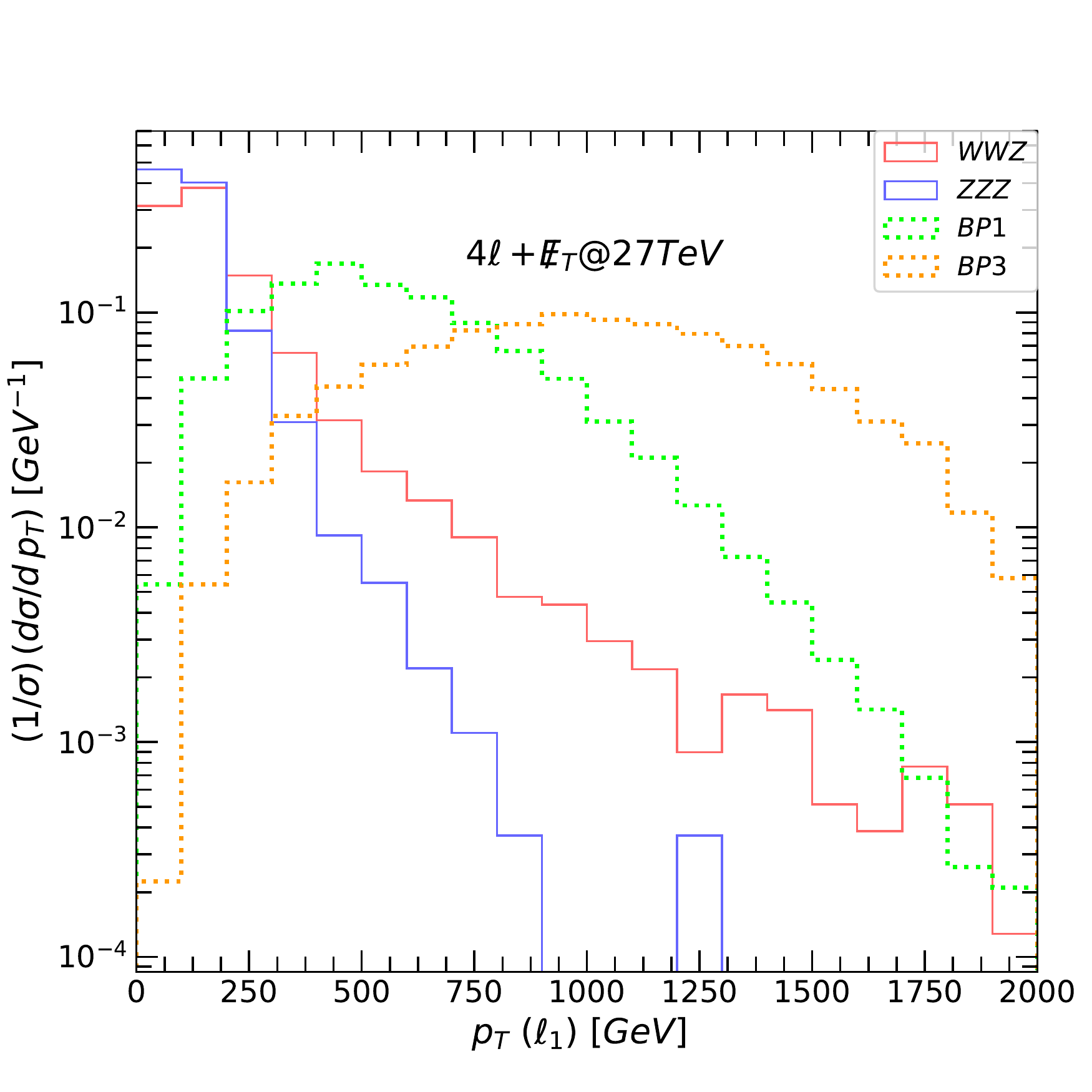} &\hspace*{-0.22cm}
			\includegraphics[height=6.5cm,width=6.5cm]{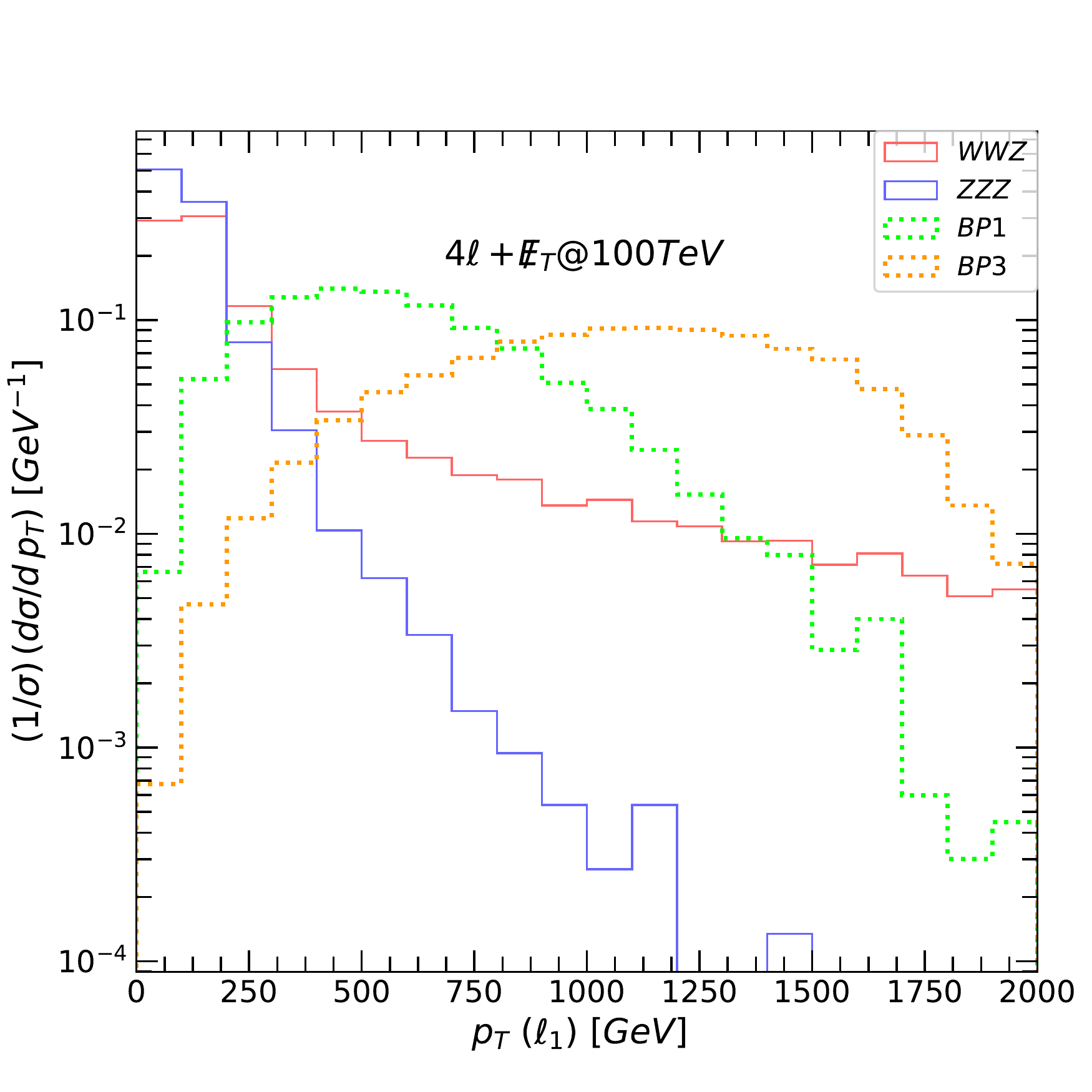} \\[-1em] \hspace*{-0.8cm}
			\includegraphics[height=6.5cm,width=6.5cm]{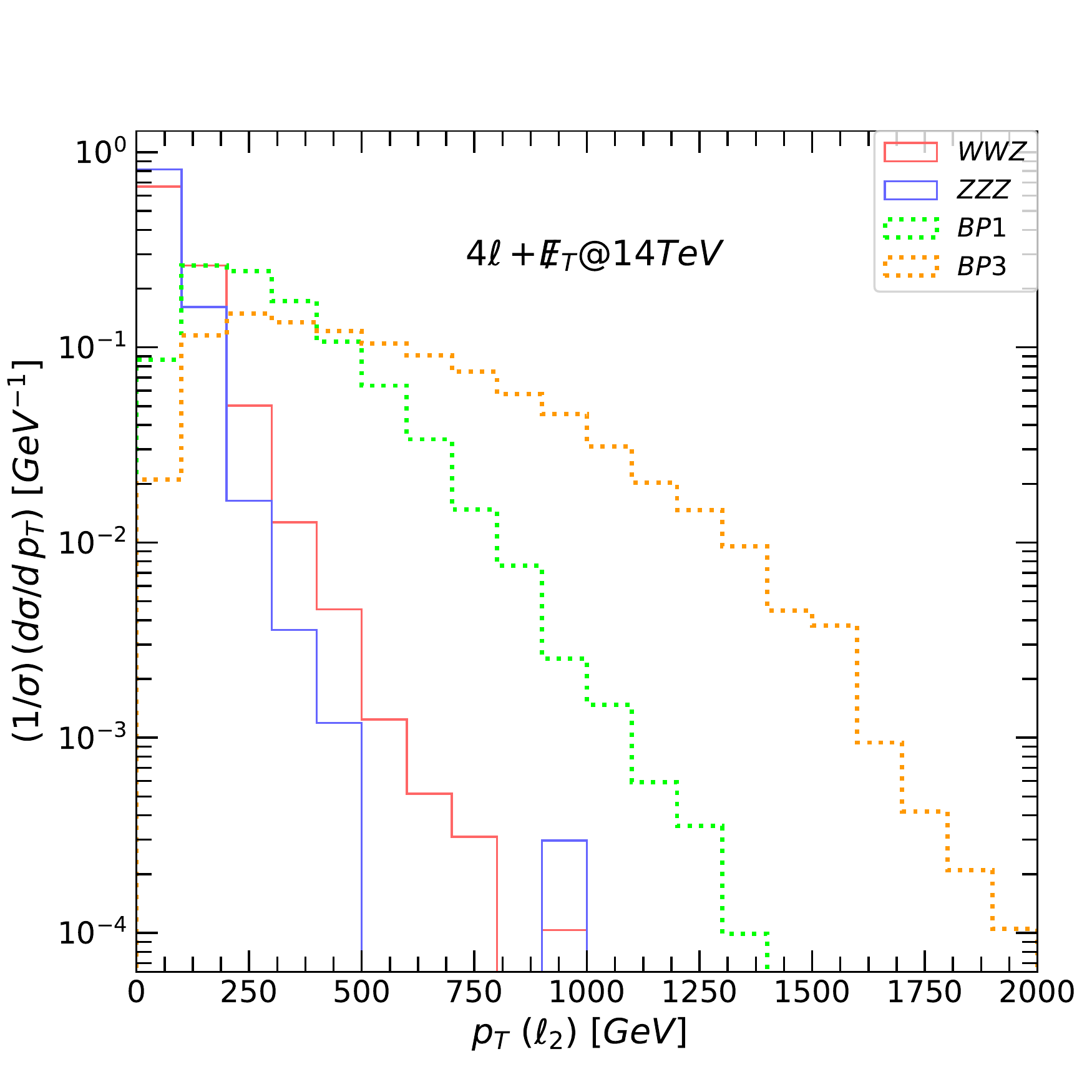} &\hspace*{-0.22cm}
			\includegraphics[height=6.5cm,width=6.5cm]{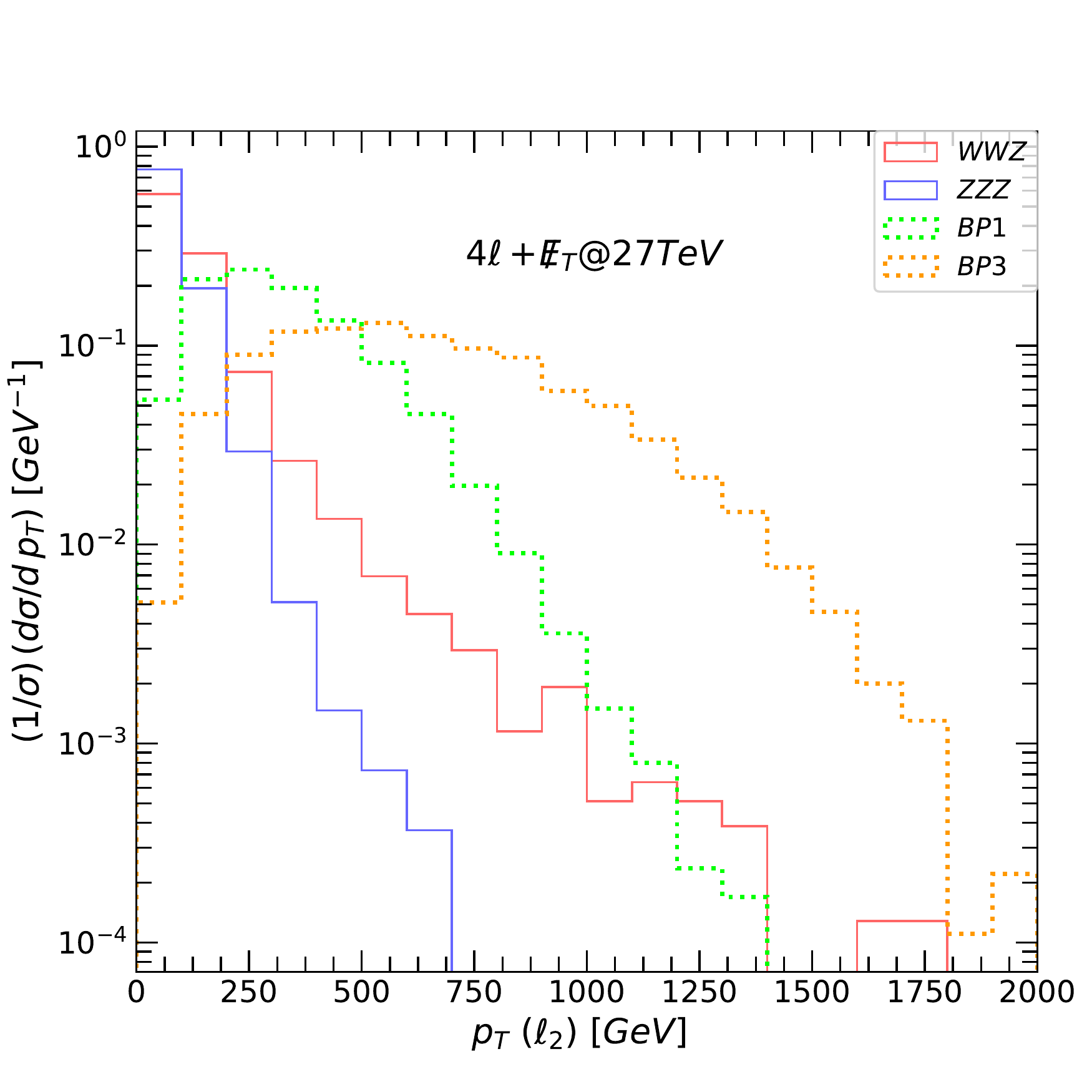} &\hspace*{-0.22cm}
			\includegraphics[height=6.5cm,width=6.5cm]{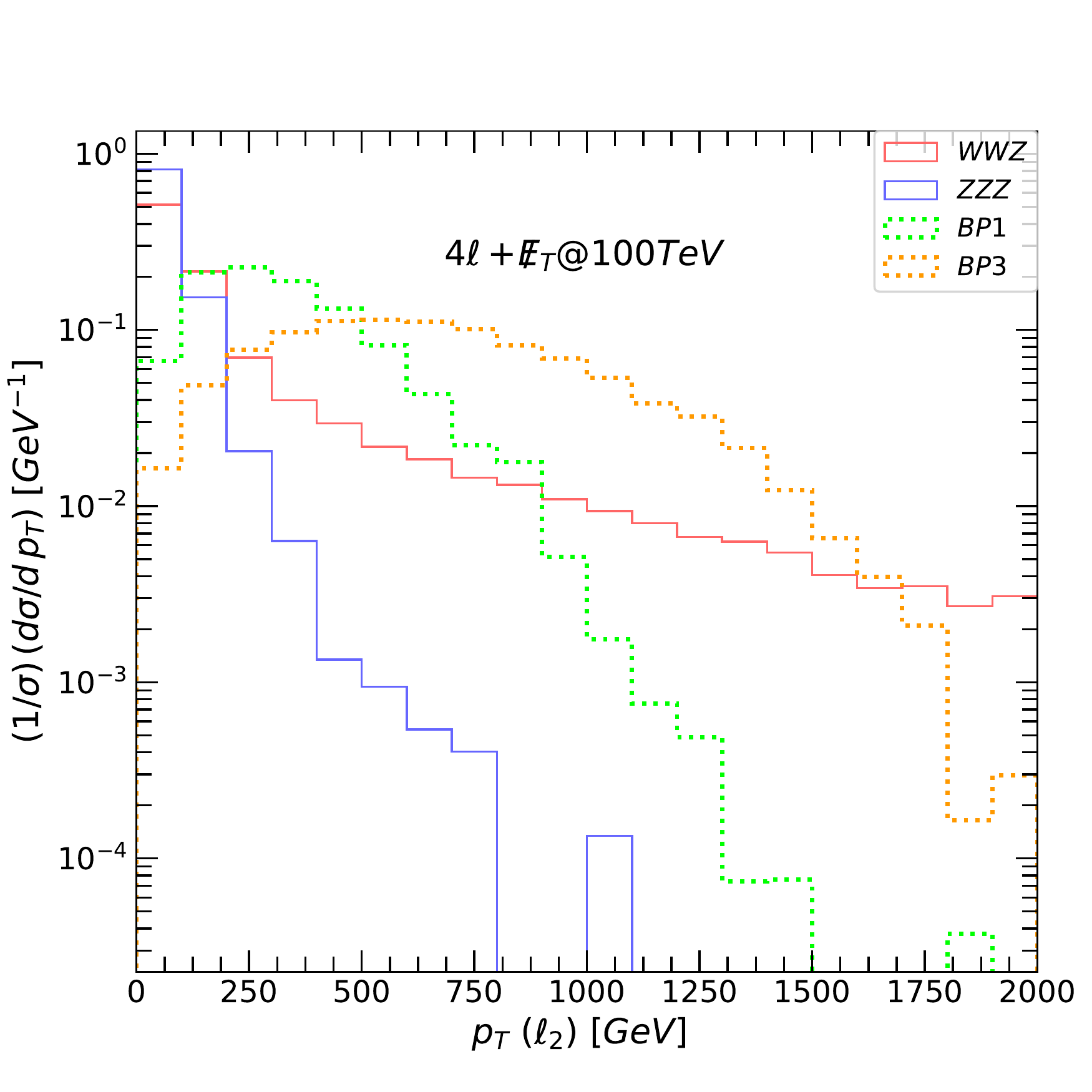}  
		\end{array}$
	\end{center}
	\vskip -0.3in
	\caption{The transverse momentum $p_T$,  for the leading lepton $\ell_1$ (top panels) and next-to-leading lepton $\ell_2$ (bottom panels), for the signal and background for the $\fourleps$ signal. (Left-hand)  signals and backgrounds at 14 TeV,  (middle)  at 27 TeV, and (right-hand) at 100 TeV. The main backgrounds (three-bosons) are indicated in solid lines while the signals are plotted in dotted lines: green for BP1,  and orange for BP3.}
	\label{fig:4lpt}
\end{figure*}
\begingroup
\setlength{\tabcolsep}{10pt} 
\renewcommand{\arraystretch}{0.95} 
\begin{table*}[htbp]
	\caption{ Signal selection strategy and cuts imposed in the  $\fourleps$ scenario at 14, 27 and 100 TeV. We give the cross-section for background and benchmark scenarios in fb. Statistical significances $\sigma_{\mathcal A}$ and $\sigma_{\mathcal B}$ of $\fourleps$ signal are given for each energy.}\label{tab:cuts-bckgr4l}
	\setlength{\extrarowheight}{2pt}
	\begin{tabular*}{\textwidth}{@{\extracolsep{\fill}} cccc}
		\hline\hline
		$\ 4\ell +\EmissT@14~\rm TeV $&\rm Background~[fb] &$ \rm BP1~[fb]$&$\rm BP3~[fb]$\\
		\hline
		$\rm No~Cut$&$3.45\times 10^{-2}$&$1.99\times 10^{-3}$&$1.86\times 10^{-3}$\\ 
		$ |\eta_i|<2.5,\ \Delta R_{12} \ge 0.5$ &$2.89\times 10^{-2}$&$5.31\times 10^{-4}$&$4.63\times 10^{-4}$\\
		$ p_T(\ell_1)>100$ GeV&$1.7\times 10^{-2}$&$5.1\times 10^{-4}$&$4.61\times 10^{-4}$\\
		$ p_T(\ell_2)>50$ GeV&$1.58\times 10^{-2}$&$5.02\times 10^{-4}$&$4.61\times 10^{-4}$\\
		$ p_T(\ell_3)>25$ GeV&$1.52\times 10^{-2}$&$4.87\times 10^{-4}$&$4.59\times 10^{-4}$\\
		$ p_T(\ell_4)>15$ GeV&$1.4\times 10^{-2}$&$4.44\times 10^{-4}$&$4.31\times 10^{-4}$\\
		$ \EmissT>400~\rm GeV$ &$1.0\times 10^{-4}$&$5.5\times 10^{-5}$&$1.3\times 10^{-4}$\\
		\hline
		$\rm Significance:\quad \sigma_{\mathcal A}$&${\cal L}=3\,{\rm ab}^{-1}$&$0.24\sigma$&$0.46\sigma$\\
		\hskip 2.1cm $\sigma_{\mathcal B}$
		& &$0.19\sigma$&$0.38\sigma$\\		
		\hline\hline
		$ 4\ell +\EmissT@27~\rm TeV $&&&\\
		\hline
		$\rm No~Cut$&$9.59\times 10^{-2}$&$3.03\times 10^{-2}$&$3.86\times 10^{-2}$\\ 
		$ |\eta_i|<2.5,\ \Delta R_{12} \ge 0.5$ &$7.52\times 10^{-2}$&$5.9\times 10^{-3}$&$4.14\times 10^{-3}$\\
		$ p_T(\ell_1)>100$ GeV&$4.7\times 10^{-2}$&$5.74\times 10^{-3}$&$4.13\times 10^{-3}$\\
		$\ p_T(\ell_2)>50$ GeV&$4.39\times 10^{-2}$&$5.68\times 10^{-3}$&$4.13\times 10^{-3}$\\
		$ p_T(\ell_3)>25$ GeV&$4.2\times 10^{-2}$&$5.52\times 10^{-3}$&$4.12\times 10^{-3}$\\
		$ p_T(\ell_4)>15$ GeV&$3.86\times 10^{-2}$&$5.08\times 10^{-3}$&$3.9\times 10^{-3}$\\
		$ \EmissT>350~\rm GeV$ &$1.1\times 10^{-3}$&$1.2\times 10^{-3}$&$2.2\times 10^{-3}$\\
		\hline
		$\rm Significance:\quad \sigma_{\mathcal A}$&${\cal L}=15\,{\rm ab}^{-1}$&$3.1\sigma$&$4.6\sigma$\\
		\hskip 2.0cm $\sigma_{\mathcal B}$ 
		& &$2.6\sigma$&$4.1\sigma$\\		
		\hline\hline
		$ 4\ell +\EmissT@100~\rm TeV $&&&\\
		\hline
		$\rm No~Cut$&$2.48$&$9.38\times 10^{-1}$&$1.64$\\ 
		$ |\eta_i|<2.5,\ \Delta R_{12} \ge 0.5$ &$6.56\times 10^{-1}$&$1.33\times 10^{-1}$&$1.16\times 10^{-1}$\\
		$ p_T(\ell_1)>100$ GeV&$4.46\times 10^{-1}$&$1.3\times 10^{-1}$&$1.15\times 10^{-1}$\\
		$ p_T(\ell_2)>50$ GeV&$4.1\times 10^{-1}$&$1.29\times 10^{-1}$&$1.14\times 10^{-1}$\\
		$ p_T(\ell_3)>25$ GeV&$3.83\times 10^{-1}$&$1.22\times 10^{-1}$&$1.01\times 10^{-1}$\\
		$ p_T(\ell_4)>15$ GeV&$3.1\times 10^{-1}$&$9.9\times 10^{-2}$&$8.44\times 10^{-2}$\\
		$ \EmissT>800~\rm GeV$ &$7.1\times 10^{-3}$&$8.0\times 10^{-3}$&$2.3\times 10^{-2}$\\
		\hline
		$\rm Significance:\quad \sigma_{\mathcal A}$&${\cal L}=30\,{\rm ab}^{-1}$&$11.0\sigma$&$23.0\sigma$\\
		\hskip 2.0cm $\sigma_{\mathcal B}$ 
		& &$9.4\sigma$&$22.0\sigma$\\		
		\hline\hline	
	\end{tabular*}
\end{table*}
\endgroup

The main decay modes of the $Z^\prime$ boson yielding  $\fourleps$ signals are
\begin{eqnarray}
	Z^{\prime} &\rightarrow& \tilde{\nu}_{\ell_R}
	\tilde{\nu}_{\ell_R}\rightarrow \fourleps, \nonumber \\
	Z^{\prime}&\rightarrow& \tilde\ell_R\tilde\ell_R\rightarrow
	\fourleps.
\end{eqnarray}
In the following figures, we first plot the relevant kinematic variables signals and background at 14, 27 and 100 TeV,  before any cuts were imposed.  We show, in Fig. \ref{fig:4lht}, 
 the total missing transverse energy $\EmissT$. For each figure  we plot, in the left-hand columns,  signals and backgrounds at 14 TeV,  the middle columns  at 27 TeV, and right-hand columns, at 100 TeV. Both these plots indicate that the signal over background, before cuts, seems most promising at higher energies, and that $\EmissT$ is overall a better variable to differentiate between signal and background.  We also note that, as for the $\twoleps$ case, the scenario BP3 is the most promising.

In Fig. \ref{fig:4lmt} we plot the  transverse mass $M_T$ 
for the leading lepton $\ell_1$ (top row panels) and next-to-leading lepton $\ell_2$ (bottom row panels) for the signal and background for the $\fourleps$ signal.
In Fig. \ref{fig:4lpt}, we show the transverse momenta of the leading lepton $\ell_1$ (top row) and next-to-leading lepton $\ell_2$ (bottom), for the signal and background. The transverse momenta of the third and fourth leptons  have similar distributions to the next-to leading lepton, and thus we do not plot them. The left-hand side panels in both figures correspond to signals and backgrounds at 14 TeV, the middle at 27 TeV, and the right-hand side for 100 TeV. The main backgrounds (three-bosons) are indicated in solid lines while the signals are plotted in dotted lines: green for BP1,  and orange for BP3\footnote{Note that $\fourleps$ signal is not realized in  BP2 as indicated in Fig.~\ref{fig:FeyndiagramsBPs}.}. These graphs show clearly that at large $M_T(\ell)$ and $p_T(\ell)$, the signal dominates the background, and the graphs justify our choice of kinematic cuts.  The distribution is rather similar for the leading leptons, and the $p_T$ observable is better than the $M_T$ at distinguishing signals from the backgrounds. In all the plots, the signal from BP3 is most promising, particularly at high energy/momenta.

Table~\ref{tab:cuts-bckgr4l} gives values for signal and background cross sections after each cut. We also show the signal significance, for both $\sigma_{\cal A}$ and $\sigma_{\cal B}$, for each benchmark, at proposed total integrated luminosity: for  14 TeV at ${\cal L}= 3$ ab$^{-1}$, for  27 TeV at ${\cal L}= 15$ ab$^{-1}$ and for  100 TeV at ${\cal L}= 30$ ab$^{-1}$.  Unlike the case of $\twoleps$, we keep the cuts constant for different centre-of-mass energies. Again, the significance for all benchmarks at 14 TeV  for observing the $Z^\prime$ boson in the $\fourleps$ final state is low.  However, at 27 TeV both  BP1 and BP3 show some promise, and we obtain large significances of $\sim 3\sigma$ or more for $\sigma_{\cal A}$. At 100 TeV, though there are many uncertainties and unknowns, and our results should be interpreted as estimates only, both BP1 and BP3 show significant promise for observability.


\subsubsection{Six lepton signal: $\sixleps$}
\label{subsubsec:6leptons}

\begin{figure*}[htbp]
	\begin{center}$
		\begin{array}{ccc}
			\hspace*{-0.8cm}
			\includegraphics[height=6.5cm,width=6.5cm]{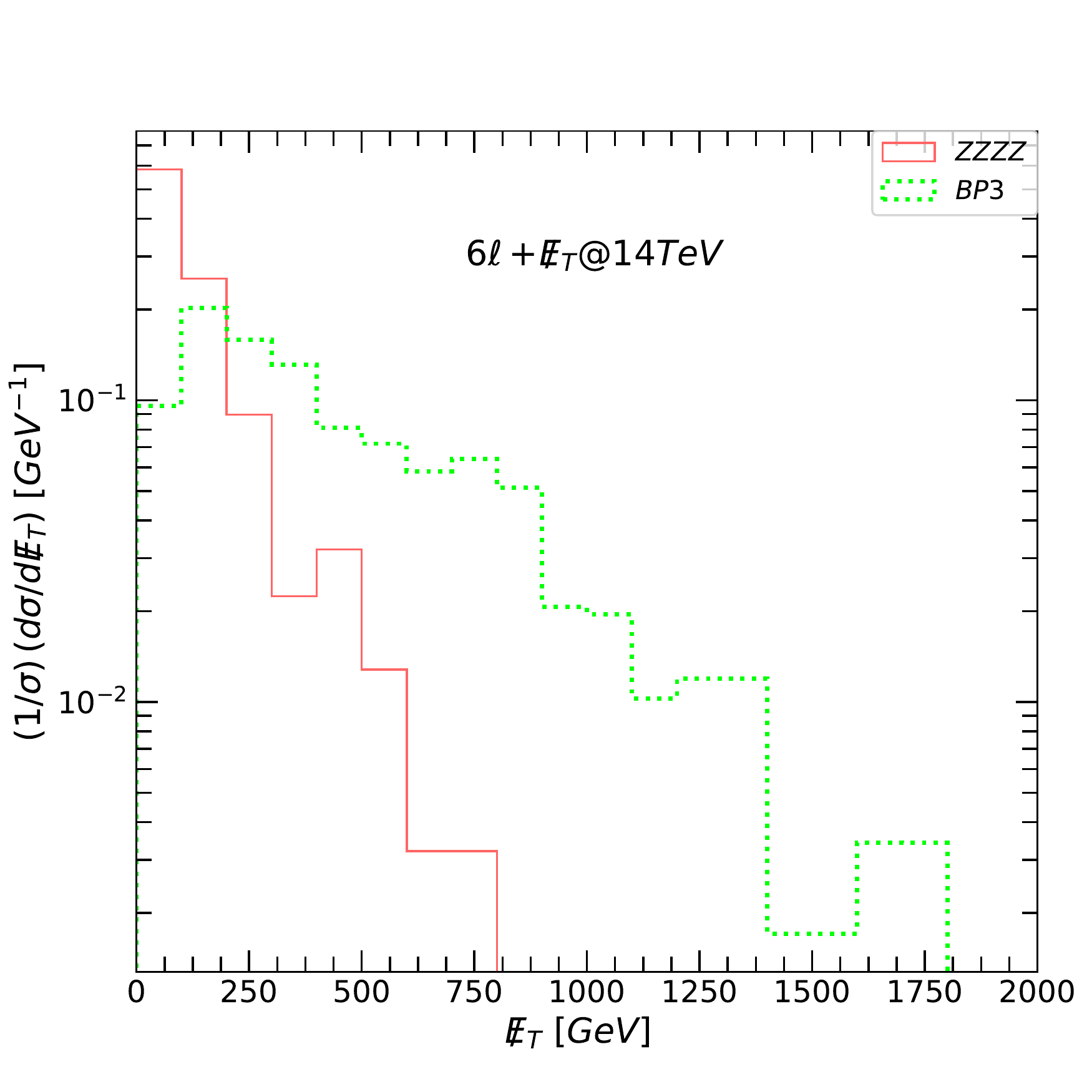} & \hspace*{-0.22cm}
			\includegraphics[height=6.5cm,width=6.5cm]{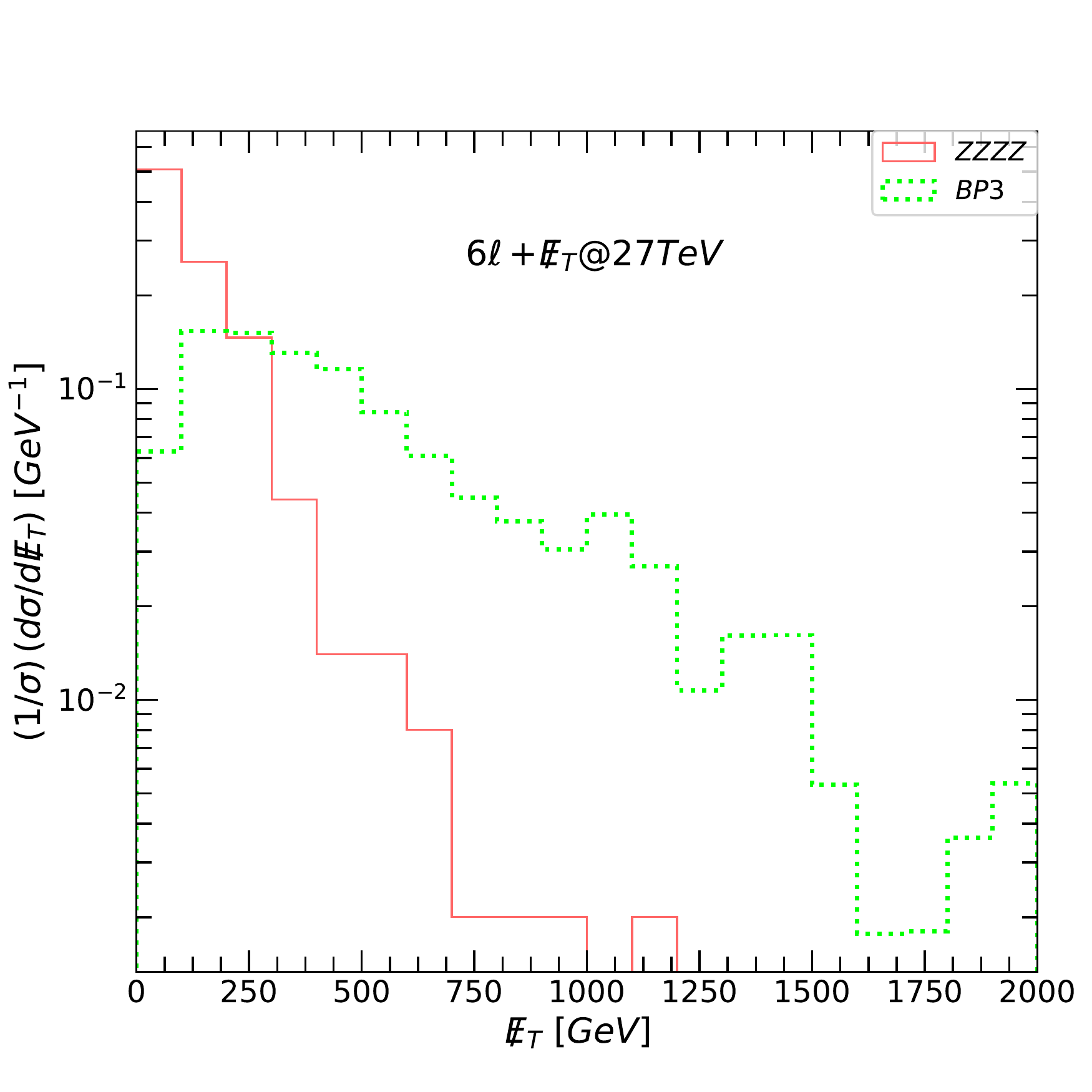} &
			\hspace*{-0.22cm}		\includegraphics[height=6.5cm,width=6.5cm]{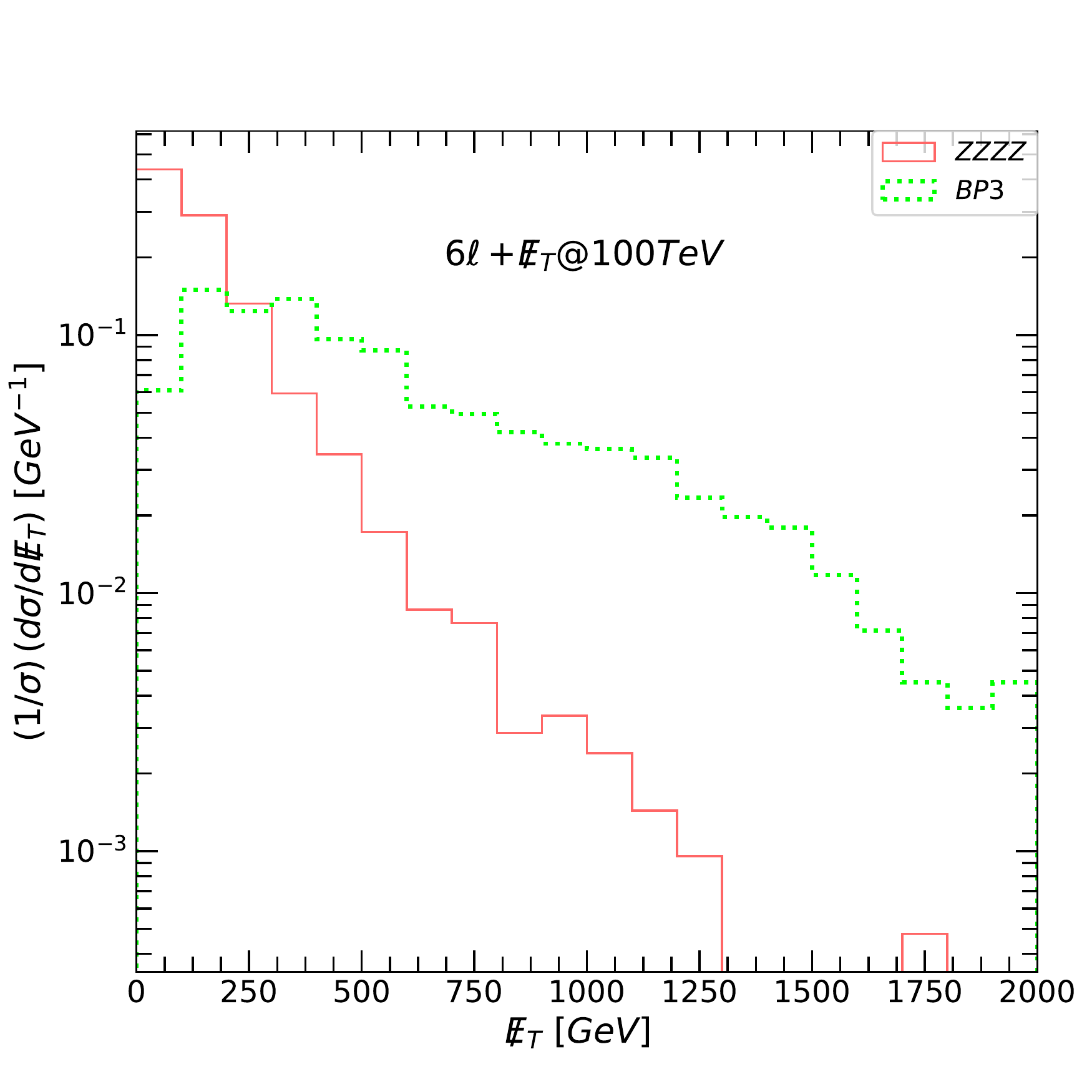} 
		\end{array}$
	\end{center}
	\vskip -0.3in
	\caption{The missing transverse energy $\EmissT$, for the signal and background for the $\sixleps$ signal. (Left-hand)  signals and backgrounds at 14 TeV,  (middle)  at 27 TeV,  and (right-hand) at 100 TeV. The main backgrounds (four-bosons) are indicated in solid lines while the  dotted green line represents the signal BP3.}
	\label{fig:6lmet}
\end{figure*}

The dominant decay mode of the $Z^\prime$ gauge boson, yielding a $\sixleps$ signal is
$$Z^{\prime}\rightarrow \tilde\ell_R\tilde\ell_R\rightarrow \sixleps \,.$$
Quite clearly, the $\sixleps$ signal requires that this decay have a non-negligible branching ratio, which occurs for the  BP3 scenario, as seen from Table~\ref{tab:crsrlc}, where $Z^{\prime}\rightarrow \tilde\ell_R\tilde\ell_R \sim 3\%$. Thus as expected, this will be the only signal of interest for the $\sixleps$ signal.
In Fig. \ref{fig:6lmet} we plot the missing transverse energy $\EmissT$, for the signal and background for the $\sixleps$ signal: (left-hand)  signals and backgrounds at 14 TeV,  (middle)  at 27 TeV,  and (right-hand) at 100 TeV. We show the main (four-bosons) backgrounds  in solid lines while the  the signal BP3 is given by a dotted green line.

\begin{figure*}[htbp]
	\begin{center}$
		\begin{array}{ccc}
			\hspace*{-0.8cm}
			\includegraphics[height=6.5cm,width=6.5cm]{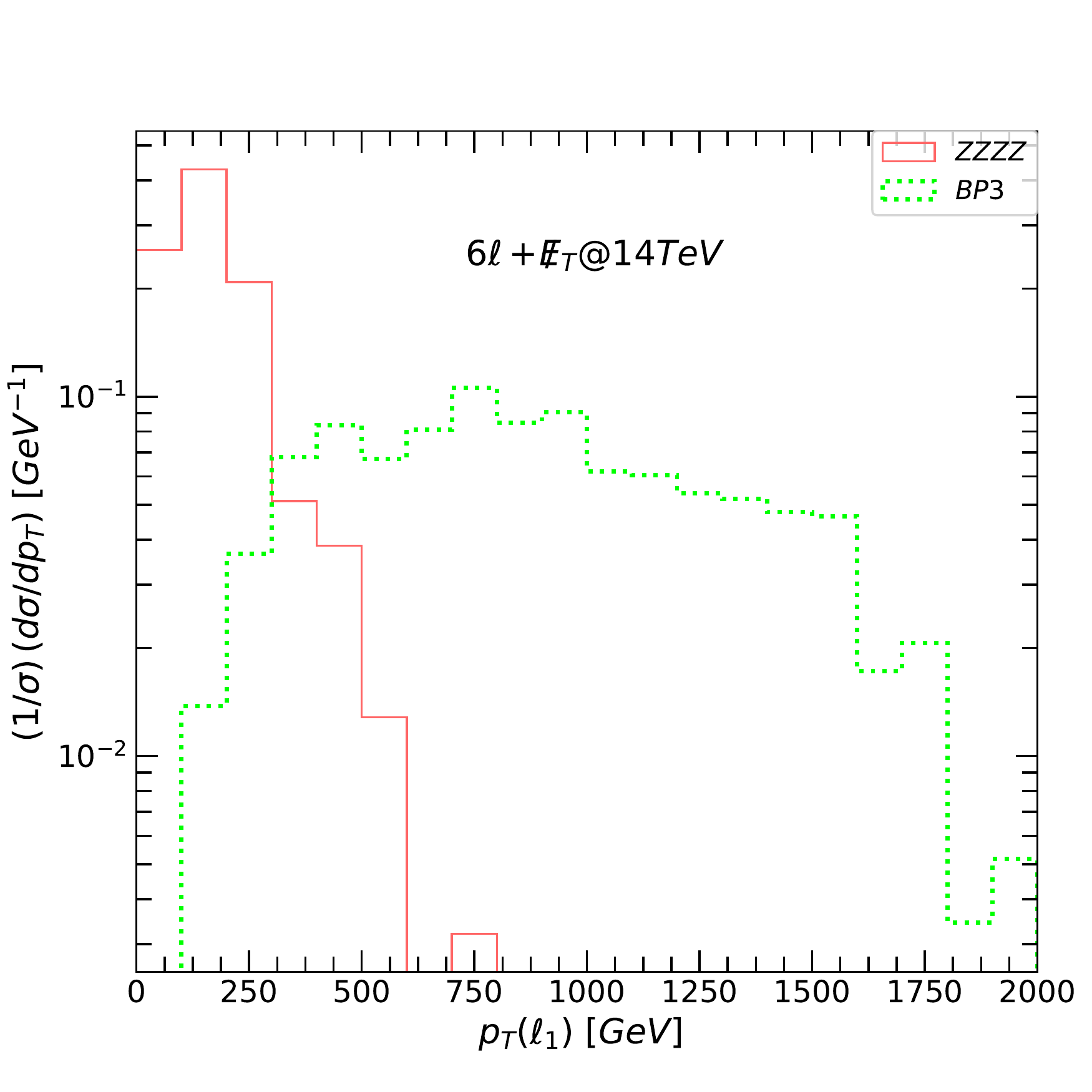} &
			\hspace*{-0.22cm}	\includegraphics[height=6.5cm,width=6.5cm]{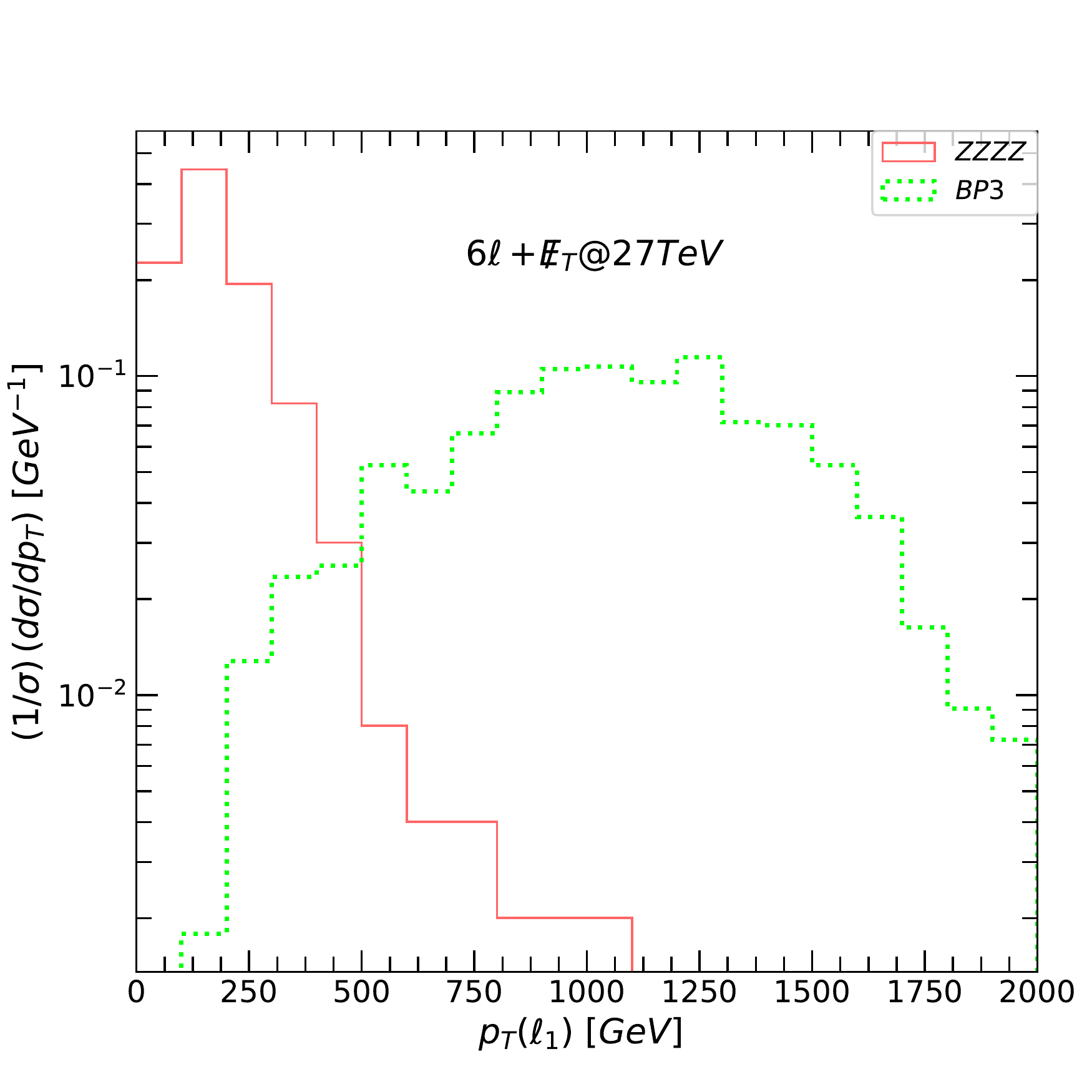} &
			\hspace*{-0.22cm}
			\includegraphics[height=6.5cm,width=6.5cm]{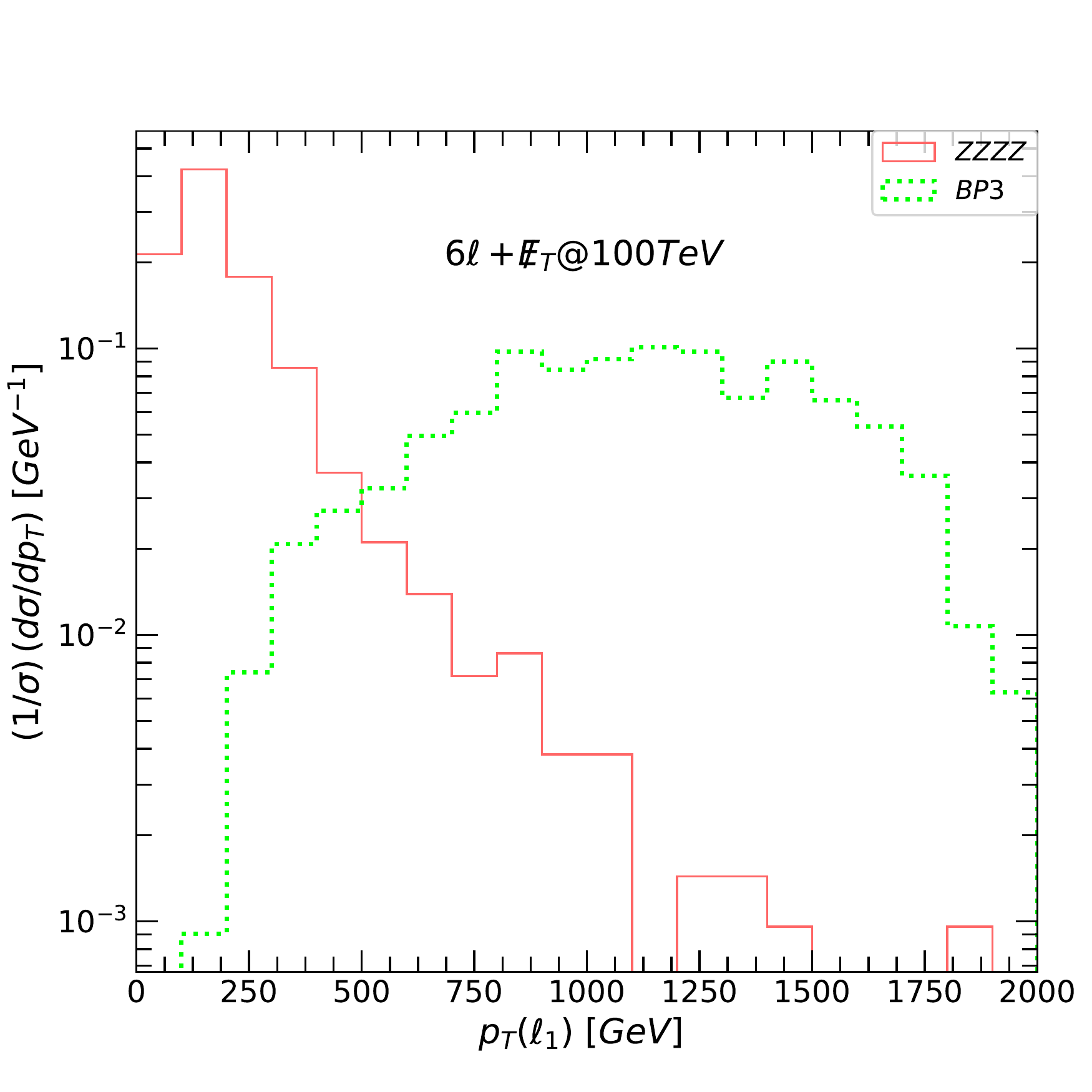} \\[-1em]
			\hspace*{-0.8cm}
			\includegraphics[height=6.5cm,width=6.5cm]{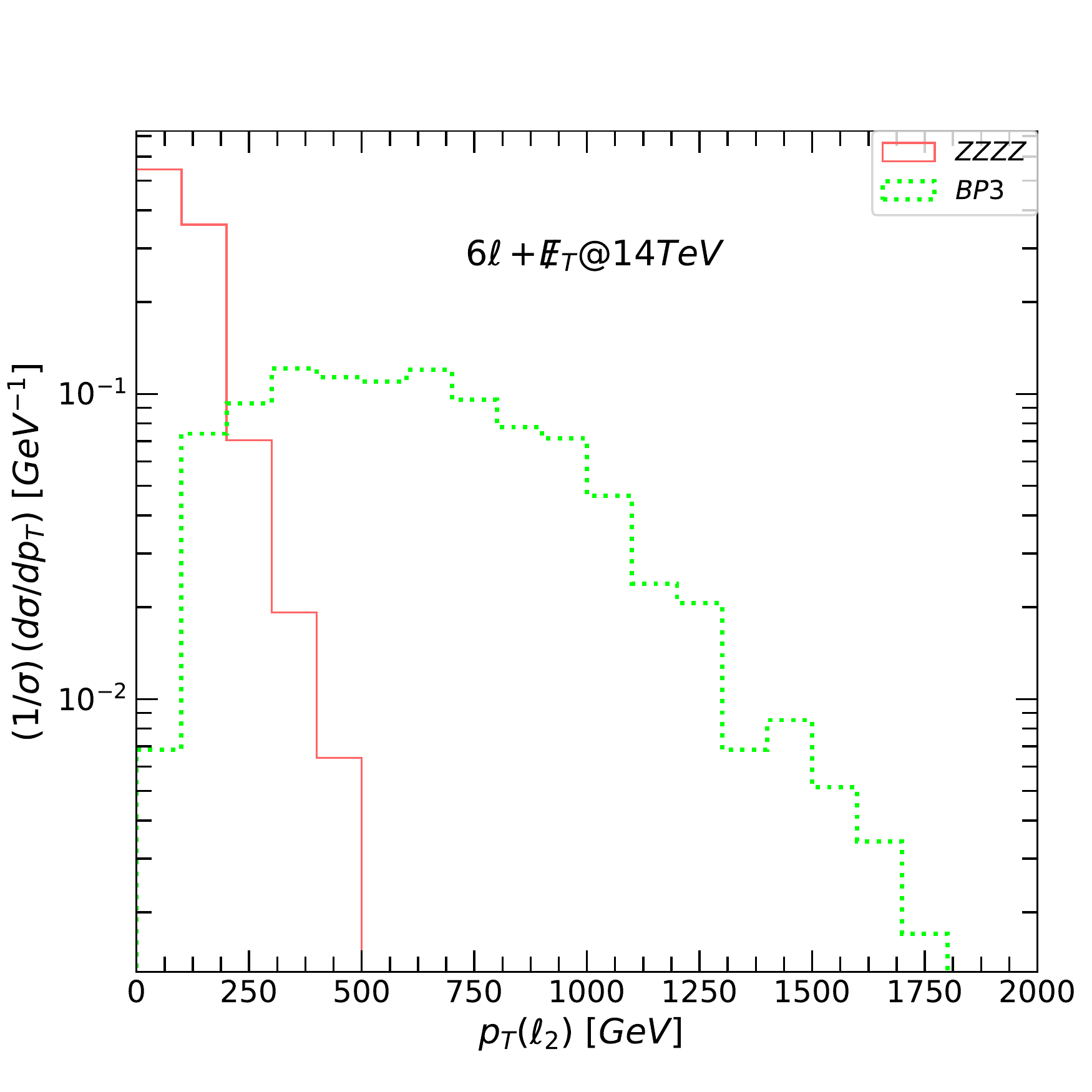} &
			\hspace*{-0.22cm}	\includegraphics[height=6.5cm,width=6.5cm]{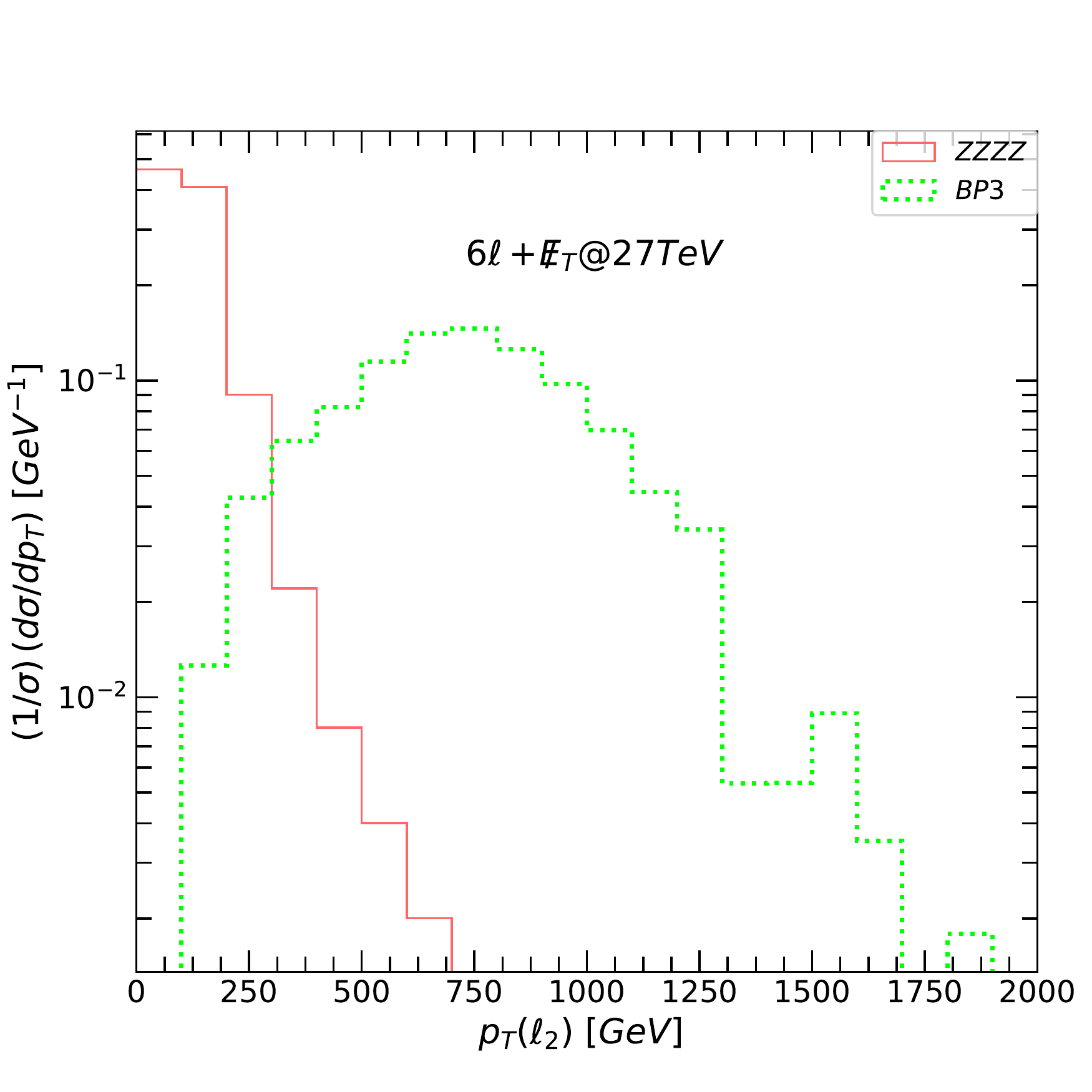} &
			\hspace*{-0.22cm}
			\includegraphics[height=6.5cm,width=6.5cm]{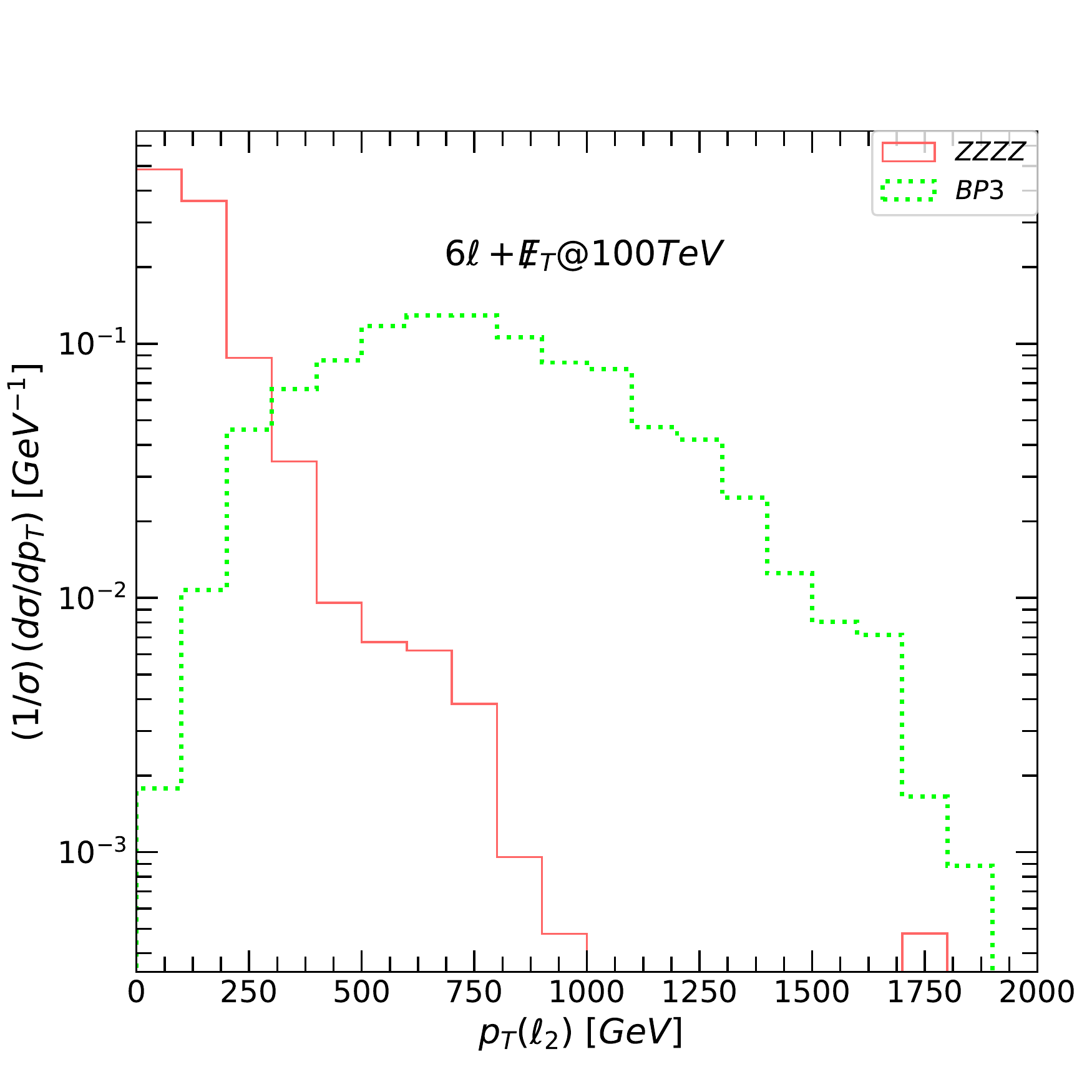} \\[-1em]
			\hspace*{-0.8cm}
			\includegraphics[height=6.5cm,width=6.5cm]{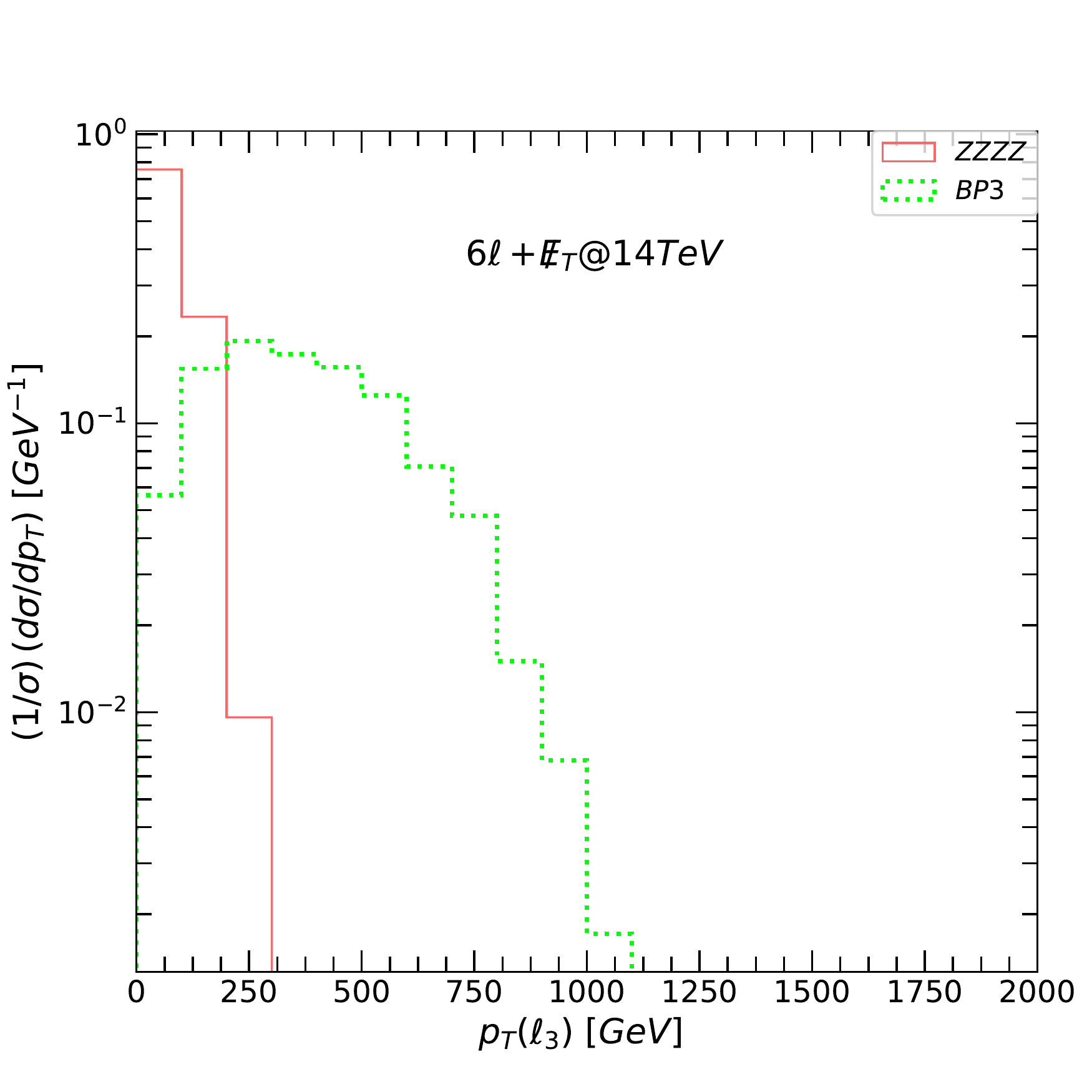} &
			\hspace*{-0.22cm}
			\includegraphics[height=6.5cm,width=6.5cm]{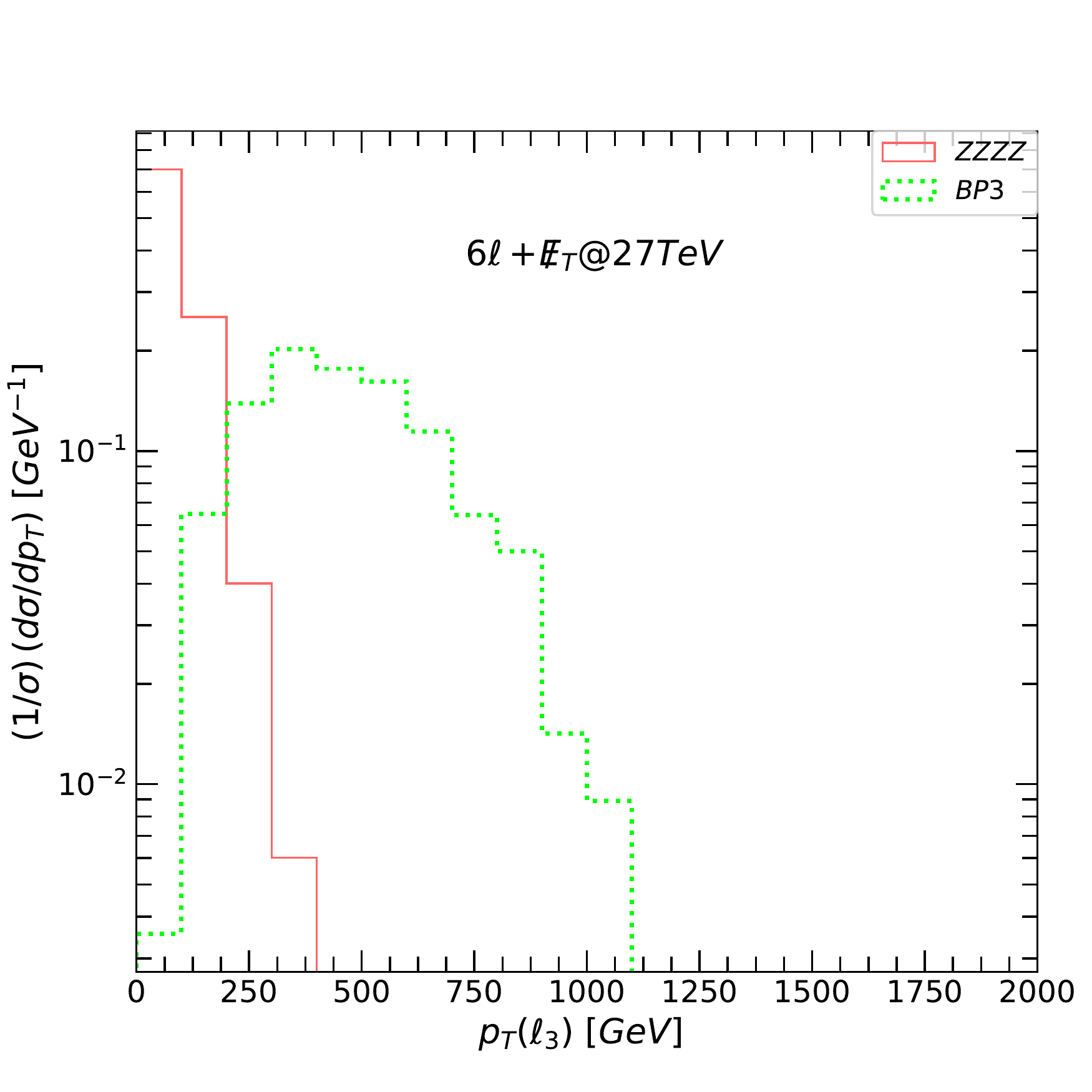} &
			\hspace*{-0.22cm}
			\includegraphics[height=6.5cm,width=6.5cm]{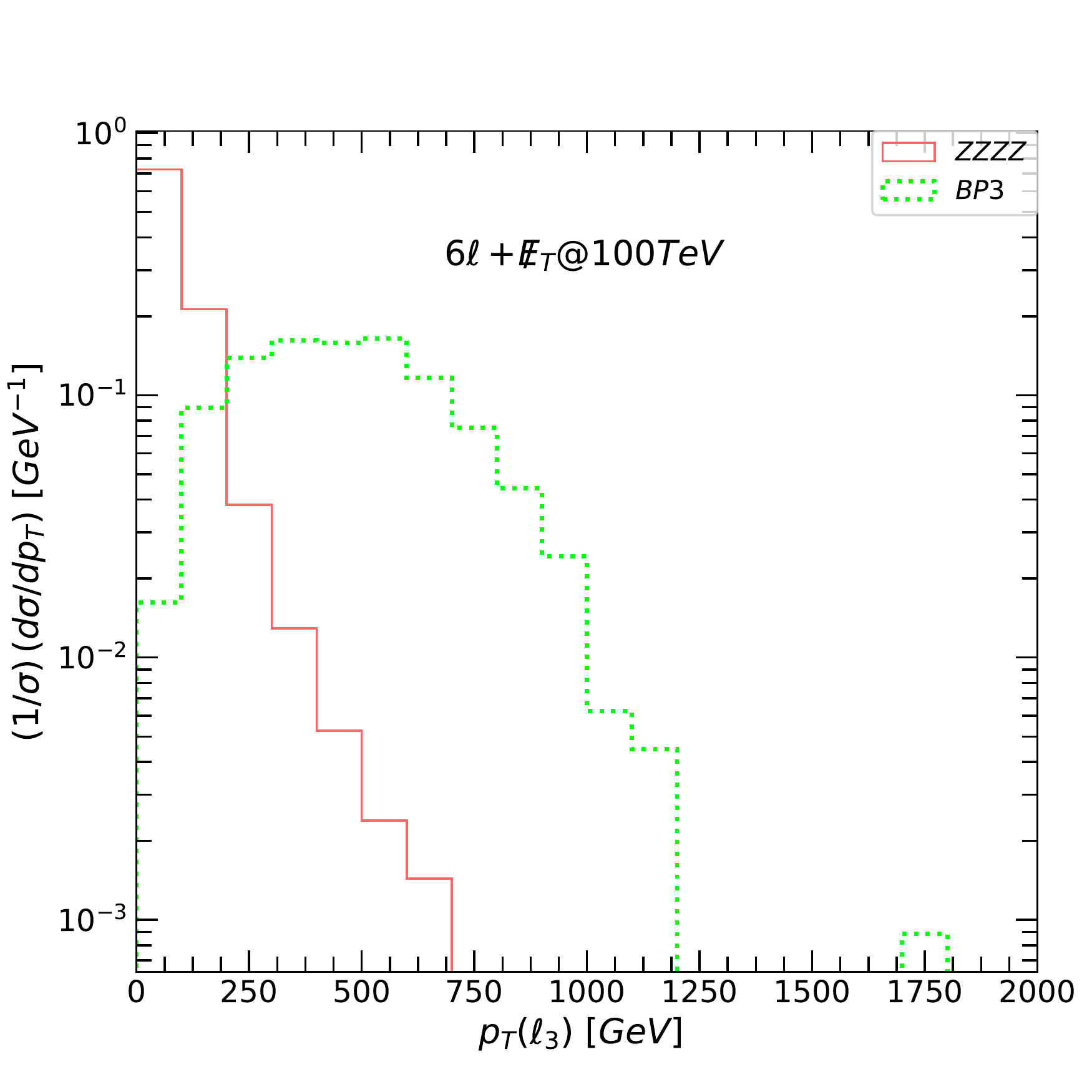} 
		\end{array}$
	\end{center}
	\vskip -0.3in
	\caption{The transverse momentum $p_T$,  for the leading lepton $\ell_1$ (top panels), next-to-leading lepton $\ell_2$ (second panels), and next-to-next to leading lepton $\ell_3$ (bottom panels)  for the signal and background for the $\sixleps$ signal. (Left-hand)  signals and backgrounds at 14 TeV,  (middle)  at 27 TeV, and (right-hand) at 100 TeV. The main backgrounds (four-bosons) are indicated in solid lines while the  dotted green line represents the signal BP3.}
	\label{fig:6lptl13}
\end{figure*}
\begingroup
\setlength{\tabcolsep}{10pt} 
\renewcommand{\arraystretch}{0.95} 
\begin{table*}[htbp]
	\caption{ Signal selection strategy and cuts imposed in the  $\sixleps$ scenario at 14, 27 and 100 TeV. We give the cross-section for background and benchmark scenarios in fb. Statistical significances $\sigma_{\mathcal A}$ and $\sigma_{\mathcal B}$ of $\sixleps$ signal are given for each energy.}\label{tab:cuts-bckgr6l}
	\setlength{\extrarowheight}{2pt}
	\begin{tabular*}{\textwidth}{@{\extracolsep{\fill}} ccc}
		\hline\hline
		$ 6\ell +\EmissT@14~\rm TeV $&\rm Background~[fb] &$ \rm BP3~[fb]$\\
		\hline
		$\rm No~Cut$&$3.12\times 10^{-7}$&$8.1\times 10^{-5}$\\ 
		$ |\eta_i|<2.5,\ \Delta R_{12} \ge 0.2$ &$2.91\times 10^{-7}$&$4.37\times 10^{-5}$\\
		$
		\left\{
		\blap{$ p_T(\ell_1)>50$ GeV\,, $ p_T(\ell_{2,3,4})>20$ \ GeV\,, \\
			$ p_T(\ell_5)>10$ GeV\,, $ p_T(\ell_6)>5$  GeV}
		\right\}
		$
		&$2.91\times 10^{-7}$&$4.37\times 10^{-5}$\\
		$ \EmissT>100~\rm GeV$ &$1.2\times 10^{-7}$&$3.8\times 10^{-5}$\\
		\hline
		$\rm Significance:\quad \sigma_{\mathcal A}$&${\cal L}=3\,{\rm ab}^{-1}$&$0.34\sigma$\\
		\hskip 2cm $\sigma_{\mathcal B}$ 
		& &$0.40\sigma$\\		
		\hline\hline
		$ 6\ell +\EmissT@27~\rm TeV $&&\\
		\hline
		$\rm No~Cut$&$8.4\times 10^{-7}$&$1.71\times 10^{-3}$\\ 
		$\ |\eta_i|<2.5,\ \Delta R_{12} \ge 0.2$ &$7.87\times 10^{-7}$&$7.03\times 10^{-4}$\\
		$
		\left\{
		\blap{$ p_T(\ell_1)>50$ GeV\,, $ p_T(\ell_{2,3,4})>20$ GeV\,, \\
			$ p_T(\ell_5)>10$ GeV\,, $ p_T(\ell_6)>5$ GeV}
		\right\}
		$
		&$7.80\times 10^{-7}$&$7.03\times 10^{-4}$\\
		$ \EmissT>100~\rm GeV$ &$3.9\times 10^{-7}$&$6.5\times 10^{-4}$\\
		\hline
		$\rm Significance:\quad \sigma_{\mathcal A}$&${\cal L}=15\,{\rm ab}^{-1}$&$3.1\sigma$\\
		\hskip 2cm $\sigma_{\mathcal B}$ 
		& &$3.7\sigma$\\		
		\hline\hline
		$6\ell +\EmissT@100~\rm TeV $&&\\
		\hline
		$\rm No~Cut$&$2.27\times 10^{-5}$&$9.52\times 10^{-2}$\\ 
		$ |\eta_i|<2.5,\ \Delta R_{12} \ge 0.2$ &$1.04\times 10^{-5}$&$4.92\times 10^{-3}$\\
		$ p_T(\ell_1)>100$ GeV&$8.27\times 10^{-6}$&$4.92\times 10^{-3}$\\
		$ p_T(\ell_2)>50$ GeV\,, $ p_T(\ell_3)>20$ GeV &$8.17\times 10^{-6}$&$4.92\times 10^{-3} $\\
		$ p_T(\ell_4)>20$ GeV&$8.07\times 10^{-6}$&$4.66\times 10^{-3}$\\
		$ p_T(\ell_5)>15$ GeV&$7.5\times 10^{-6}$&$3.65\times 10^{-3}$\\
		$ p_T(\ell_6)>5$ GeV&$6.77\times 10^{-6}$&$2.88\times 10^{-3}$\\
		$ \EmissT>100~\rm GeV$ &$3.9\times 10^{-6}$&$2.3\times 10^{-3}$\\
		\hline
		$\rm Significance:\quad \sigma_{\mathcal A}$&${\cal L}=30\,{\rm ab}^{-1}$&$8.3\sigma$\\
		\hskip 2cm $\sigma_{\mathcal B}$ 
		& &$9.7\sigma$\\		
		\hline\hline	
	\end{tabular*}
\end{table*}
\endgroup

In Fig. \ref{fig:6lptl13} we plot transverse momentum $p_T$  for the leading lepton $\ell_1$ (top panels), the next-to-leading lepton $\ell_2$ (second panels), and the next-to-next to leading lepton $\ell_3$ (bottom panels)  for the signal and background for the $\sixleps$ signal. The transverse momentum $p_T$ plots  for the fourth,  fifth,  and sixth lepton are similar and we do not show them here.  As before the left side panels indicate signals and backgrounds at 14 TeV,  the middle  at 27 TeV, and the right side at 100 TeV. The main backgrounds (four-bosons) are indicated in solid lines while the  dotted green line represent the signal BP3. As expected, the leading lepton $p_T$ distribution is most promising in distinguishing this signal from background, with other lepton $p_T$ distributions slightly less so.

Similar to the $\twoleps$ and  $\fourleps$ signals, the effects of  various cuts on cross sections are listed in Table~\ref{tab:cuts-bckgr6l}. We also show the signal significance, for both $\sigma_{\cal A}$ and $\sigma_{\cal B}$, for BP3, at proposed total integrated luminosity: for  14 TeV at ${\cal L}= 3$ ab$^{-1}$, for  27 TeV at ${\cal L}= 15$ ab$^{-1}$ and for  100 TeV at ${\cal L}= 30$ ab$^{-1}$. The signal significance can be around $3\sigma$ at 27 TeV and even greater than $ 8\sigma$ at 100 TeV.

\section{Summary and Conclusion
	\label{sec:conc}}

In this work we have analyzed the LHC, HE-LHC and FCC-hh discovery prospects of a new
neutral gauge boson ($Z^{\prime}$) through its supersymmetric decay
modes, as a promising signal for supersymmetry in an extended gauge structure. We have assumed that the $Z^\prime$ originates from an additional $U(1)^\prime$ symmetry, and we decoupled its mass scale from that of supersymmetry breaking, assumed to be the same as the scale of breaking $U(1)^\prime$. This allows the $Z^\prime$ boson to be heavy, as indicated by lower limits $M_{Z^\prime} \gsim 4$ TeV from the experimental searches, while  the electroweakinos remain light. For this, we relied on the secluded $U(1)^\prime$ model, where three additional singlet superfields are added to the model. Unlike the VEV of the singlet scalar field which breaks $U(1)^\prime$ and affects the mass of supersymmetric particles, the VEVs of the additional scalars enter only in the expression for the $Z^\prime$ mass. 

This scenario provides a fertile ground for analyzing $Z^\prime$ decays into chargino, neutralino, slepton and sneutrino pairs. As LHC is particularly sensitive to events containing one or more leptons, we looked for production of $Z^\prime$ followed by decays into multileptons plus missing energy. For this, we devised three benchmarks (BP1, BP2 and BP3) where branching ratios into some supersymmetric particles are enhanced. For instance in BP1, the decay into sneutrinos and chargino is enhanced, in BP2 it is the decay into chargino and neutralino pairs, while for BP3, the decay into right sleptons and lightest staus is important. The benchmarks are chosen also so they satisfy collider and relic density constraints.

We proceed by analyzing the observability of the signals at $\sqrt{s}$ = 14, 27 and 100 TeV, looking separately at $\twoleps$, $\fourleps$ and $\sixleps$ signals. Throughout our benchmarks, the ratio $\Gamma_{Z^\prime}/M_{Z^\prime}$ remains under 10\%, so we can treat 
$Z^{\prime}$ as a narrow resonance.  For each signal, we perform a Monte Carlo simulation of the signal and background, and devise cuts to disproportionately suppress the latter.  We present  the results before and after the cuts,  and calculate the significance in two different ways. 

Overall, our findings indicate that the probability of observing  $Z^{\prime}$ through supersymmetric decays at 14 TeV is not good, even at high total integrated luminosity ${\cal L}=3$ ab$^{-1}$. This occurs across all $\twoleps$, $\fourleps$ and $\sixleps$ signals and for all benchmarks. This confirms past analyses for $\twoleps$ \cite{Frank:2012ne,CidVidal:2018eel}, which indicated that, unless $Z^\prime$ is leptophobic, and thus much lighter, the signal significance is small. However, that is not necessarily so at 27 or 100 TeV and across all signals. At 27 TeV, benchmark BP3 gives a significance well above 5$\sigma$ in $\twoleps$ signal. While for the $\fourleps$ signal, we obtain significances of 3-4$\sigma$ for both BP1 and BP3, and much higher at 100 TeV. For the $\sixleps$ signal, only BP3, were the $Z^\prime$ decay into right sleptons is important, gives any significant contributions. The significance at 27 TeV with total integrated luminosity ${\cal L}=15$ ab$^{-1}$ is 3-4$\sigma$, and can reach 8-9$\sigma$ at 100 TeV with total integrated luminosity ${\cal L}=30$ ab$^{-1}$.

Of course, analyses at 27 TeV are plagued by uncertainties, and those at 100 TeV can be interpreted as merely estimates. However, our analysis shows that HE/HL-LHC and FCC-hh can be promising grounds for observing consequences of both supersymmetry and extended gauge symmetry, of which an additional neutral gauge boson is one of the simplest examples. A heavy $Z^\prime$ boson, accompanied by a light electroweakino sector would also be indicative of a $U(1)^\prime$ model with a secluded sector, as this set-up facilitates the splitting of the two scales.

\begin{acknowledgements}
The work of M.F. has been partly supported by NSERC through grant number SAP105354.  M.F. is grateful to a TUBITAK-BIDEB 2221 grant which enabled this collaboration, even if the spread of COVID-19 prevented a visit to METU.
\end{acknowledgements}


\bibliographystyle{spphys}       
\bibliography{Zsecluded}   

\end{document}